\documentclass[twocolumn,trackchanges]{aastex63}

\usepackage{comment}

\newcommand{\quotes}[1]{``#1''}

\shorttitle{Spiral Disk at center of Perseus BCG}
\shortauthors{Yeung, Ohyama, \& Lim}

\graphicspath{{./}{}}
\received{2021 September 8}
\revised{2022 January 11}
\accepted{2022 January 21}

\begin{document}

\title{Recent Formation of a Spiral Disk Hosting Progenitor Globular Clusters at the center of the \\ Perseus Brightest Cluster Galaxy: I. Spiral Disk}

\author[0000-0002-5697-0001]{Michael C. H. Yeung}
\altaffiliation{Present address: Max-Planck-Institut für extraterrestrische Physik, Giessenbachstraße, 85748 Garching, Germany}
\affiliation{Department of Physics, The University of Hong Kong, Pokfulam Road, Hong Kong}
\email{myeung@mpe.mpg.de}

\author[0000-0001-9490-3582]{Youichi Ohyama}
\affiliation{Academia Sinica, Institute of Astronomy and Astrophysics, 11F of Astronomy-Mathematics Building, No.1, Section\,4, Roosevelt Rd., Taipei 10617, Taiwan, Republic of China}

\author[0000-0003-4220-2404]{Jeremy Lim}
\affiliation{Department of Physics, The University of Hong Kong, Pokfulam Road, Hong Kong}
\submitjournal{\apj}

\begin{abstract}
We address the nature and origin of a spiral disk at the center of NGC\,1275, the giant elliptical galaxy at the center of the Perseus cluster, that spans a radius of $\sim$5\,kpc.
By comparing stellar absorption lines measured in long-slit optical spectra with synthetic spectra for single stellar populations, we find that fitting of these lines requires two stellar populations: (i) a very young population that peaks in radial velocity at $\pm 250 {\rm \, km \, s^{-1}}$ of the systemic velocity within a radius of $\sim$720\,pc of the nucleus, a $1\sigma$ velocity dispersion significantly lower than $140 {\rm \, km \, s^{-1}}$, and an age of $0.15 \pm 0.05$\,Gyr; and (ii) a very old population having a constant radial velocity with a radius corresponding to the systemic velocity, a much broader velocity dispersion of $\sim$$250 {\rm \, km \, s^{-1}}$, and an age of around 10\,Gyr.  We attribute the former to a post-starburst population associated with the spiral disk, and the latter to the main stellar body of NGC\,1275 along the same sight line.  If the spiral disk is the remnant of a cannibalized galaxy, then its progenitor would have had to retain an enormous amount of gas in the face of intensive ram-pressure stripping so as to form a total initial mass in stars of $\sim$$3 \times 10^9 \, M_\sun$.  More likely, the central spiral originally comprised a gaseous body accreted over the distant past from a residual cooling flow, before experiencing a starburst $\sim$0.15\,Gyr ago to form its stellar body.  
\end{abstract}



\section{Introduction}\label{sec:intro}

\begin{figure*}
\edit1{\includegraphics[width=\textwidth]{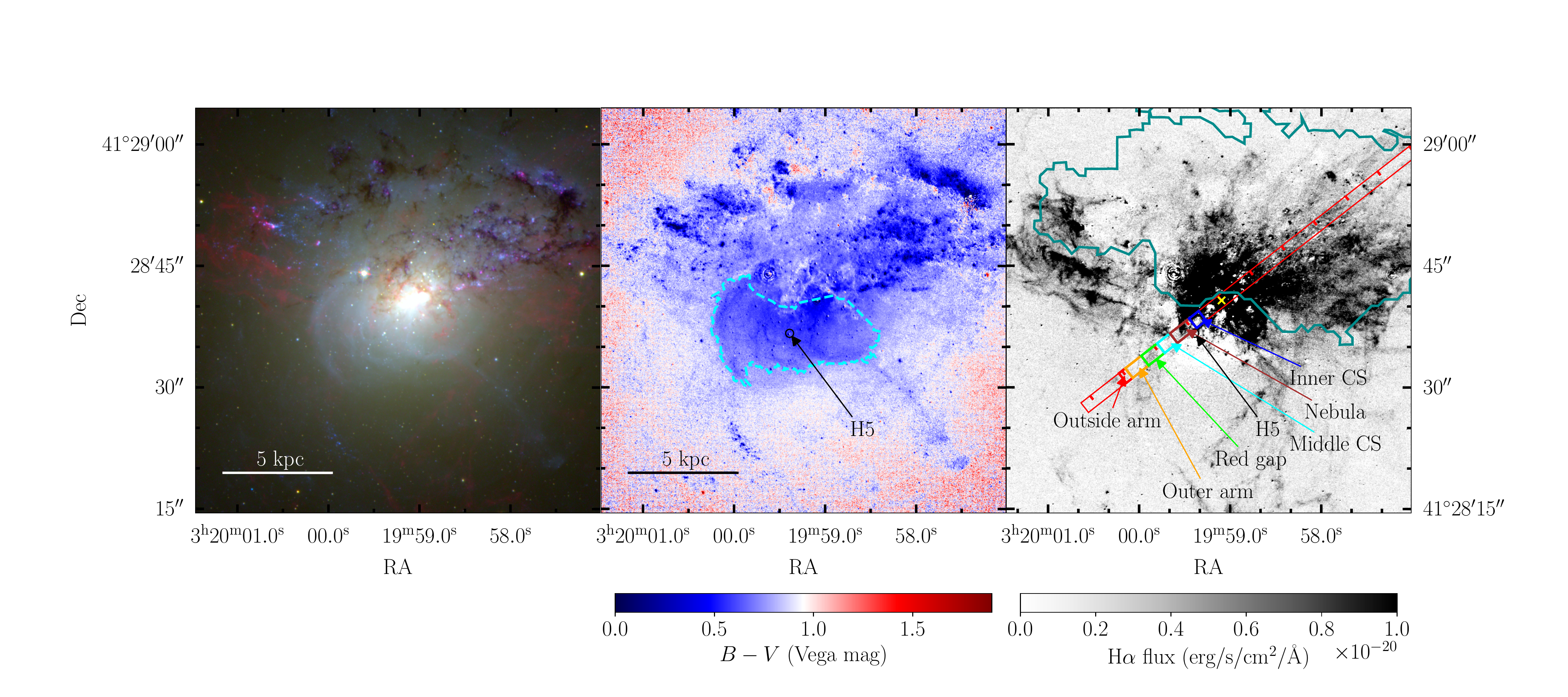}}
\caption{$BVR$ (left panel), $B-V$ (middle panel), and $\rm H\alpha + [N\,II]$ (right panel) images of the central region of NGC\,1275 constructed from images in the $B$ (F435W),$V$ (F550M), and $R$ (F625W) bands taken with the Hubble Space Telescope/Advanced Camera for Surveys.  The central spiral, comprising spiral arms superposed on an approximately circular disk, stands out in sharp relief owing to its higher surface brightness (left panel) and bluer colors (middle panel) than its surroundings.  Filamentary dust seen in silhouette on the northern side of NGC\,1275 (left panel) belongs to the High Velocity System (HVS), a distorted spiral galaxy falling toward NGC\,1275; the outer extent of this galaxy as detected in $\rm H\alpha + [N\,II]$ \citep{HVS} is indicated by the teal contour (right panel).  The dashed cyan contour (middle panel) defines the visibly dust-free region of the central spiral, used to compute its initial stellar mass as described in Section\,\ref{minor merger}.  The $\rm H\alpha + [N\,II]$ image (right panel) is dominated by the emission-line nebula associated with NGC\,1275, although there is a relatively small contribution from ionized gas associated with the HVS.  The slit of width $1\farcs5$ employed in the spectroscopic observation is indicated in the right panel, with a tick mark every 5\arcsec: it crosses the active galactic nucleus (AGN) in NGC\,1275, as indicated by a yellow cross, at a position angle of 128\degr. The locations of the six subregions selected for analyses are shown by rectangles having different colors and labeled accordingly (right panel); their radial positions with respect to the center of NGC\,1275, marked by its AGN, are listed in Table\,\ref{tab:region}.  The star cluster H5, lying just $0\farcs4$ outside the slit, is circled and labeled in the middle and right panels.}
\label{fig:HST_img}
\end{figure*}

NGC\,1275, the Brightest Cluster Galaxy (BCG) of the Perseus cluster, must be among the most intensively scrutinized and yet remains one the most enigmatic galaxies in the sky. Its most prominent peculiarity at optical wavelengths is a complex and filamentary emission-line nebula, the innermost portion of which can be seen in Figure\,\ref{fig:HST_img} (right panel), that extends to over 50\,kpc from the center of the galaxy.  This at least partially atomic nebula has an ionized counterpart in X-rays \citep{Fabian2006,Sanders}, a molecular counterpart in the rotational-vibrational lines of molecular hydrogen \citep{Jeremy2012}, as well as in the rotation lines of CO \citep{Salome06,SMA_Jeremy,Salome08b,Salome08a,Ho_jeremy,salome2011}. Although likely originating from cooling of the intracluster X-ray emitting gas that is in large part reheated by the action of an active galactic nucleus (AGN) in NGC\,1275 (henceforth, the gas component that cools to form the nebula is referred to as a residual cooling flow), the excitation, and hence physical properties, of this nebula is not fully understood, nor the physical processes that led to its complex structure.  Several thousand compact and massive star clusters -- progenitor globular clusters -- have been found associated with, especially, the outer part of the nebula: these stars clusters have formed at an essentially constant rate with time for, at least, the past 1\,Gyr \citep{Jeremy}.

The second most obvious peculiarity of NGC\,1275 at optical wavelengths is a complex network of dust that is elongated roughly east-west and lying northeast of the center of the galaxy, as shown in Figure\,\ref{fig:HST_img} (left and middle panels).  This dust network has a counterpart in optical emission lines, which is redshifted by $3000 \rm \, km \, s^{-1}$ with respect to NGC\,1275 and hence is known as the High Velocity System (HVS).  The HVS has been established to constitute an infalling spiral galaxy that is suffering intense ram-pressure stripping in what is likely to be its first passage through the Perseus cluster, taking it at high speeds close to the cluster center and placing it just in front of NGC\,1275 at the present time \citep{HVS}.

The third peculiarity of NGC\,1275 at optical wavelengths was first noticed in images taken at high angular resolutions with the Hubble Space Telescope (HST).  Discovered by \citet{star_clus_ID} and clearly seen for the first time in images taken by \citet{Carlson}, the center of NGC\,1275 hosts spiral arms superposed on a roughly circular disk, as can be seen in Figure\,\ref{fig:HST_img} (left and middle panels).  These features, henceforth referred to as the central spiral, stands out in sharp relief owing to its higher surface brightness and bluer color than its surroundings.  The central spiral appears to be closely centered on the nucleus of NGC\,1275, and has a radius of $\sim$15\arcsec\ or about 5\,kpc.  Numerous semicircular arcs can be seen around the central spiral visible out to a radius of $\sim$11\,kpc, as pointed out by \citet[][see their Figure\,8]{conselice} and even more clearly by \citet[][see their Figure\,19]{penny}.  Both \citet{conselice} and \citet{penny} attributed the semicircular arcs to partial shells tracing past interactions or mergers.  Thus motivated in part, \citet{conselice} attributed the central spiral to a merger remnant, and its relatively blue color to a recent episode of star formation.

The nature of the central spiral is the subject of this manuscript.  In our work, we use archival spectroscopic data at optical wavelengths to deduce the kinematics and ages of the stellar population associated with the central spiral for the first time.  In Section\,\ref{sec:archival}, we describe the data used in our work and the manner by which we processed the data to compare with models incorporating stellar-population synthesis.  The method used to derive and the results for the stellar kinematics of the central spiral are reported in Section\,\ref{sec:stellar kinematics}.  The methods used to derive and the results for, the age of its stellar population is reported in Section\,\ref{sec:stellar population}.  In Section\,\ref{sec:discussion}, we use the results obtained to provide an interpretation for the nature of the central spiral, which as we shall demonstrate faces considerable difficulties if attributed to the remnant of a cannibalized galaxy.  Finally, in Section\,\ref{sec:conclusion}, we provide a concise summary of the most important points of our work.

An especially crucial parameter for the purposes of our study is the systemic velocity of NGC\,1275.  We regard the most reliable and precise measurement of this parameter to have been made by \citet{Riffel}.  Based on CO bandhead features in the near-infrared (around $2.3\,\mu$m) within a radius of 1\farcs25 of the nucleus, they report a (heliocentric) systemic velocity for NGC\,1275 of $5284 \pm 21$\,km\,s$^{-1}$.  Thus, throughout this paper, we adopt a nominal systemic velocity of $5284 \rm \, km \, s^{-1}$ for NGC\,1275, a value that we found to be consistent with the radial velocities of selected stellar absorption lines measured in our work.  For comparison, using Ca triplet features at wavelengths around $8600$\,{\AA}, \citet{nelson} derived a systemic velocity of $5253 \pm 44$\,km\,s$^{-1}$ toward the nucleus of NGC\,1275.  Perhaps the next most reliable measurement is that made by \citet{scharwchter}, who used an integral field unit aided by adaptive optics to map a $\rm 1\,kpc \times 1\,kpc$ region around the center of NGC\,1275.  They detected a rotating disk seen in both the H$_2$\,1--0\,S(1)\, line at 2.122\,$\mu$m and the $[\mathrm{Fe\,II}]$ line at 1.644\,$\mu$m within a radius of $\sim$50\,pc centered on the nucleus of the galaxy.  They quote a redshift of $z = 0.017559$, corresponding to a systemic velocity of $5266 \rm \, km \, s^{-1}$, for NGC\,1275, although it is not clear to us whether this value was adopted or derived based on a best-fit model for the disk.

As in our previous studies, adopting a redshift of $z = 0.01756$ for NGC\,1275 and $H_0 = 70 \rm \, km \, s^{-1} \, Mpc^{-1}$, the distance to this galaxy is 74\,Mpc.  At this distance, $1\arcsec = 360 \rm \, pc$.

\section{Archival Data and Data Reduction}\label{sec:archival}

\subsection{Archival Data}\label{subsec:obs}

We retrieved, from the Keck Observatory Archive (KOA), data taken with the Low Resolution Imaging Spectrometer \citep[LRIS;][]{LRIS} on the Keck I telescope comprising long-slit optical spectroscopy crossing the nucleus of NGC\,1275 (proposal ID: C07L). The observation was conducted on 2004 February 27, during which the LRIS was configured in the dual-beam mode with a slit width of $1\farcs5$.  Blue and red channels were formed by feeding light from NGC\,1275 through the dichroic beam splitter D560, which reflects and transmits light below and above 5696\,{\AA}.  The bluer light was directed to the 600/4000 grism to form spectra spanning 3300--5880\,{\AA}, and the redder light to the 831/8200 grating to form spectra spanning 6830--8730\,{\AA}.  The spectral resolution in the blue channel is 6.1\,{\AA} at the atmospheric [OI]$\lambda$5577\,{\AA} line at full-width half-maximum (FWHM), corresponding to a velocity resolution of 330\,km\,s$^{-1}$; the spectral resolution in the red channel is 4.6\,{\AA} at 8500\,{\AA}, corresponding to a velocity resolution of 160\,km\,s$^{-1}$.  The CCD pixels were not binned on-chip.  

A point-like source, corresponding to the AGN in NGC\,1275, was acquired on the slit at a position angle (PA) of $128{\degr}$, as shown in Figure\,\ref{fig:HST_img}.  The slit cuts through the central spiral on the southeastern side, and the HVS on the northwestern side, of the nucleus of NGC\,1275.  It passes close to---about $0\farcs4$ from---a blue star cluster referred to as H5 previously studied by \citet{brodie} through optical slit spectroscopy also using the LRIS; in Section\,\ref{subsec:kinematics:velocity_fields}, we compare the radial velocity inferred by \citet{brodie} toward this star cluster with that we infer for a position close to this star cluster.  Seven exposures, each of duration 600\,s, were taken while performing telescope dithering along the slit between each exposure.  Atmospheric seeing based on the standard star data was about $1\farcs0$ at FWHM at a wavelength of 4000\,{\AA}, similar to the measured FWHM of the AGN at the same wavelength.

The instrument internal flats were taken immediately after the NGC\,1275 exposures at the same telescope position as the last galaxy exposure.  Next, a spectroscopic standard star, BD+75\,325, was observed at the same slit PA but at larger air mass (1.8) without changing the instrument configuration.  The atmospheric dispersion corrector (ADC) was not available at the time of the observation, leading to a problem in the spectral-response calibration owing to significant loss of bluer light from the star (see Section\,\ref{subsec:calibration_issue} for how we dealt with this problem).  Arc lamp (Hg, Cd, Ne, and Ar) images were taken immediately after the standard star spectra at the same instrument and telescope orientation.

\begin{deluxetable*}{cccccc}[htb]
\tablecaption{Radial locations of six subregions defined along and beyond the central spiral.\label{tab:region}}
\tablehead{
\colhead{Inner} & \colhead{Nebula} & \colhead{Middle} & \colhead{Red} & \colhead{Outer} & \colhead{Outside}\\
\colhead{Central spiral} & \colhead{} &\colhead{Central spiral} & \colhead{Gap} & \colhead{Arm} & \colhead{Outer arm}
}
\startdata
($-3\farcs0$)--($-4\farcs5$) & ($-4\farcs5$)--($-7\farcs5$) & ($-7\farcs5$)--($-9\farcs5$) & ($-9\farcs5$)--($-12\farcs0$) & ($-12\farcs0$)--($-14\farcs5$) & ($-14\farcs5$)--($-15\farcs5$)\\
\enddata
\tablecomments{Locations measured with respect to the center of NGC\,1275 as denoted by its active galactic nucleus.}
\end{deluxetable*}

The work reported here is confined to the spectrum taken in the blue channel, covering 3700--5520\,{\AA}, that as we shall show reveals a post-starburst stellar population in the central spiral, the most prominent features from which are Balmer absorption lines.  In particular, we focus on the higher-order Balmer absorption lines (primarily $\rm H8 = H\zeta$\footnote{H$\zeta$ is called \quotes{H6} in \citet{Dokkum,Demarco,Nantais}.}, but including also $\rm H9 = H\eta$ and H$\epsilon$) because the lower-order lines are strongly contaminated by the corresponding emission lines from the filamentary nebula in NGC\,1275 \citep[see, e.g.,][for examplar spectra of the nebular emission lines]{conselice,Hatch06,Fabian,Gendron}.  We also analyze a number of metal absorption features (\ion{Ca}{2} H and K, G4300, Mg\,$b$, Fe$\lambda$5270\,{\AA}, and Fe$\lambda$5335\,{\AA}) in the blue channel to further characterize the post-starburst stellar population in the central spiral, as well as to characterize the underlying old stellar population associated with NGC\,1275.

\begin{figure*}
\centering
\includegraphics[width=\textwidth]{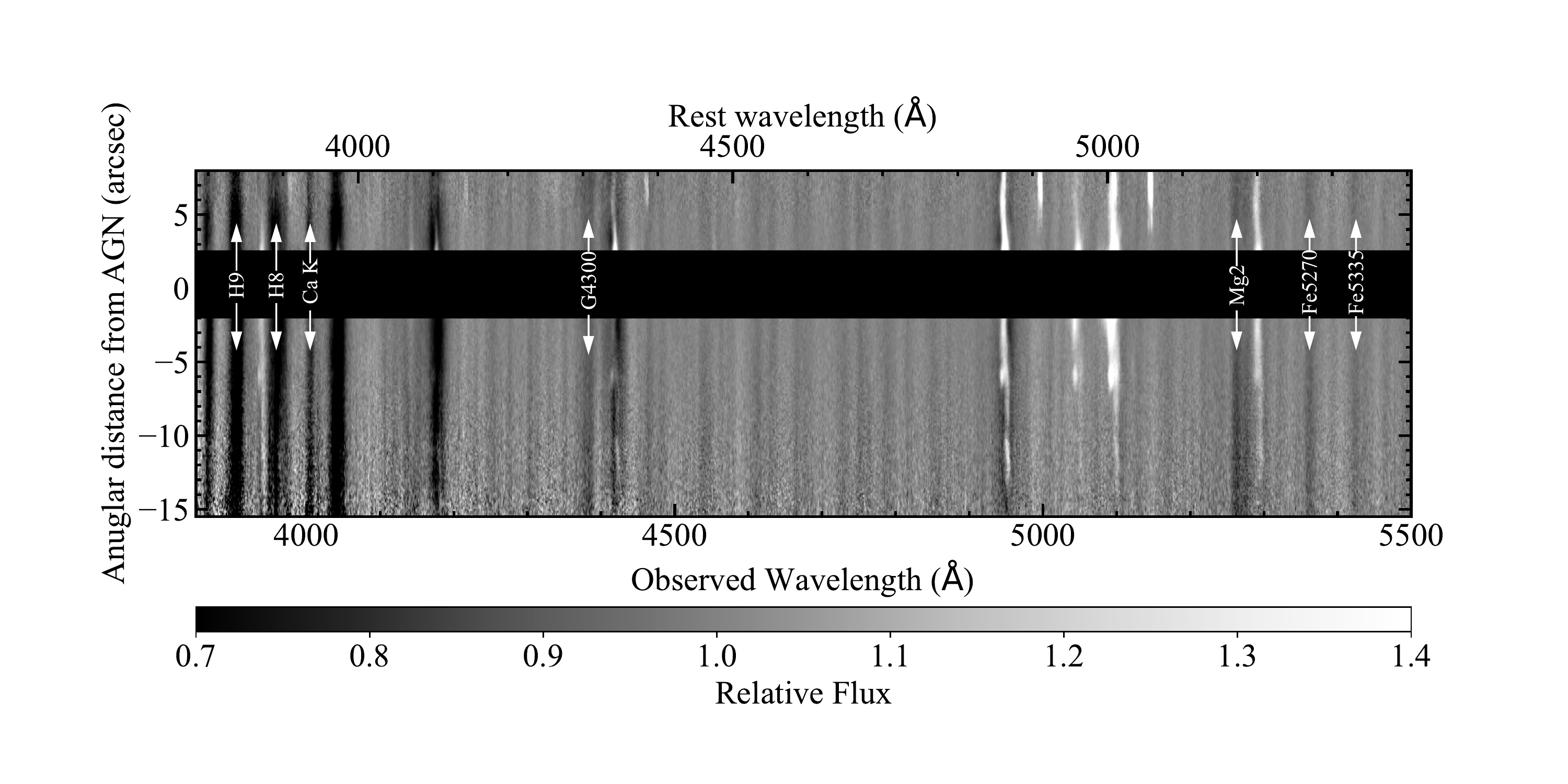}
\\
\vspace{-0.6cm}
\caption{Continuum-normalized spectra derived from long-slit spectroscopy toward the central region of NGC\,1275 as depicted in Figure\,\ref{fig:HST_img}.  The vertical axis denotes slit position, such that the southeastern side has negative positions and the northwestern side positive positions. The upper horizontal axis is the rest wavelength at the systemic velocity of NGC\,1275 ($5284 \rm \, km \, s^{-1}$), and lower horizontal axis is observed wavelength.  At each slit position, the stellar continuum was measured over selected wavelength ranges free of detectable spectral lines, and then extrapolated across the spectral lines so as to normalize the stellar continuum at all wavelengths.  Darker shades correspond to absorption and lighter shades to emission, with relative strengths as depicted by the bar at the bottom of the spectra. The stellar absorption features that we analyzed are indicated by arrows and labeled.  A black strip masks the central region, excluded from our analysis, contaminated by bright continuum and line emission from the active galactic nucleus in NGC\,1275.  On the northwestern side of the nucleus (positive positions), emission lines from the High Velocity System that is redshifted by $3000 \rm \, km \, s^{-1}$ with respect to NGC\,1275 can be seen.}
\label{fig:spec_2d}
\end{figure*}

At this point, and for later reference, we define the following subregions along the slit as indicated in Figure\,\ref{fig:HST_img} and listed in Table\,\ref{tab:region}.  Positions are measured with respect to the AGN in NGC\,1275, such that positive positions lie to the northwest and negative positions to the southeast of the center.  The \quotes{outer arm} refers to the most prominent blue arm at $-13\arcsec$,  
which defines the visible outer boundary of the central spiral.
Beyond the \quotes{outer arm} and extending out to $-15\farcs5$, beyond which the signal-to-noise ratio (S/N) of the spectra becomes too low for our purposes, is the region referred to as \quotes{outside the outer arm}.
Immediately inner of the outer arm is the \quotes{red gap}, defined based on its locally redder $B-V$ color (see middle panel of Figure\,\ref{fig:HST_img} and lower-rightmost panel of Figure\,\ref{fig:pos_index_color}), located at $-11\arcsec$.  Inner of the red gap, and at progressively closer distances to the center, are the \quotes{middle central spiral}, \quotes{nebula}, and \quotes{inner central spiral}.  The \quotes{nebula} region at $-6\arcsec$ coincides with the tip of a nebular filament as judged by the relatively high H$\alpha/B$ flux ratio at this location (see more details on diagnostics and the degree of nebular contamination along the slit in Appendix\,\ref{appendix:nebula_contami}).

\subsection{Data Reduction}\label{subsec:reduction}

We performed standard image processing for long-slit optical spectroscopy utilizing CCD detectors, including bias (overscan) subtraction, flat-fielding, wavelength calibration using both arc and sky lines, and frame stacking.  Preliminary calibration of the wavelength scale was carried out using the arc images, before correcting for the zero-point offset caused by instrument flexure between the target object and the arc exposures using bright sky lines recorded on the target exposures.  The estimated rms accuracy in wavelength calibration from fitting the arc lines is 0.13\,{\AA}, corresponding to 10\,km\,s$^{-1}$ at 4000\,{\AA}.  After flat-fielding, we subtracted the sky as measured at the outermost region of the slit on the northwestern side of the center.
%
%
The uncertainty in the flux measurements was estimated based on the standard deviation in fluxes among seven individual exposure frames.  

\subsection{Spectral Calibration Difficulties and Mitigation}\label{subsec:calibration_issue}

Owing to loss of blue light from the standard star as the ADC was not available during the observation, along with the appearance of an optical ghost from the red channel, spectral-response calibration of the blue channel proved impossible.
To extract reliable information from such spectra, we (i) limited our study to portions of the spectra where the optical ghost is negligible, and (ii) adopted an analysis technique that does not require spectral-response calibration.  Below we briefly describe the issues encountered and analysis technique adopted, with more details given in Appendix\,\ref{appendix:calibration_issue}.

Contamination in the blue channel by the ghost spectrum from the red channel is especially problematic when the ghost spectrum arises from the bright AGN of NGC\,1275.  Fortunately, the slit position in the blue channel thus contaminated is located $\simeq 8\arcsec$ northwest of the center, where the visibility of the central spiral is affected anyway by the HVS.  At a given slit position, the ratio in counts between the ghost spectrum and the actual blue spectrum increases toward shorter wavelengths, and becomes especially severe at wavelengths shortward of 4000\,{\AA}.  The shortest wavelength spectral line analyzed was therefore H9, which at the redshift of NGC\,1275 lies at a wavelength of 3903\,{\AA}.  In addition to the steps taken above, to further mitigate the effects of the ghost spectra we derived radial velocities for the central spiral through cross-correlation of the measured spectral lines with stellar spectral templates as described in Section\,\ref{subsec:kinematics:crosscorrelation}.

As we were not able to perform accurate spectral-response calibration, we decided to work only on continuum-normalized spectra for the cross-correlation analysis as well as for measurements of line indices.  To this end, we carefully selected wavelength ranges free of spectral features to define the continuum.  Figure\,\ref{fig:spec_2d} shows the spectra after continuum normalization, following which the continuum break index at 4000\,{\AA}, $D$(4000) \citep{Bruzual}, could no longer be used as a diagnostic of the stellar population(s).  On the other hand, because the cross-correlation relies only on localized stellar spectral features, the radial velocities thus derived are not sensitive to any spectral-response calibration.  To interpret line indices, we compared the measured indices in the normalized spectra with model spectra subjected to the same continuum normalization as described in Appendix\,\ref{subsubsec:population:indices:measurement}.

\begin{figure*}
\centering
\includegraphics[width=\textwidth]{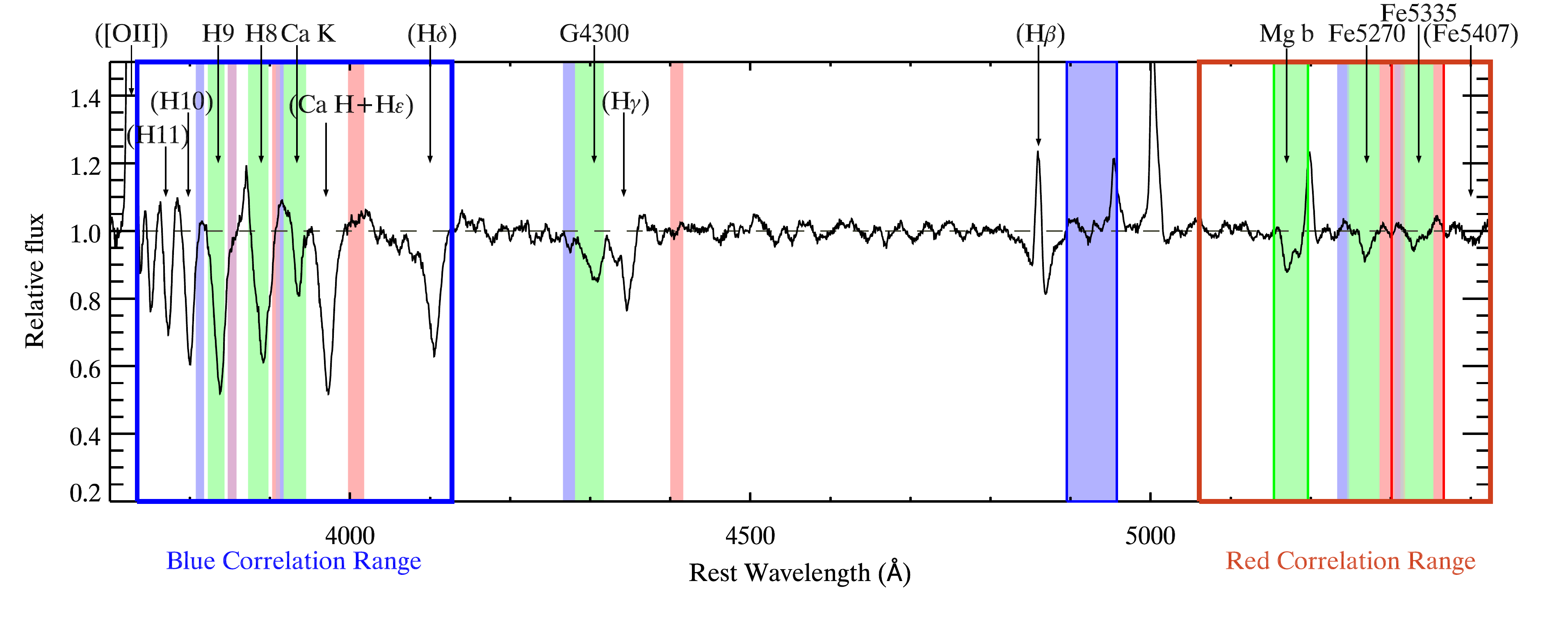}
\\
\vspace{-0.3cm}
\caption{Examplar continuum-normalized spectrum at the inner central spiral to depict the wavelength ranges used for the cross-correlation analysis (to derive stellar radial velocities) and for measuring line indices (to derive stellar ages).  The blue correlation range is encompassed by the blue rectangle, and the red correlation range by the red rectangle, both of which lie in the blue channel of the Low Resolution Imaging Spectrometer.  Spectral lines labeled within brackets either may be contaminated by Balmer emission lines from the nebula associated with NGC\,1275, those for which flanking bandpasses to determine the pseudo-continuum for deriving line indices are difficult to define, or are too weak to provide meaningful measures (e.g., Fe\,5407).  For the remaining spectral lines or features labeled, line indices (see Appendix\,\ref{subsubsec:population:indices:measurement}) were computed using line bandpasses, as indicated by the green strips, each of which is bracketed by a blue strip at a shorter wavelength and a red strip at a longer wavelength that together define the pseudo-continuum; magenta strips indicate where a blue and a red strip overlap.  The Mg\,$b$ bandpass is indicated by the green strip having a thick green outline, and its flanking bandpass at a shorter wavelength indicated by the blue strip having a thick blue outline; the corresponding flanking bandpass at a longer wavelength is indicated by a unfilled red box (thinner line than red correlation range) to avoid overlap with the Fe bandpasses.
\label{fig:index_window}}
\end{figure*}

\section{Stellar Kinematics}\label{sec:stellar kinematics}
In this section, we show that the line centers and linewidths of the absorption lines selected for study cannot be explained by just a single stellar population.  Rather, a combination of (at least) two stellar populations having different kinematics are required.

\subsection{Cross-correlation Analysis}\label{subsec:kinematics:crosscorrelation}

Following continuum normalization, we selected two wavelength ranges within the blue channel for cross-correlation analysis, as illustrated in Figure\,\ref{fig:index_window} for the measured spectrum averaged over the inner central spiral.  These wavelength ranges span (i) 3850--4200\,{\AA} to include the Balmer H8--H11, Ca\,K, blended Ca\,H and H$\epsilon$, and H$\delta$ lines, henceforth referred to as the blue correlation range; and (ii) 5150--5520\,{\AA} to include the Mg\,$b$, Fe5270, Fe5335, and Fe5407 lines, henceforth referred to as the red correlation range.  
The selected wavelength ranges exclude [$\mathrm{O\,II}$]$\lambda$3727\,{\AA}, [$\mathrm{O\,III}$]$\lambda$4959\,{\AA}, and [$\mathrm{O\,III}$]$\lambda$5007\,{\AA}, emission lines that could potentially arise (at least in part) from the nebula associated with NGC\,1275.  In addition, the H$\gamma$ and lower-order Balmer absorption lines are excluded owing to contamination by the corresponding emission lines from the nebula.  Within the two wavelength ranges considered, weaker emission lines potentially produced by the nebula, namely [$\mathrm{Ne\,III}$]$\lambda3869$\,{\AA} (near H8) and [$\mathrm{N\,I}$]$\lambda5200$\,{\AA} (near Mg\,$b$), were removed by masking out regions of the spectra encompassing these lines.  The division of the blue channel into the two selected wavelength ranges proved convenient because, as it turned out, the spectra toward the central spiral are produced by two distinct stellar populations -- having different radial velocities -- that each dominates over one of the two selected wavelength ranges: (i) a relatively young (post-starburst) population, associated with the central spiral, that is preferentially traced by the higher-order Balmer absorption lines in the blue correlation range; and (ii) a relatively old population, associated with the main body of NGC\,1275, that is preferentially traced by the metal absorption lines in the red correlation range.

We adopted the MILES \citep{miles_first,miles} single stellar population (SSP; stars sharing the same age and metallicity) spectral library (v11) to construct stellar spectral templates to compare with the observed spectra.  This library, which is widely used for stellar population analyses, provides high-quality model spectra at a higher spectral resolution.  For constructing the spectral templates, we adopted a Kroupa initial mass function (IMF) over the stellar mass range $0.1$--$100\,M_{\odot}$ \citep{Kroupa}, the BaSTI isochrones \citep{Pietrinferni1,Pietrinferni2}, and a scaled solar metallicity\footnote{The scaled solar metallicity corresponds to [M/H]$=+0.06$ and [$\alpha$/Fe]=0.0.  This scaling includes corrections for the effect of changing [$\alpha$/Fe] as a function of [M/H] following the Galactic pattern in the empirical MILES stellar spectrum library (see \citealt{miles} for more details).},
henceforth referred to as simply \quotes{solar metallicity} or $Z=Z_\odot$, unless explicitly mentioned as $Z=1.5\,Z_\odot$ or \quotes{$\alpha$-enhanced}.

From trial and error, we found that spectral templates incorporating only one SSP could not provide a good fit to the blue and red correlation ranges simultaneously at any of the positions examined. Instead, different SSP spectral templates -- having different ages, radial velocities, and velocity dispersions -- are required to provide good fits to the different correlation ranges (see also Section\,\ref{subsec:population:SSP}). From these preliminary fits, we later constructed two-SSP composite spectral templates that present the best fits to the blue and red correlation ranges simultaneously (see Section\,\ref{subsec:population:composite}).

At this stage of the analysis, we emphasize that the radial velocities derived from our cross-correlation analysis are not greatly sensitive to the adopted values in age and metallicity so long as the model spectral template approximately reproduces the major characteristics of the observed spectral features.  For example, we found that the deep Balmer absorption lines in the blue correlation range are produced primarily by main-sequence A-type stars weighted toward earlier subtypes; the Balmer absorption lines of these stars exhibit little dependence on metallicity and the assumed isochrone.  In the red correlation range, both Mg\,$b$ and Fe are produced primarily by an old stellar population, and exhibit little age dependence.

\subsection{Radial Velocity Dependencies}\label{subsec:kinematics:velocity_fields}

\begin{figure*}
\centering
\includegraphics[width=\textwidth]{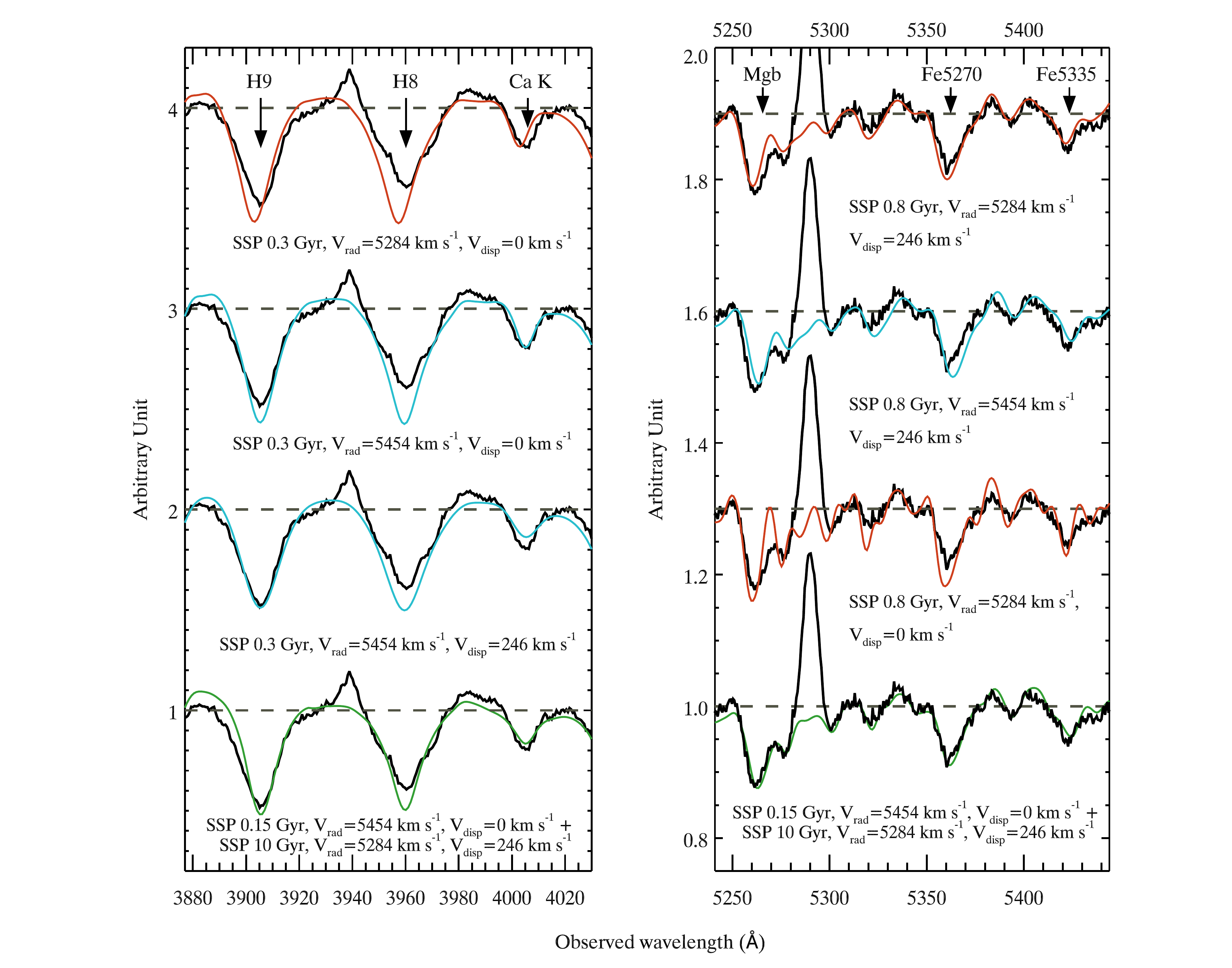}
\caption{Measured spectrum at the inner central spiral (black) restricted to a selected portion of the blue (left column) and red (right column) correlation ranges spanning stellar absorption lines of interest as labeled. The same spectrum is plotted multiple times in each column, with each offset vertically from another for clarity. Model spectra of the adopted solar-metallicity single stellar population (SSP) are overlaid on the measured spectra either at a radial velocity of $5454 \rm \, km \, s^{-1}$ (blue) or at the systemic velocity of NGC\,1275 of $5284 \rm \,km \, s^{-1}$ (red); all are convolved to the instrumental spectral resolution, although only selected model spectra are convolved also by a stellar velocity dispersion of $246 \rm \, km \, s^{-1}$.  The measured spectrum in the blue correlation range, in particular the Ca\,K line profile, is best reproduced by the model spectrum in blue overlaid on the second spectrum. By contrast, the measured spectrum in the red correlation range, in particular the Fe line profiles, is best reproduced by the model spectrum in red overlaid on the first spectrum.  Nebula emission lines of [$\mathrm{Ne\,III}$]$\lambda$3869\,{\AA} (redshifted to 3940\,{\AA}) and [$\mathrm{N\,I}$]$\lambda$5200\,{\AA} (redshifted to 5290\,{\AA}), visible but not labeled, are not considered in the model fits.  The best-fit two-SSP composite model having $Z=1.5 \, Z_\odot$ (green) is overlaid on the fourth spectrum, for which the younger SSP has an age of 0.15\,Gyr and a radial velocity (at this position) of $5454 \rm \, km \, s^{-1}$, and the older SSP has an age of 10\,Gyr and a radial velocity corresponding to the systemic velocity of NGC\,1275 of $5284 \rm \,km \, s^{-1}$ (see Section\,\ref{subsec:population:composite}).  The model spectra of the two SSPs are convolved to the instrumental spectral resolution, but only the spectrum of the older SSP is convolved to a stellar velocity dispersion of $246 \rm \, km \, s^{-1}$.}
\label{fig:velocity_fields:profile_comp}
\end{figure*}

Figure\,\ref{fig:velocity_fields:profile_comp} (left column) shows a limited portion of the blue correlation range spanning just the Ca\,K, H8, and H9 lines, extracted over the inner central spiral.  The same spectrum is plotted multiple times, each offset vertically from the other for clarity, and will henceforth be referred to by the order of their appearance from top to bottom.  The relatively deep H8 and H9 lines, by contrast with the relatively shallow Ca\,K lines, demand a relatively young stellar population, in keeping with the relatively blue colors of the central spiral.  Adopting solar metallicity, we found through trial and error that a stellar population having an age of 0.3\,Gyr roughly reproduces the relative absorption depths of the Balmer and Ca\,K absorption lines, where we erred on the side of reproducing the Ca\,K absorption depth at the price of overpredicting the Balmer absorption depths, as the latter could be partially filled in by the corresponding emission lines from the nebula in NGC\,1275; see Appendix\,\ref{appendix:nebula_contami}. 

We now show that this stellar population must have a radial velocity significantly different from the systemic velocity of NGC\,1275 so as to reproduce the line centers of the H8, H9, and Ca\,K absorption lines. Referring to the first spectrum in the blue correlation range (Figure\,\ref{fig:velocity_fields:profile_comp}, left column), we overlay in red an SSP spectral template having solar metallicity, an age of 0.3\,Gyr, and a radial velocity of $5284 \rm \,km \, s^{-1}$ corresponding to the systemic velocity of NGC\,1275, that has been convolved to the instrumental spectral resolution of 6.1\,{\AA}.
This spectral template is systematically blueshifted with respect to the line centers of all three of the stellar absorption lines shown in the blue correlation range. In the second spectrum, we overlay in blue the same SSP spectral template except now having a higher radial velocity of $5454 \rm \, km \, s^{-1}$, that is once again convolved to the instrumental spectral resolution.
This spectral template presents a clearly superior fit to the line centers of all three of the stellar absorption lines in the blue correlation range, albeit overpredicting (for the reasons mentioned above) the depths of the H8 and H9 lines.

The same spectral template as that overlaid in blue on the second spectrum, but convolved additionally by a stellar velocity dispersion of $246 \rm \, km \, s^{-1}$, is overlaid in blue in the third spectrum.  This velocity dispersion is the mean value measured by \citet{Heckman} over the radial range $\sim$5\arcsec--$25\arcsec$ on either side of the center of NGC\,1275; their measurements encompassed three different slit orientations spanning the wavelength range 5165--5180\,{\AA} (i.e., around Mg $b$), from which they found a constant velocity dispersion as a function of radius within measurement uncertainties.  Note that the radial range covered by their measurements extends far beyond the central spiral, and therefore traces the velocity dispersion of the old stellar population comprising the main body of NGC\,1275.  By comparison with the spectral template overlaid on the second spectrum, the spectral template overlaid on the third spectrum underpredicts the depth of the Ca\,K line profile; a spectral template having the same parameters but tuned to match the depth of the Ca\,K line would then overpredict its width. These comparisons demonstrate that the stellar absorption lines in the blue correlation range (Figure\,\ref{fig:velocity_fields:profile_comp}, left column) originate from a stellar population that is much younger, has a radial velocity different from the systemic velocity of NGC\,1275, and a narrower velocity dispersion than the old stellar population comprising the main body of NGC\,1275.


Figure\,\ref{fig:velocity_fields:profile_comp} (right column) shows a limited portion of the corresponding red correlation range (i.e., also extracted over the inner central spiral) spanning just the Mg\,$b$, Fe5270, and Fe5335 lines, all good indicators of metallicity. Like before, the same spectrum is plotted multiple times, each offset vertically from the other for clarity. Because the old stellar population in massive elliptical galaxies is known to be very metal rich (especially at the inner region of these galaxies), it would be reasonable to expect that the metal absorption lines preferentially trace this population. We therefore overlay in red an SSP spectral template having solar metallicity, a radial velocity of $5284 \rm \,km \, s^{-1}$ corresponding to the systemic velocity of NGC\,1275, and a stellar velocity dispersion of $246 \rm \, km \, s^{-1}$ corresponding to that measured by \citet{Heckman} for the main stellar body of NGC\,1275, which is convolved to the instrumental spectral resolution to the first spectrum in the red correlation range (Figure\,\ref{fig:velocity_fields:profile_comp}, right column). We found that an age of at least 0.8\,Gyr, significantly older than that required to fit the blue correlation range, is necessary to reproduce the relative absorption depths of the three metal lines considered in this range.  
This spectral template provides a good fit to the Fe5270 and Fe5335 lines, albeit not the Mg\,$b$ line. For comparison, the spectral template overlaid in blue on the second spectrum has all the same parameters except a radial velocity of $5454 \rm \,km \, s^{-1}$, corresponding to the best-fit radial velocity in the blue correlation range.  This spectral template provides a clearly inferior fit to the Fe5270 and Fe5335 lines and no improvement in fitting the Mg\,$b$ line, demonstrating that the metal lines in the red correlation range have a different radial velocity than the Balmer and Ca\,K absorption lines in the blue correlation range.

The SSP spectral template overlaid in red on the third spectrum also has the same parameters as that overlaid in red on the first spectrum, except convolved only to the instrumental spectral resolution but not the stellar velocity dispersion.  This spectral template also provides a poorer fit to (overpredicts the depth when designed to fit the wings of) the Fe5270 and Fe5335 lines, while providing no improvement in fitting the Mg\,$b$ line.

In this way, we found that two stellar populations -- having different ages, radial velocities, and stellar velocity dispersions -- are required to fit the measured spectra at all slit positions along the central spiral. Next, adopting the ages and velocity dispersions thus determined for each stellar population, we derive the best-fit radial velocities of the younger and older populations to the spectrum in the red and blue correlation ranges, respectively, as a function of position along the slit. Before proceeding, we emphasize once again that although the cross-correlation analysis provides relatively accurate determinations of the radial velocities of the two stellar populations, this technique is relatively insensitive to their exact ages; in Section\,\ref{subsec:population:composite}, we derive more exacting ages by fitting each spectrum with two stellar populations simultaneously, from which we show that the age of the older stellar population is compatible with that of old stellar populations in present-day elliptical galaxies having ages of $\sim$10\,Gyr.  

\begin{figure*}
\centering
\gridline{\fig{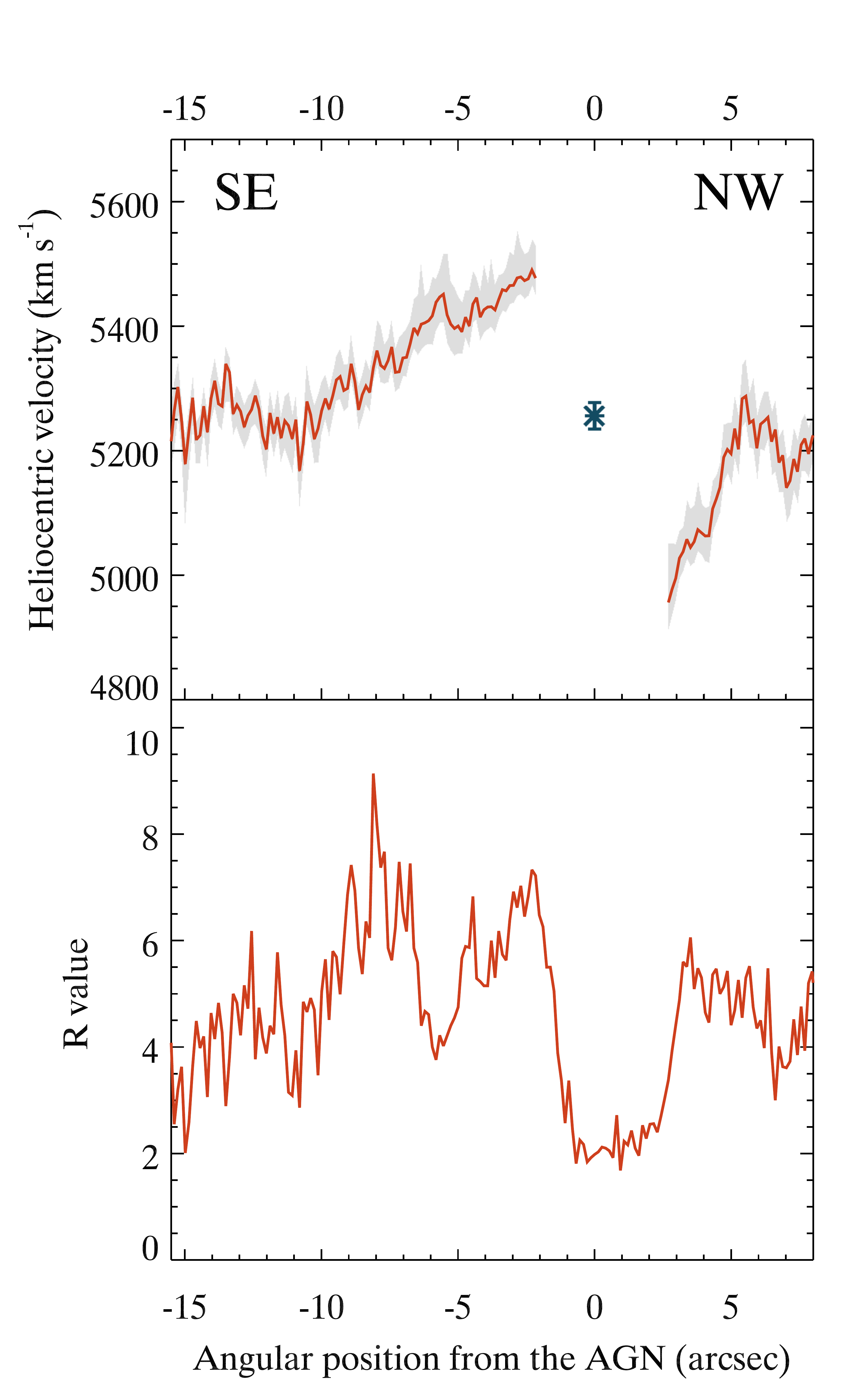}{0.45\textwidth}{}
              \fig{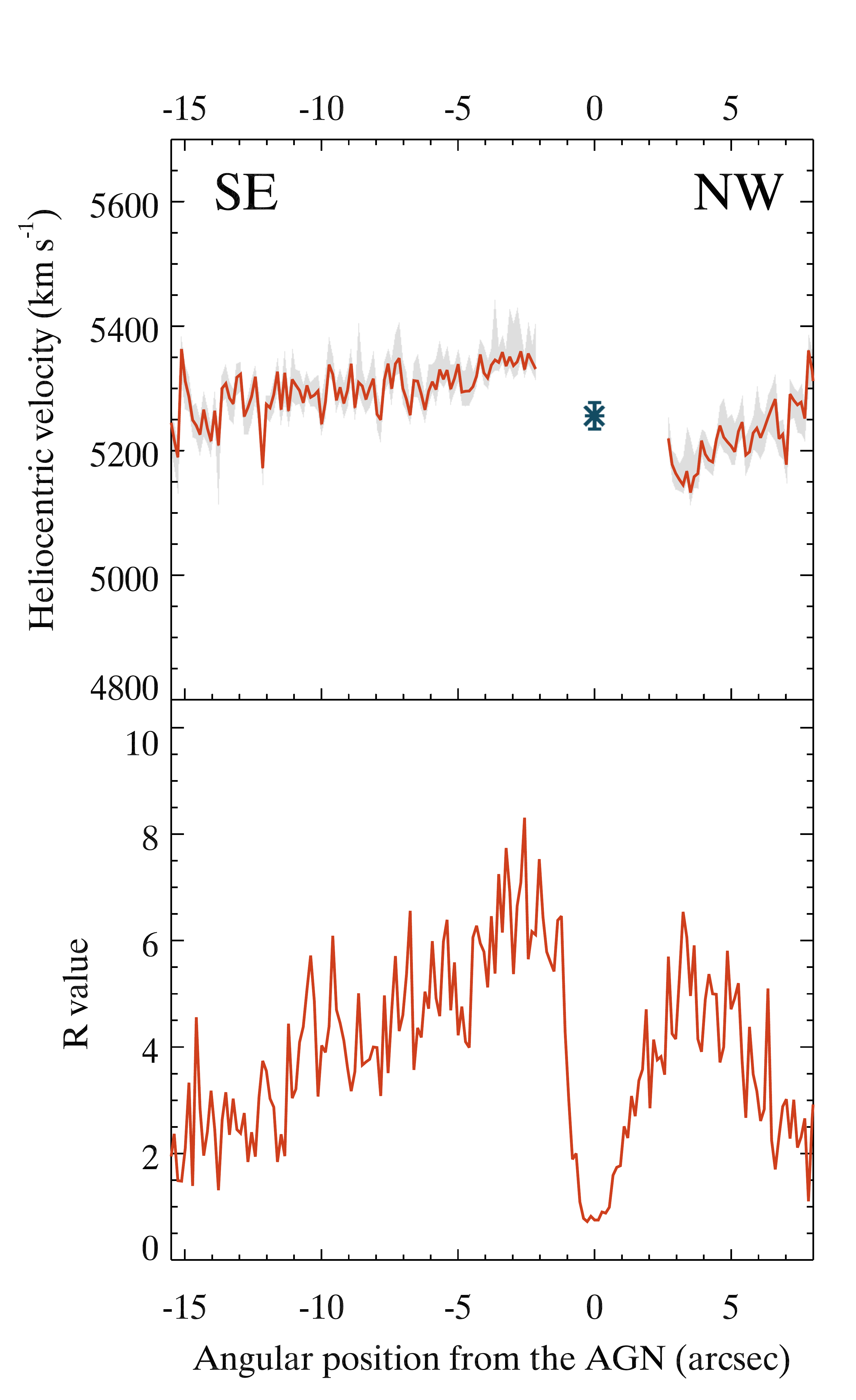}{0.45\textwidth}{}}
\vspace{-0.7cm}
\caption{Heliocentric velocities inferred for the younger stellar population from the blue correlation range (upper-left panel), and those inferred for the older stellar population from the red correlation range (upper-right panel), as a function of radius from the center of NGC\,1275.  The gray band indicates the $\pm 1\sigma$ uncertainty of the measurements.  The asterisk indicates the systemic velocity of NGC\,1275 and its associated uncertainty.  As explained in the text, the velocity gradient seen in the older stellar population is actually generated by the younger stellar population, which contributes to the stellar absorption lines in the red correlation range.  The correlation significance, $R$, of the cross-correlation analysis is shown separately for the blue (lower-left panel) and red (lower-right panel) correlation ranges.  Heliocentric velocities close to the center of NGC\,1275 are not shown owing to the small values of $R$ over this range, nor those at larger radial distances lying beyond the range encompassed by these plots owing also to the small values of $R$ there.}
\label{fig:rv}
\end{figure*}

Figure\,\ref{fig:rv} shows the best-fit radial velocities in the blue (left column) and red (right column) correlation ranges separately, along with the corresponding correlation significance, $R$,  as a function of slit position.  The inferred radial velocities have typical $1\sigma$ measurement uncertainties of 30\,km\,s$^{-1}$ close to the center where $R \simeq 5$, and 50\,km\,s$^{-1}$ further away from the center where $R \simeq 3$.  The stellar velocity field (upper row) can be measured inwards as close as $-2\farcs0$ southeast and $+2\farcs5$ northwest of the center, within which emission lines from the AGN of NGC\,1275 not accounted for in our spectral templates cause a dip in the correlation significance to $R \lesssim 2$, as can be seen in Figure\,\ref{fig:rv} (lower row).  Beyond this region, $R$ generally decreases with radial distance outwards as the S/N drops.  We terminate measurements of the stellar velocity field where the correlation significance dips to $R < 3$, corresponding to radial distances beyond $15\farcs5$ on the southeastern side of the center, but only $8\arcsec$ on the northwestern side of the center owing to overlap with the HVS as well as the optical ghost of the AGN on this side; spectral lines associated with the HVS are, of course, also not accounted for in our spectral templates.

The radial velocity of the younger stellar population (upper-left panel) reaches maximal values of about $\pm 250 \rm \,km\,s^{-1}$ from the systemic velocity of NGC\,1275 at the closest measurable angular distance from the center.  The stellar radial velocity then decreases gradually outwards toward both the southeastern and northwestern sides of the center.  On the southeastern side, where measurements can be made as far out as $-15\farcs5$, the stellar radial velocity decreases to that of the systemic velocity of NGC\,1275 at a slit position of about $-10\arcsec$, corresponding to the red gap and beyond.  Note that the dip in the $R$ value at a slit position of $-6\arcsec$ coincides with a nebular filament.  We shall henceforth attribute this stellar radial velocity dependence to that of the central spiral.

By contrast, the older stellar population (upper-right panel) exhibits a gentler change in radial velocity with radius from the center of NGC\,1275.  The radial velocity of this population reaches maximal values at the closest measurable angular distance from the center of NGC\,1275, just like for the younger stellar population; the maximal radial velocities reached by the older stellar population of about $\pm 100 \rm \,km\,s^{-1}$ from the systemic velocity of NGC\,1275, however, is lower (in absolute values) than the corresponding maximal values reached by the younger stellar population.  On the southeastern side of the center, the radial velocity decreases linearly with radial distance outwards to the systemic velocity of NGC\,1275 at a slit position of about $-15\arcsec$.  Both the radial velocity and velocity dispersion of the older stellar population that we determined from the red correlation range is consistent with that measured by \citet{Heckman} through optical slit spectroscopy at nearly the same PA of $110^{\circ}$ (see their Figure\,1) in the radial range over which both measurements overlap.  As will be explained in Section\,\ref{subsec:population:composite}, however, the trend in radial velocity inferred for the older stellar population in the red correlation range is actually generated by the younger stellar population: because the latter contributes more weakly to the stellar absorption lines in this correlation range, its velocity gradient is moderated by the constant radial velocity of the older stellar population.

The slit shown in Figure\,\ref{fig:HST_img} passes only $0\farcs4$ away from H5, a relatively young and massive star cluster found by \citet{star_clus_ID} located $5\farcs3$ southeast of the center of NGC\,1275 at a PA of 140\degr.  \citet{brodie} measured the radial velocity of both H5 and stars (which dominate the light) external to H5 along the same sight line by cross-correlating A-type stellar templates with Balmer absorption lines as measured from 3900\,{\AA} to 4700\,{\AA} (encompassing the Balmer transitions from H$\gamma$ to H9).  They found a radial velocity for stars external to H5 of $5318 \pm 47$\,km\,s$^{-1}$; weighted toward stars exhibiting strong Balmer absorption, this radial velocity more closely traces that of the younger rather than older stellar population.
By comparison, at the same slit position, we measured a radial velocity of $5418 \pm 46 $\,km\,s$^{-1}$ for the younger stellar population that dominates the light in the blue correlation range, being almost consistent with the measurement uncertainties.

\section{Stellar Populations}\label{sec:stellar population}
Our cross-correlation analysis (Section\,\ref{subsec:kinematics:velocity_fields}) demands two stellar populations having different kinematics to fit the measured line centers of the stellar absorption lines in the blue and red correlation ranges separately. Here, we show that setting aside agreement with the line centers, just one stellar population alone cannot fit the depths of the stellar absorption lines in both the blue and red correlation ranges simultaneously (Section\,\ref{subsec:population:SSP}).

In Appendix\,\ref{subsec:line indices}, we show that the relative strengths of spectral line indices provide the strongest constraints on the ages of the stellar populations necessary to reproduce the measured spectra.  The spectral lines used for constructing line indices are indicated by the green bands in Figure\,\ref{fig:index_window}, and the methods used for computing line indices are explained in Appendix\,\ref{subsubsec:population:indices:measurement}.  When the line indices are considered individually, we find that invoking just a single stellar population requires (i) relatively young stars (multiple 0.1\,Gyr old) to reproduce, primarily, the H8, H9, and Ca\,K indices; and (ii) an increasing age with radius outwards from the center of NGC\,1275 so as to explain radial variations in the line indices (Appendix\,\ref{subsubsec:population:indices:spatial_trends}).  
When considered together, however, just one stellar population, albeit with different ages at different locations, cannot reproduce all the line indices simultaneously (Appendix\,\ref{subsubsec:population:indices:index_index}).  Instead, a radial variation in the relative contribution from two stellar populations having different ages is able to to reproduce the measured line indices (Appendix\,\ref{subsubsec:2-SSP:index_index}), and, as we demonstrate below, also the measured line profiles at different positions along the slit (Section\,\ref{subsec:population:composite}).

\subsection{Single Stellar Population Model with Variable Age}\label{subsec:population:SSP}

We first examine how well model spectra based on a SSP that is allowed to have different ages at different slit positions are able to fit the depths of the stellar absorption lines in both the blue and red correlation ranges simultaneously.  For all these comparisons, as well as those to follow in Section\,\ref{subsec:population:composite} based on a combination of two SSPs, we consider fits to the H8 and H9 lines to be satisfactory if the model line profiles are deeper than the measured line profiles, as the latter may be partially filled in by the corresponding emission lines from the nebula associated with NGC\,1275.
%

\begin{figure*}
\centering
\includegraphics[width=\textwidth]{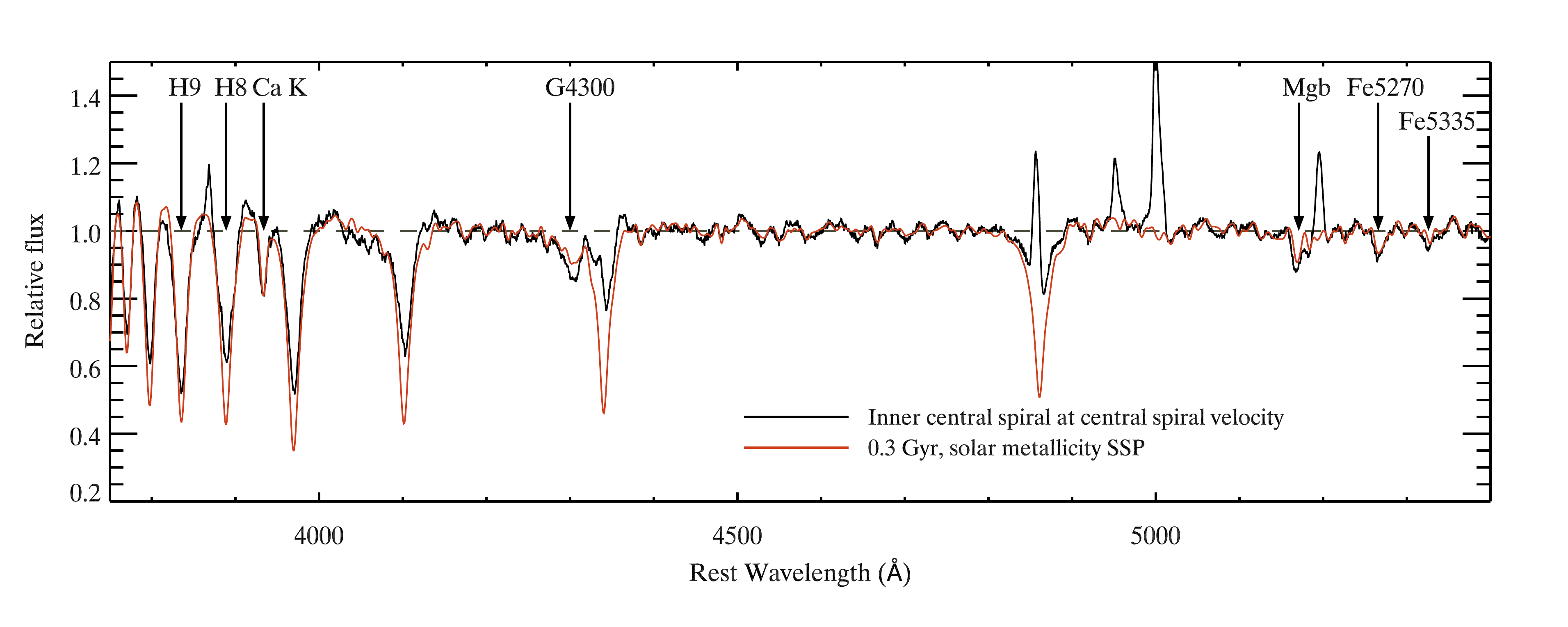}
\includegraphics[width=\textwidth]{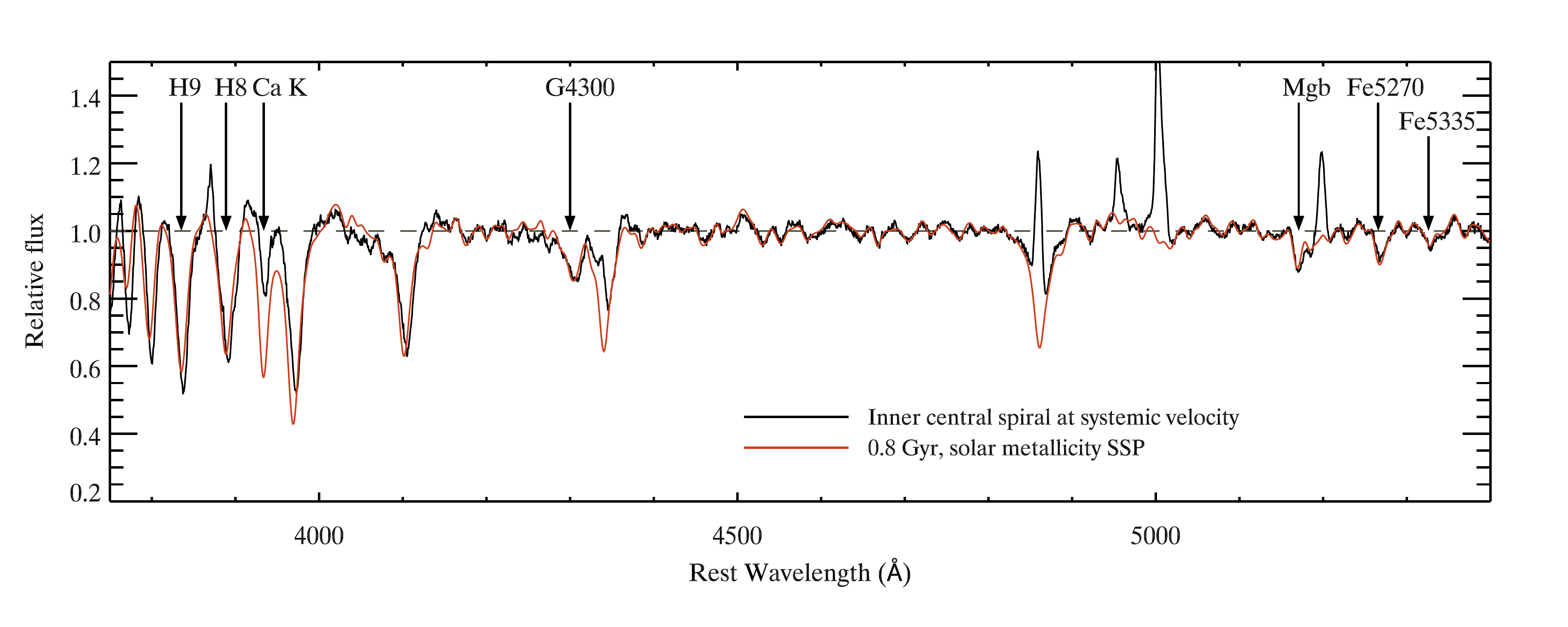}
\\
\vspace{-0.5cm}
\caption{Normalized spectrum in black averaged over the subregion corresponding to the inner central spiral, plotted at the rest wavelength corresponding to the radial velocity at this subregion (upper panel) and the systemic velocity of NGC\,1275 (lower panel).  Overlaid are model spectra in red for the adopted solar-metallicity single stellar population smoothed to the instrumental spectral resolution and having (i) an age of 0.3\,Gyr, and (ii) an age of 0.8\,Gyr, further smoothed to a velocity dispersion of $246 {\rm \, km \, s^{-1}}$ (corresponding to that of the old stellar population comprising the main body of NGC\,1275; lower panel).}
\label{fig:spec-4}
\end{figure*}

Figure\,\ref{fig:spec-4} shows the measured spectra (colored black) averaged over the subregion corresponding to the inner central spiral, plotted at the rest wavelength corresponding to the radial velocity at the inner central spiral (upper panel) and at the systemic velocity of NGC\,1275 (bottom panel), respectively. In the upper panel, a model spectrum (colored red) is overlaid corresponding to the adopted solar-metallicity SSP with an age of 0.3\,Gyr (based on the measured H8/Ca\,K index ratio, a sensitive indicator of age, at this location; see Appendix\,\ref{subsubsec:population:indices:spatial_trends} and Figure\,\ref{fig:sing_calib}), smoothed to the instrumental spectral resolution. This model spectrum reproduces the shallow Ca\,K and (overestimates) the even deeper H8 line profiles, but underestimates the G4300, Mg\,$b$, and Fe line profiles. By contrast, a model spectrum (colored red) is overlaid corresponding (in the lower panel) to the adopted solar-metallicity SSP having an older age of 0.8\,Gyr, smoothed to both the instrumental spectral resolution and a velocity dispersion of $246 {\rm \, km \, s^{-1}}$ (corresponding to that of the old stellar population comprising the main body of NGC\,1275). This model spectrum reproduces the G4300, Mg\,$b$, and Fe line profiles, but slightly underestimates the H8 and H9 line profiles as well as severely overestimating the Ca\,K line profile.

\begin{figure*}
\centering
\includegraphics[width=\textwidth]{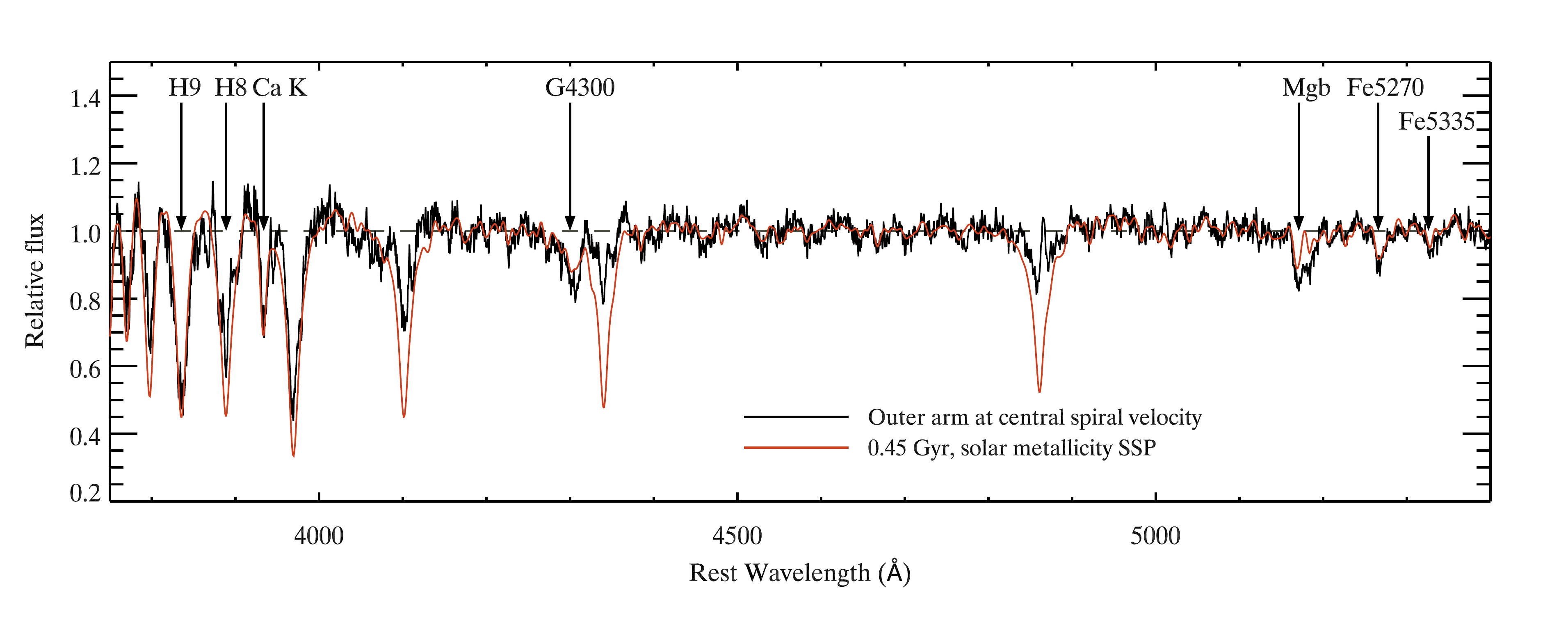}
\includegraphics[width=\textwidth]{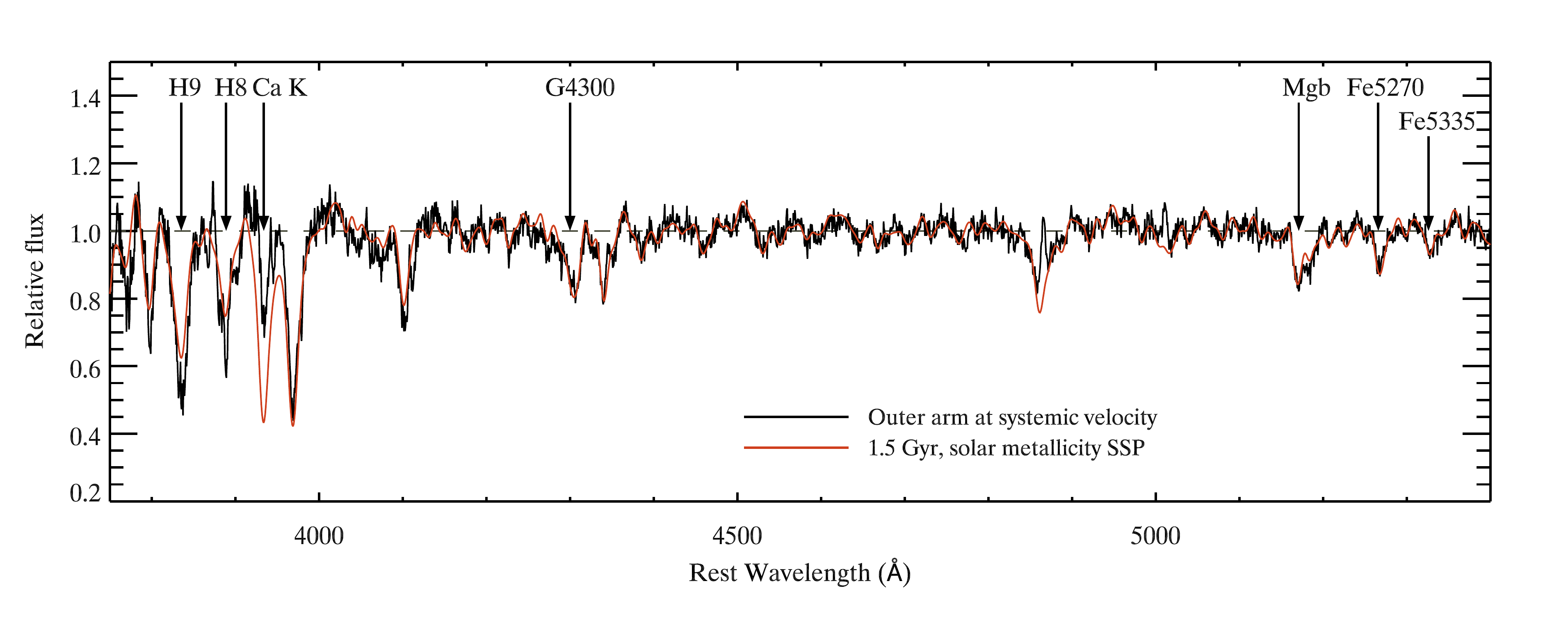}
\vspace{-1cm}
\caption{Same as Figure\,\ref{fig:spec-4}, but for the subregion corresponding to the outer arm and plotted at the radial velocity specified in the corresponding panels of Figure\,\ref{fig:spec-4}. Model spectra are for a 0.45\,Gyr old (upper panel) and 1.5\,Gyr old (lower panel) solar-metallicity single stellar population, each convolved to a spectral resolution described in Figure\,\ref{fig:spec-4}.}
\label{fig:spec-13}
\end{figure*}

Figure\,\ref{fig:spec-13} shows the measured spectra (colored black) averaged over the subregion corresponding to the outer arm, plotted at the rest wavelength corresponding to the radial velocity at the outer arm (upper panel) and at the systemic velocity of NGC\,1275 (bottom panel) respectively. A model spectrum (not shown) based on the adopted solar-metallicity SSP having an age of 0.8\,Gyr, smoothed to the instrumental spectral resolution, severely overestimates the Ca\,K line profile. Adopting instead an age of 0.45\,Gyr (upper panel), the model spectrum (colored red) reproduces the Ca\,K and (slightly overestimates the) H8 line profiles, but somewhat underestimates the G4300 and Fe line profiles and more severely underestimates the Mg\,$b$ line profile. Reproducing just the G4300, Fe, and Mg\,$b$ line profiles requires the adopted solar-metallicity SSP to have an age of 1.5\,Gyr (lower panel), and the resulting model spectrum (colored red) to be smoothed to the instrumental spectral resolution as well as to a velocity dispersion of $246 {\rm \, km \, s^{-1}}$ (corresponding to that of the old stellar population comprising the main body of NGC\,1275). Clearly, at either the inner central spiral or the outer arm, a single SSP cannot simultaneously reproduce the stellar absorption line profiles in the blue and red correlation ranges---a point already demonstrated by our cross-correlation analysis to derive radial velocities---as well as in G4300.  

\subsection{Two-Single Stellar Population Composite Model}\label{subsec:population:composite}
Our analyses of the line kinematics (Section\,\ref{subsec:kinematics:velocity_fields}), line indices (Appendix\,\ref{subsec:line indices}), and depths of the stellar absorption lines (Section\,\ref{subsec:population:SSP}) all make clear that (at least) two stellar populations are involved in producing the observed stellar absorption lines, namely (i) a younger population associated with the central spiral, and (ii) an older stellar population presumably associated with the main body of NGC\,1275.  We therefore developed a two-SSP composite model comprising a younger stellar population superposed on an older stellar population, represented by two SSPs having different physical parameters.  Suspecting that the apparent age gradient in the younger population is caused by an outwardly increasing contribution from the older population, we fixed the age of each SSP along the slit and allow the fractional flux contributed by each population to change along the slit in steps of 10\%.  At a given slit position, the fractional flux contributed by each population varies with wavelength owing to their different continua slopes or shapes: the values henceforth quoted for this quantity are defined (unless stated differently) just shortward in wavelength of the Mg\,$b$ line at $5095\pm55$\,{\AA}.  In this way, we generated a range of spectra ranging from those dominated by the younger to those dominated by the older population.
For simplicity, we assume solar metallicity ($Z=Z_\odot$) for the younger population, and either $Z=Z_\odot$ or $1.5\,Z_\odot$ for the older population as BCGs are at least as metal enriched as typical elliptical galaxies of a comparable mass \citep{von_der_Linden}.

In addition to supersolar metallicities, elliptical galaxies show an $\alpha$ enhancement that increases with their stellar velocity dispersions (and hence masses), such that massive elliptical galaxies typically have [$\alpha/$Fe]$\sim0.3$ (e.g., \citealt{Trager1,Trager2,Thomas,Sanchez}).
BCGs are known to show even stronger $\alpha$ enhancement than elliptical galaxies of comparable masses, with typically [$\alpha/$Fe]$=0.41\pm0.1$ \citep{von_der_Linden,loubser09}.
Such an abundance pattern naturally enhances, if the total metallicity (in both the $\alpha$- and the non-$\alpha$ elements) is a constant, the Mg2 index owing to an enhanced Mg abundance (an $\alpha$ element) while reducing the Fe (a non-$\alpha$ element) line indices.  For this reason, we also explored an $\alpha$-enhanced old population having [$\alpha/$Fe]$=0.4$, the only $\alpha$-enhancement option available in the MILES SSP library.  Because super-solar metallicity models having this level of $\alpha$ enhancement, for example \{$Z=1.5\,Z_\odot$, [$\alpha/$Fe]$=0.4$\}, predict too deep a Mg\,$b$ profile to reproduce the measurements, we explore only \{$Z=Z_\odot$, [$\alpha/\rm{Fe}]=0.4$\} when $\alpha$ enhanced.  For brevity, we shall henceforth refer to \{$Z=1.5\,Z_\odot$, [$\alpha/\rm{Fe}] =0.0$\} as $Z=1.5\,Z_\odot$, and \{$Z=Z_\odot$, [$\alpha/\rm{Fe}]=0.4$\} as $Z=Z_\odot$ $\alpha$-enhanced.

\begin{figure*}
\centering
\includegraphics[width=\textwidth]{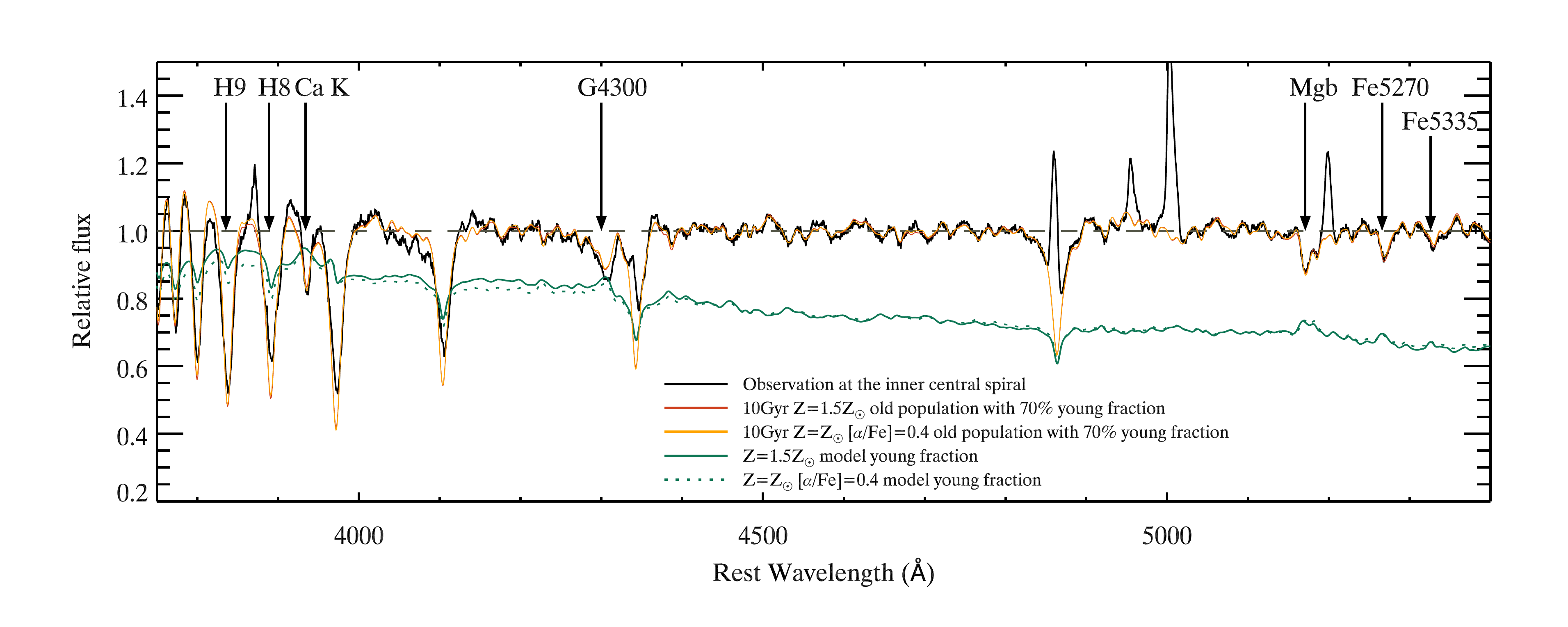}
\includegraphics[width=\textwidth]{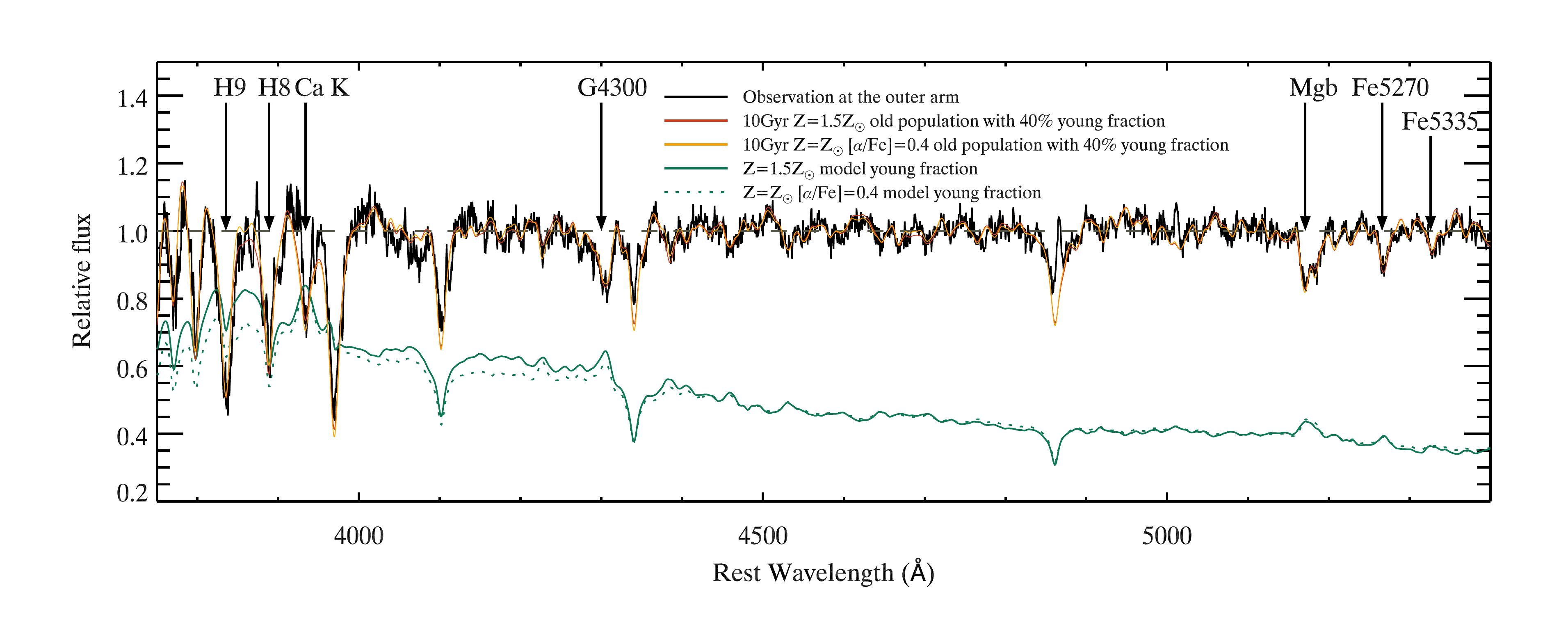}
\caption{Measured continuum-normalized spectra (black), plotted at the rest wavelength corresponding to the systemic velocity of NGC\,1275, for the two subregions corresponding to the inner central spiral (upper panel) and outer arm (lower panel).  Overlaid on the measured spectra are model spectra based on the best-fit two-single stellar populations (SSP) composite model having an age of 0.15\,Gyr for the younger SSP and 10\,Gyr for the older SSP, and for $Z=1.5\,Z_\odot$ (red) and $Z=Z_\odot$ $\alpha$-enhanced (orange).  The model spectra of the younger SSP is shifted in velocity with respect to the older SSP in accordance with its radial velocity at a given location, as shown in Figure\,\ref{fig:rv}.  The model fractional contribution in flux of the younger SSP, which varies with wavelength, is shown for both $Z=1.5\,Z_\odot$ (solid green) and $Z=Z_\odot$ $\alpha$-enhanced (dotted green).
\label{fig:composite_spec}}
\end{figure*}

In Appendix\,\ref{subsubsec:2-SSP:index_index}, we explain how, through extensive trial and error, the two-SSP composite models were tuned to match the index--index measurements (plot of one line index versus another) thus yielding ages of 0.15\,Gyr and 10\,Gyr for the younger and older stellar populations, respectively.  To demonstrate how well the continuum-normalized spectra generated from the best-fit two-SSP composite model reproduce the measured spectra, Figure\,\ref{fig:composite_spec} (upper panel) shows the measured spectra at the inner central spiral as in Figure\,\ref{fig:spec-4}, and Figure\,\ref{fig:composite_spec} (lower panel) the measured spectra at the outer arm as in Figure\,\ref{fig:spec-13}.  Model spectra are shown for both $Z=1.5  \, Z_\odot$ (red line) and $Z=Z_\odot$ $\alpha$-enhanced (orange line), with the radial velocity of the younger SSP set to its inferred value at a given location (see Figure\,\ref{fig:rv}, upper-left panel), and the radial velocity of the older SSP set to the systemic velocity of NGC\,1275.  The fractional contribution in flux from the younger SSP (age 0.15\,Gyr) as a function of wavelength at each location is indicated for both $Z=1.5  \, Z_\odot$ (green line) and $Z=Z_\odot$ $\alpha$-enhanced (dotted green line).   A closer visual inspection of how well the model spectrum for $Z=1.5  \, Z_\odot$ fits the selected stellar absorption lines in the blue and red correlation ranges is shown in the bottom spectra of Figure\,\ref{fig:velocity_fields:profile_comp}.


\begin{deluxetable*}{lCCCCCC}
\tablecaption{Best-fit fractional flux for the younger Single Stellar Population (SSP) in the Two-SSP composite models, and the corresponding model predictions for their line indices and $B-V$ color.\label{tab:composite_best-fit}}
\tablehead{
\colhead{} & \colhead{Outside}& \colhead{Outer} & \colhead{Red} & \colhead{Middle} & \colhead{Nebula} & \colhead{Inner} \\
\colhead{} & \colhead{Outer arm}& \colhead{Arm} & \colhead{Gap} & \colhead{Central spiral} & \colhead{} & \colhead{Central spiral}
}
\startdata
\multicolumn{7}{c}{$Z=1.5\,\rm{Z_\odot}$ composite model} \\\hline
Young flux fraction\tablenotemark{a} & 0.4 & 0.4 & 0.5 & 0.6 & 0.6 & 0.7 \\
H8 ({\AA}) & 6.28 & 6.28 & 6.72 & 7.05 & 7.05 & 7.31 \\
H9 ({\AA}) & 6.46 & 6.46 & 6.62 & 6.74 & 6.74 & 6.84 \\
Ca\,K ({\AA}) & 5.27 & 5.27 & 4.32 & 3.54 & 3.54 & 2.90 \\
G4300 ({\AA}) & 2.46 & 2.46 & 2.20 & 1.98 & 1.98 & 1.78 \\
Mg2 (mag) & 0.110 & 0.110 & 0.098 & 0.087 & 0.087 & 0.076 \\
$\langle\rm{Fe}\rangle$ ({\AA}) & 2.11 & 2.11 & 1.94 & 1.76 & 1.76 & 1.58 \\
$D$(4000) & 1.57 & 1.57 & 1.49 & 1.43 & 1.43 & 1.38 \\
$B-V$ (mag) & 0.73 & 0.73 & 0.63 & 0.54 & 0.54 & 0.45 \\\hline
\multicolumn{7}{c}{$Z=\rm{Z_\odot}$ $\alpha$-enhanced composite model} \\\hline
Young flux fraction\tablenotemark{a} & 0.4 & 0.4 & 0.5 & 0.6 & 0.6 & 0.7 \\
H8 ({\AA}) &5.76 & 5.76 & 6.31 & 6.75 & 6.75 & 7.11 \\
H9 ({\AA}) &6.65 & 6.65 & 6.76 & 6.85 & 6.85 & 6.92 \\
Ca\,K ({\AA}) &5.27 & 5.27 & 4.32 & 3.54 & 3.54 & 2.90 \\
G4300 ({\AA})&2.33 & 2.33 & 2.11 & 2.00 & 2.00 & 1.81 \\
Mg2 (mag) & 0.112 & 0.112 & 0.100 & 0.089 & 0.089 & 0.077 \\
$\langle\rm{Fe}\rangle$ ({\AA}) &1.67 & 1.67 & 1.56 & 1.46 & 1.46 & 1.35 \\
$D$(4000) & 1.53 & 1.53 & 1.46 & 1.41 & 1.41 & 1.37 \\
$B-V$ (mag) & 0.66 & 0.66 & 0.58 & 0.50 & 0.50 & 0.42 \\
\enddata
\tablenotetext{a}{The young population flux fraction at 5095\,{\AA}.}
\end{deluxetable*}

As is apparent, the best-fit two-SSP composite model can explain most of the features exhibited by the stellar absorption lines in a consistent manner.  At the inner central spiral, the deep H8 and H9 but relatively shallow Ca\,K and G4300 lines are produced primarily by the younger stellar population, whereas the Mg\,$b$ and Fe features are produced primarily by the older stellar population.
The same is true at the outer arm, except at this location the younger stellar population contributes less, and hence the older stellar population contributes more, to the total flux. Table\,\ref{tab:composite_best-fit} lists the relative contribution in flux from the younger stellar population at the different subregions along the slit, as defined in Table\,\ref{tab:region}.  Overall, the fractional contribution to the flux from the younger population decreases outwards (from 0.7 at the inner central spiral to 0.4 at the outer arm) so as to explain increasingly deeper Ca\,K, Mg2, and $\langle$Fe$\rangle$ lines, and the shallower H8 line, outwards.  This trend also can explain the increasingly redder $B-V$ color radially outwards along the slit as described next in Section\,\ref{subsec:population:HST_colour}.  The composite model produces only a shallow and slowly changing $D$(4000) index ($=1.4$--$1.6$), in good agreement with measurements of this index by \citet{johnstone} as described in Appendix\,\ref{subsubsec:population:indices:spatial_trends}.
%

Despite the good general agreement, there are minor discrepancies between the model spectra based on the best-fit two-SSP composite models and the measured spectra.  First, neither models can explain the deeper H9 than H8 in the outer regions beyond $-6\arcsec$ southeast of the nucleus, and the anticorrelation between these two line indices at these locations as described in Appendix\,\ref{subsubsec:population:indices:spatial_trends} and Appendix\,\ref{subsubsec:population:indices:index_index}.  As explained in Appendix\,\ref{appendix:h9}, however, this unexpected behavior is probably caused by Mg\,I absorption overlapping with the H9 line, and so we did not tweak the composite models so as to better reproduce the H9 line.  Secondly, the H8 index is slightly but consistently overestimated in the best-fit two-SSP composite models, more so for $Z=1.5\,Z_\odot$ than $Z=Z_\odot$ $\alpha$-enhanced.  The H8 line, however, may be partially filled by the corresponding emission from the nebula associated with NGC\,1275.  Finally, the $\langle$Fe$\rangle$ index is well reproduced by the $Z=1.5\,Z_\odot$ but not the $Z=Z_\odot$ $\alpha$-enhanced composite model, which underestimates the index by $\sim$0.3--0.4\,{\AA} in all subregions.



\subsection{Broadband Color}\label{subsec:population:HST_colour}

We examined the broadband color of the central spiral and the regions beyond as an independent indicator of the stellar population(s) involved.  For this purpose, we used the HST Advanced Camera for Surveys images in $B$ (F435W) and $V$ (F550M) bands obtained by \citet{Fabian} and reprocessed by \citet{Jeremy}.  The two selected filters avoid all strong emission lines from the nebula associated with NGC\,1275 apart from the [$\mathrm{O\,II}$]$\lambda3727\,${\AA} line, which lies in the $B$ band.  This line is visible just bluewards of the blue correlation range in the spectrum shown in Figure\,\ref{fig:index_window}, so bright that it extends vertically far beyond the limits of the ordinate.  We determined its equivalent width from the measured spectra, and found that the [$\mathrm{O\,II}$]$\lambda3727\,${\AA} line adds at most just $\sim$0.03\,mag to the $B$ band and can therefore be safely neglected for the purpose of deriving broadband colors.  We corrected the brightnesses of the images for a Galactic extinction of 0.67\,mag in the $B$ band and 0.5\,mag in the $V$ band, as in \citet{Jeremy}.   Because no obvious dusty features are visible on the southeastern side of the nucleus encompassing the central spiral and beyond (Figure\,\ref{fig:HST_img}), we applied no internal dust-reddening correction. 

The measured $B-V$ color (in the Vega system) is shown as a function of radius from the center of NGC\,1275 in Figure\,\ref{fig:pos_index_color}, and listed in Table\,\ref{tab:indices_measured} averaged over each subregion.  As can be seen, the color becomes monotonically redder outwards from the inner central spiral ($B-V=0.57$) to the outer arm ($B-V=0.75$).  The color outside the outer arm (at $-15\arcsec$) is even redder ($B-V=0.86$), although remaining bluer than would be expected for an old stellar population having an age beyond $\sim$2\,Gyr (see Figure\,\ref{fig:sing_calib}) let alone 10\,Gyr, whereby $B-V=1.0$--$1.2$ for the two metallicities examined (see Figure\,\ref{fig:comp_calib}).  The increasing redder color outwards, even beyond the central spiral, is consistent with earlier studies showing that NGC\,1275 becomes increasingly bluer inwards beginning at radii much larger than the central spiral (e.g., \citealt{Romanishin1,Romanishin2,McNamara,Norgaard2}).

Our best-fit two-SSP composite model can reproduce the observed trend of redder color outwards as the relative contribution to the flux from the younger stellar population decreases outwards.  In either the $Z=1.5\,Z_\odot$ or $Z=Z_\odot$ $\alpha$-enhanced cases, however, the predicted colors (Table\,\ref{tab:composite_best-fit}) are bluer than the measured colors (Table\,\ref{tab:indices_measured}) within the central spiral as well as beyond the outer arm.  The $Z=1.5\,Z_\odot$ two-SSP composite model is systematically offset by $\sim$0.10\,mag, and therefore performs slightly better than the $Z=Z_\odot$ $\alpha$-enhanced two-SSP composite model that is systematically offset by $\sim$0.15\,mag.  The latter has a bluer color than the former owing to the combined effects of lower metallicity as well as the reduced continuum blanketing effect mostly by Fe in the $B$ band so as to conserve the total metal abundance when subjected to $\alpha$-enhancement \citep{miles}.

Given the relatively small color offset and the fact that we did not simultaneously fit the measured spectra and $B-V$ colors, we consider the composite models to reproduce the measured colors adequately well.  Including more free parameters, such as a radial variation in age for the younger stellar population, may well improve the fit to both the spectra and colors, but is beyond the immediate goal and scope of this work. 

Finally, we return to the previously mentioned reports that NGC\,1275 exhibits bluer colors inwards than would be expected for normal elliptical galaxies beginning at a radius much larger than the size of the central spiral.  Beyond the outer arm of the central spiral, the H8 line is much deeper and the H8/Ca\,K index ratio much larger than would be expected for an $\sim$10\,Gyr population characteristic of the main stellar body of NGC\,1275; e.g., at $-15\arcsec$, the H8 line has an index of $\sim$4\,{\AA} and the H8/Ca\,K index ratio a value of $\sim$1.0, compared with a H8 index of $\sim$1\,{\AA} (Figure\,\ref{fig:sing_calib}) and a H8/Ca\,K index ratio of $\sim$0.1 (Figure\,\ref{fig:comp_calib}) predicted by the SSP model.  Both broadband colors and stellar absorption lines therefore indicate a contribution from relatively young stars even beyond the central spiral (we are referring here to a spatially diffuse population of young stars, not the young and very compact stellar clusters distributed throughout NGC\,1275 as reported by \citealt{Jeremy}), a subject we plan to return to soon in a future study.

\section{Interpretation and Discussion}\label{sec:discussion}

Any interpretation of the central spiral needs to reconcile the following observational facts as established in our work.  First, although the central spiral appears to have the same systemic velocity as NGC\,1275, it is kinematically decoupled from the main stellar body of NGC\,1275 comprising an old population having a likely age around 10\,Gyr.  
The central spiral has a (projected) rotation velocity that reaches values as high as $\sim250 \rm \, km \, s^{-1}$ at $\simeq 2\arcsec$ (720\,pc) on either side of the nucleus of NGC\,1275, and exceeds $50 \rm \, km \, s^{-1}$ out to a radius of $\sim$9\arcsec\ (3.2\,kpc) before dropping to the systemic velocity of NGC\,1275 at a radius of $\sim$12\arcsec (4.3\,kpc).  By contrast, \citet{Heckman} report no radial variations in the (projected) rotation velocity of stars comprising the main body of NGC\,1275, such that the average rotation velocity is  $48 \pm 12 \rm \, km \, s^{-1}$ over the radial range 5\arcsec--25\arcsec (2--9\,kpc).  Furthermore, the stellar velocity dispersion of the central spiral is significantly lower than $140 \rm \, km \, s^{-1}$ at $1\sigma$, whereas \citet{Heckman} report a velocity dispersion for the main stellar body of NGC\,1275 that remains constant at $\sim250 \rm \, km \, s^{-1}$ over the radial range 5\arcsec--25\arcsec (2--9\,kpc).  We also find the same stellar velocity dispersion for the older population (attributed to the main body of NGC\,1275) along all sight lines toward the central spiral.  

Second, the age of the stellar population associated with the central spiral is very different from that of the main stellar body of NGC\,1275.  Stars in the central spiral have a nominal age of only $\sim$0.15\,Gyr, and are unlikely to lie much beyond the range 0.1--0.2\,Gyr based on our attempts at finding best-fit two-SSP composite models to the H8/Ca\,K index ratio.  Finally, albeit with a lesser degree of certainty, the central spiral appears to have settled into the gravitational potential of NGC\,1275.  As can be seen in Figure\,\ref{fig:HST_img}, despite the presence of dust features associated with the HVS on the northern (and especially northwestern) side of the nucleus, the central spiral can be clearly picked out as a bluish and roughly circular disk that is closely centered on the nucleus of NGC\,1275.  Thus, not only is the kinematics of the central spiral centered on the systemic velocity of NGC\,1275, the center of the central spiral also appears to coincide with the nucleus of NGC\,1275.

\subsection{Remnant of Minor Merger?}\label{minor merger}

The most immediately obvious explanation for the central spiral is that it is the remnant of a minor merger \citep[e.g.,][]{conselice}, whereby the dominant galaxy, NGC\,1275, has cannibalized a relatively cool-gas-rich (i.e., rich in atomic and/or molecular gas) galaxy.  Such a minor merger could also provide a natural explanation, as proposed by \citet{conselice}, for the semicircular arcs visible around the central spiral out to a radius of $\sim$11\,kpc (see Figure\,8  of \citet{conselice} and Figure\,19 of \citet{penny}).  

In this scenario, one would need to invoke a burst of star formation in the central spiral about 0.15\,Gyr ago, forming the large bulk, if not the entirety, of its stellar body.  Now, the merger timescale---time between close physical contact and for the merger remnant to settle into the center of NGC\,1275---would have had to be much longer than just 0.15\,Gyr.  In that case, why has the progenitor galaxy of the merger remnant not undergone multiple episodes of star formation as it experienced strong gravitational perturbations during its infall into the center of NGC\,1275?  One could argue that stars formed in earlier starburst episodes have faded away sufficiently so that only those produced in the most recent starburst remain detectable or dominate the light; indeed, by using Balmer (albeit also incorporating Ca\,K) lines to trace the stellar population in the central spiral, we preferentially pick up those that produce the strongest Balmer absorption lines, namely any with an age of a few 100\,Myr.  Alternatively, the progenitor galaxy could have been ripped apart and hence halted star formation as it physically merged with NGC\,1275; after reassembling at the center of NGC\,1275, the merger remnant experienced a starburst to produce the post-starburst population observed today.

A serious problem with this scenario is the gas mass the merger remnant would have had to retain so as to fuel its recent starburst.  To compute the total stellar mass formed during this starburst, we estimate a flux density in the $B$ band of $\sim$$1\times10^{-14}\,{\rm erg\,s^{-1}\,cm^{-2}\AA^{-1}}$ over an area encompassing the dust-free regions of the central spiral contained within the contour shown in Figure\,\ref{fig:HST_img} (middle panel).  Roughly $\sim$60\% of this flux density is contributed by the younger population (with the older stellar population contribution the remainder; see Section\,\ref{subsec:population:HST_colour}), which if comprising a SSP with an age of $\sim$0.15\,Gyr would have had to have an initial stellar mass of $\sim$$2 \times 10^9\,M_\odot$.  Assuming the other approximately one-third of the central spiral silhouetted behind the HVS contributes an equal fractional share by area, the initial stellar mass formed during this starburst is therefore $\sim$$3 \times 10^9\,M_\odot$.
Even at an unrealistic near 100\% efficiency in converting gas to stars, the gas mass required is comparable to all of the molecular hydrogen gas in the Milky Way; for a star-formation efficiency of 1\%--10\% (the range that is more typically inferred for star-forming galaxies), the required gas mass is $\sim$$3 \times 10^{10}$--$3 \times 10^{11} \, M_\sun$.  In addition, if the progenitor galaxy of the central spiral experienced multiple episodes of star formation during its infall into the center of NGC\,1275, its original gas mass would have had to be even higher.  

Could the progenitor galaxy have retained such a large gas mass in its orbit close to the center of the Perseus cluster, during which time it experiences intense ram-pressure stripping owing to the high density of the intracluster medium along with its high orbital velocity near pericenter?  This galaxy would have presumably made multiple orbits through pericenter before being slowed down sufficiently by dynamical friction---leading to a decay in both its pericentric and, especially, apocentric distance---so as to merge with NGC\,1275.  We find this aspect of the proposed scenario especially troubling.  On the other hand, if correct, then this scenario would imply that a surprising large amount of gas can survive ram-pressure stripping even up to the point whereby cluster member galaxies merge with the central cluster galaxy.

\subsection{Disk Created by Residual Cooling Flow?}\label{cooling flow}

One way around the troubling aspect raised in the scenario proposed above (Section\,\ref{minor merger}) is to attribute the gas for fueling the most recent starburst in the central spiral to another source: the most obvious such source is a residual cooling flow, held responsible for the emission-line nebula associated with NGC\,1275.  This nebula is multiphase, comprising ionized, atomic, and molecular components as mentioned in Section\,\ref{sec:intro}.  If the conversion between CO luminosity and H$_2$ gas mass is reliable \citep[see concerns raised by][]{Jeremy2017}, the component traced in CO, which would normally constitute the reservoir for star formation, dominates the mass of the emission-line nebula.  The total mass of H$_2$ gas traced in CO well exceeds $10^{10} \, M_\sun$ over the entire nebula \citep{Salome06}, with $\sim10^9\, M_\sun$ located within $\sim$10\arcsec\ ($\sim$4\,kpc) of the center of NGC\,1275 \citep{Ho_jeremy}.

If the gas for fuelling a recent starburst in the central spiral originated from the nebula associated with NGC\,1275, should there not be a kinematic relationship between the two structures at similar spatial scales?  
As can be seen in Figure\,\ref{fig:composite_spec}, at a given position along the slit, the radial velocity of the nebular gas (as measured in Balmer emission lines) is obviously different from that of stars in the central spiral.   At scales much smaller than the innermost radius that we are able to probe in the central spiral (i.e., $< 2\arcsec$), observations in H$_2$ and [$\mathrm{Fe\,II}$] at near-infrared wavelengths by \citet{scharwchter} revealed a rotating disk seen in both lines within a radius of $\sim$0\farcs14 ($\sim$50\,pc) that is centered on the nucleus of NGC\,1275; observations in CO, HCN, and HCO$^+$ by \citet{Nagai} 
revealed, presumably, the same disk with a radius of 0\farcs3 ($\sim$100\,pc).  This disk, which has an inclination angle of perhaps $\sim$$45\degr$ (see Section\,\ref{SMBH}), exhibits a velocity gradient in the east--west direction, orthogonal to the radio jets from the AGN in NGC\,1275.  The velocity fields of these disks do not join in any obvious manner with that of the central spiral: the radial velocities of the inner disks decrease outwards with radius, reaching near the systemic velocity of NGC\,1275 at their outer radii, far beyond which the (projected) rotation velocity of the central spiral is $250 {\rm \, km \, s^{-1}}$.  Thus, there does not seem to be a kinematic relationship between the central spiral and the nebular gas on similar spatial scales, or the inner disk(s) on a much smaller spatial scale.  

The lack of any kinematic relationship between the central spiral and nebular gas, however, should not come as a surprise.  The dynamical (orbital) timescale of the central spiral at a representative radius of 5\arcsec\ ($\sim$2\,kpc) is $\gtrsim10^8$\,yr, and longer still at larger radii.  By comparison, the dynamical (infall) timescales at a similar radius from the center of NGC\,1275 is much shorter at $\lesssim10^7$\,yr \citep[see][]{SMA_Jeremy}.  Thus, if the gas that fuelled the starburst in the central spiral originated from a residual cooling flow, this gas must have been accreted $\gg 10^8$\,yr ago so as to have had time to settle around the center of NGC\,1275.  Similarly, the central spiral and the disk enclosed within, as found by \citet{scharwchter} and \citet{Nagai}, have very different dynamical timescales, and therefore likely formed over different timescales if not also at very different epochs.  

In the above scenario, the central spiral therefore comprises two components: (i) a stellar body corresponding to the merger remnant; and (ii) a gaseous body accreted from a residual cooling flow.  The semicircular arcs around the central spiral therefore comprise the historical signature of a minor merger between the progenitor of the central spiral and NGC\,1275; the accretion of gas from a residual cooling flow over time fuelled a starburst in the central spiral about 0.15\,Gyr ago.  Of course, this complication of requiring the stellar and gaseous bodies to have different origins is not necessary: a gaseous disk may have first formed from a residual cooling flow, and then undergone a starburst about 0.15\,Gyr ago to form the stellar body observed today.  In this simpler scenario, the semicircular arcs around the central spiral do not constitute the signature of a past merger.  Indeed, these arcs are relatively blue (\citealt{conselice} and middle panel of Figure\,\ref{fig:HST_img}), suggesting that they are composed of relatively young stars rather than the disrupted body of a cluster member that might be expected to be composed primarily of relatively old stars (as the cluster member, largely stripped of gas, would have ceased star formation long ago).  Thus, rather than the historical signature of a past merger, the semicircular arcs around the central spiral may constitute stellar trails generated, perhaps, by the tidal disruption of super star clusters (SSCs) that wandered too close to the center of NGC\,1275 along their orbits.

\subsection{Progenitor Globular Clusters}\label{GCs}
\begin{figure*}
    \centering
    \includegraphics[width=\textwidth]{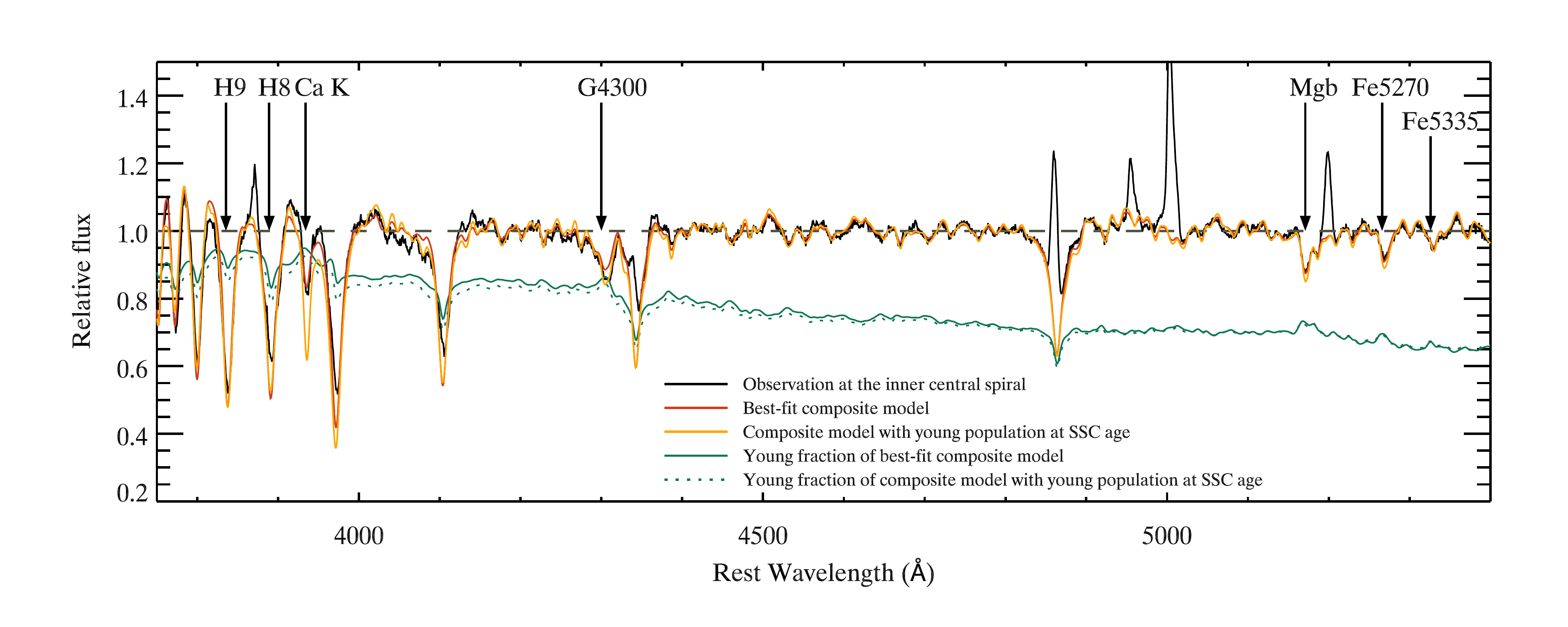}
    \includegraphics[width=\textwidth]{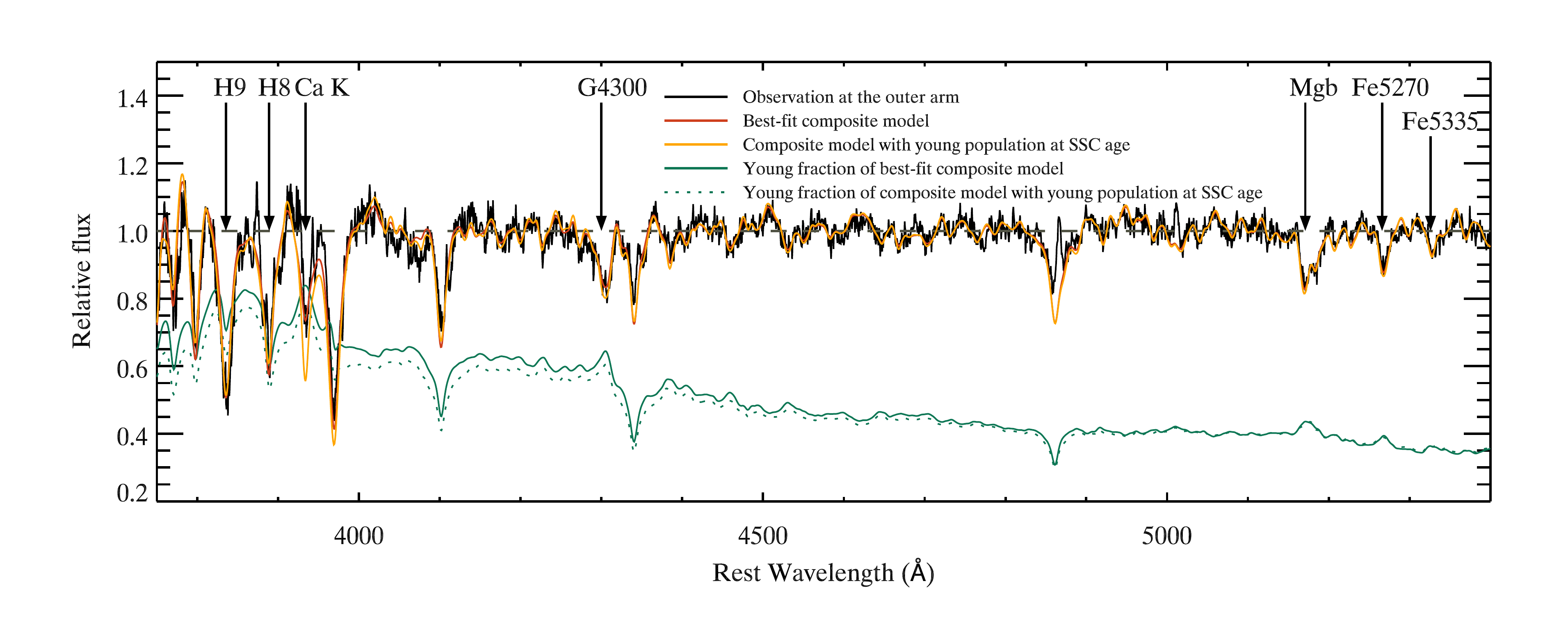}
    \caption{Same as Figure\,\ref{fig:composite_spec}, but overlaid on the measured spectra (yellow) are the composite model spectra with the young stellar population of $0.5\,{\rm Gyr}$, the age of the super star clusters (SSCs). The best-fit $Z=\rm{1.5\,Z_\odot}$ composite model spectra (red) identical to the one shown in Figure\,\ref{fig:composite_spec} are also overlaid on both panels for direct comparison. The fractional contribution to the flux from the younger population is 70\% for both model spectra at the inner central spiral and 40\% at the outer arm.}
    \label{fig:SSC}
\end{figure*}

NGC\,1275 hosts an enormous population of massive and compact star clusters \citep[see][and references therein]{Jeremy} that seemingly resemble globular clusters except for their relative youth (and perhaps also metallicity, which remains undetermined).  The first of these SSCs, comprising an especially luminous population lying within $\sim$5\,kpc of the center of NGC\,1275---the same spatial extent as the central spiral---were discovered by \citet{star_clus_ID} nearly three decades ago.  In Paper\,II \citep{paper2} of this series, we show that this central population of SSCs has different global physical properties than the much larger population of SSCs farther out, lying up to a radial distance of $\sim$30\,kpc from the center of NGC\,1275.  By comparison with the outer population of SSCs, the central population of SSCs has (i) a maximal luminosity ($\sim$$10^7 \, L_\sun$) and mass ($\sim$$10^7\, M_\sun$) that is an order of magnitude higher; (ii) a shallower luminosity and mass function; and (iii) an age of $0.5 \pm 0.1 \, \rm Gyr$, whereas the outer population of the SSCs span ages of $\sim$1\,Myr up to at least $\sim$1\,Gyr. The central SSC population is about 0.1\,mag redder in $B - V$, consistent therefore with its somewhat older age, than the central spiral.  In Figure\,\ref{fig:SSC}, we show the continuum-normalized spectra of a two-SSP composite model having the same parameters as those of the best-fit $Z=1.5\,Z_\odot$ models in Figure\,\ref{fig:composite_spec} except for an age of 0.5\,Gyr (instead of 0.15\,Gyr) for the younger stellar population, the age of the central SSC population.  As can be seen, these spectra greatly overpredict the depth of the Ca\,K line, indicating that the younger stellar population is more youthful than 0.5\,Gyr.

Informed by the theoretical simulations of \citet{Brockamp} for the survivability of newly formed globular clusters in present-day galaxies, we argue that the SSCs hosted by the central spiral have survived an early short but violent phase of strong tidal disruption near the center of NGC\,1275, during which both relatively low-mass and relatively larger SSCs were preferentially disrupted.  This brief but highly destructive phase would naturally explain why the mass function of the central population of SSCs is shallower than the mass function of the SSCs lying farther out.  Furthermore, as shown by \citet{Carlson2001}, the SSCs within the central $\sim$5\,kpc of NGC\,1275 exhibit a range of smaller radii than the SSCs farther out, suggesting that larger SSCs also have been preferentially disrupted.  The remaining SSCs associated with the central spiral are therefore those that are able to withstand the strong tidal fields near the center of NGC\,1275, and are losing mass gently over time owing to two-body relaxation.  In just $\gtrsim 0.5$\,Gyr, these SSCs will exhibit broadband colors resembling globular clusters, and be difficult to distinguish from genuine globular clusters.

As also shown in Paper\,II, the total initial stellar mass of the SSCs projected against the central spiral is $\sim$$3 \times 10^8 \, M_\sun$ over the mass range $\sim$$10^5 \, M_\sun$ (the detection threshold) to $\sim$$10^7 \, M_\sun$ (the most massive of the stellar clusters), corresponding to about one-tenth the initial stellar mass of the central spiral.  Owing to the preferential disruption of lower-mass SSCs, it is possible that the mass function of the central population of SSCs at birth was similar to that of the outer population of SSCs.  If so, then over the mass range $\sim$$10^5 \, M_\sun$ to $\sim$$10^7 \, M_\sun$ as before, the total initial stellar mass of the central SSC population would have been $\sim$$1 \times 10^9 \, M_\sun$, about one-third the initial stellar mass of the central spiral.

\subsection{Mass of Supermassive Black Hole}\label{SMBH}

The mass of the central supermassive black hole (SMBH) in NGC\,1275 has been estimated using a variety of methods.  Based on integral-field spectroscopy of the central $3\arcsec \times 3\arcsec$ region of NGC\,1275 in the  H$_2$\,1--0\,S(1) 2.122\,$\mu$m and [$\rm{Fe\,II}$] 1.644$\,\mu$m lines, \citet{scharwchter} detected a rotating disk that is visible in both lines within a radius of 0\farcs14 (50\,pc).  The (projected) rotation axis of this disk is closely aligned with the radio jets from the AGN in NGC\,1275.  For a disk inclination of $45^\circ \pm 10^\circ$, the best-fit model to the H$_2$\,1-0\,S(1) velocity field implies a central mass of $(8^{+7}_{-2}) \times 10^8 \rm \, M_\odot$, of which about half is estimated to reside in the disk and the other half accounted for by the SMBH.  \citet{Nagai} detected a disk in molecular gas as traced in both CO, HCN, and HCO$^+$ with a radius of $\sim0\farcs28$ ($\sim$100\,pc).  The (projected) rotation axis of this disk also is closely aligned with the radio jets from the AGN in NGC\,1275.  The best-fit model to the measurements yield a disk inclination of about $45^\circ$ and a mass for the SMBH of about $1 \times 10^9 \, M_\odot$.  Finally, by fitting CO absorption bandheads over the wavelength range 2.28--2.37$\,\mu$m, \citet{Riffel} deduced a stellar velocity dispersion of $265 \pm 26 {\rm \, km \, s^{-1}}$ over a radial range 0\farcs75--1\farcs25 (270--450\,pc).  They attribute this dispersion to the random orbits of stars about a central SMBH with a mass of $(1.1^{+0.9}_{-0.5}) \times 10^9 \, M_\odot$.

All the aforementioned works determine a SMBH mass of $M_{\rm BH} \simeq 1 \times 10^9  \, M_\odot$ for NGC\,1275, placing this galaxy neatly on the $M_{\rm BH}$--$\sigma_*$ relationship (whereby $\sigma_*$, in the case of elliptical galaxies, is simply their stellar velocity dispersion) for local galaxies \citep[e.g., see Figure\,1 of][]{sani}.  By contrast, using scaling relations between H$\beta$ linewidths and 5100\,{\AA} continuum luminosity from the AGN, \citet{Koss} derived a mass for the SMBH in NGC\,1275 of $\sim1.4 \times 10^7 \, M_\sun$.  Using scaling relations between Pa$\beta$ and X-ray luminosity from the AGN, \citet{Onori} derive a SMBH mass of $\sim(2.9 \pm 0.4) \times 10^7 \, M_\sun$ for NGC\,1275.  Both these SMBH mass estimates are about 2 orders of magnitude smaller than those estimated using gas and star kinematics on scales of tens of parsecs around the the nucleus of NGC\,1275 as mentioned above.  If correct, the SMBH in NGC\,1275 would then have a mass lying well below that expected from the $M_{\rm BH}$-$\sigma_*$ relationship even after taking into the account the intrinsic dispersion in this relationship \citep[see, e.g., discussion in][]{sani}.

We measure a (projected) rotation velocity of $\sim$$250 \rm \, km \, s^{-1}$ at a radius of $\sim$2\arcsec ($\sim$720\,pc) from the center of NGC\,1275.  The enclosed mass is $\sim$$1 \times 10^{10} \, M_\sun$/(sin $i$), where $i$ is the unknown inclination of the central spiral.  The mass of stars (and, if any, gas) belonging to the central spiral interior to a radius of $\sim$2\arcsec\ is not known, but is unlikely to greatly exceed about $10^9 { \, M_\sun}$, the estimated mass of stars in the central spiral between $\sim$2\arcsec--12\arcsec.  As discussed above, however, a much smaller disk having a different geometry has been detected interior to the innermost region measurable for the central spiral which makes any stricter constraint on the SMBH mass based on our work difficult to ascertain.

\section{Summary and Conclusions}\label{sec:conclusion}

The innermost region of NGC\,1275, spanning a radius of $\sim$5\,kpc, is especially blue, and exhibits spiral arms superposed on a roughly circular disk that is apparently centered on the nucleus of the galaxy (Figure\,\ref{fig:HST_img}).  Beyond the central spiral, semicircular arcs can be seen spanning a region extending radially outwards to $\sim$11\,kpc (see Figure\,8 of \citealt{conselice} and Figure\,19 of \citealt{penny}).  These features, apart from being bluer than the main stellar body of NGC\,1275, resemble tidal disturbances in the outer envelopes of elliptical galaxies undergoing a minor merger, thus motivating \citet{conselice} to attribute the central spiral to the remnant of a minor merger.   

To better elucidate the nature of the central spiral, we retrieved archival data from the KOA comprising long-slit optical spectroscopy taken with the LRIS \citep{LRIS} on the Keck I telescope.  The slit crosses the nucleus of NGC\,1275 at a PA of $128\degr$, enabling us to study the stellar kinematics and stellar populations toward the central spiral at an angular resolution of 1\arcsec (360\,pc).  Selecting stellar absorption lines that are least affected by blending as well as contamination by emission lines from the nebula associated with NGC\,1275, we found the H8 and H9 Balmer lines, along with the Ca\,K line, to peak in radial velocity at $\pm 250 {\rm \, km \, s^{-1}}$ of the systemic velocity at $\simeq 2\arcsec$ (720\,pc) on either side of the nucleus of NGC\,1275, and to slowly decrease outwards in radial velocity to reach the systemic velocity of NGC\,1275 at a radius of $\sim$12\arcsec\ (4.3\,kpc) on the southeastern side of the nucleus (measurements on the northwestern side being affected by the HVS, a distorted spiral galaxy falling toward NGC\,1275).  The widths of these stellar absorption lines are not appreciably broadened beyond the instrumental spectral resolution, which has a width at $1\sigma$ of $140 \rm \, km \ s^{-1}$.  The stellar population that produces these lines has a nominal age of 0.15\,Gyr, and is unlikely to have ages lying beyond the range $\sim$0.1--0.2\,Gyr.  We estimate a total stellar mass for this population, extrapolated over the entire central spiral, of $\sim10^9  \, M_\sun$.  This population is responsible for the relatively blue broadband colors of the central spiral, and contributes a decreasing fraction of the light toward the central spiral as a function of radius outwards.

By contrast, after accounting for the contribution from the $\sim$0.15\,Gyr population, we find the Mg\,$b$, Fe5270, and Fe5335 stellar absorption lines to have a radial velocity at, or close to, the systemic velocity of NGC\,1275 over the central spiral irrespective of radial location along the slit.  Furthermore, the widths of these absorption lines are appreciably broadened beyond the instrumental spectral resolution, and are consistent with a velocity dispersion of $\sim$$250 \rm \, km \, s^{-1}$ as previously measured by \citet{Heckman} over the radial range $\sim$2--9\,kpc, coinciding with and extending far beyond the central spiral.  The stellar population that produces these lines has an age around 10\,Gyr, and contributes an increasing fraction of the light toward the central spiral as a function of radius outwards.  We attribute this population to the main stellar body of NGC\,1275 that is dominated by very old stars.

Any interpretation of the central spiral needs to reconcile the following observational facts as established in our work: (i) the central spiral is kinematically decoupled from the main stellar body of NGC\,1275; (ii) its (dominant) stellar population has an age of only about 0.15\,Gyr, compared with an age of around 10\,Gyr for the old stellar population in the main body of NGC\,1275; and (iii) despite being decoupled both in kinematics and in stellar age, the central spiral appears to have settled into the center of NGC\,1275.  

We propose two alternative explanations for the nature of the central spiral.  First, it may indeed be the remnant of a minor merger as proposed by \citet{conselice}.  The troubling aspect of this interpretation is the very large cool-gas mass that the merger remnant would have had to possess so as to form $\sim$$3 \times 10^9 \, M_\sun$ in stars just 0.15\,Gyr ago.
Depending on the star-formation efficiency, the progenitor of the central spiral would have had to retain an enormous amount of gas (e.g., at least $\sim$$10^{10}$--$10^{11} \, M_\sun$ for a star-formation efficiency of 1\%--10\%) at the time it was slowed down sufficiently by dynamical friction to merge with NGC\,1275.  Before then, the progenitor galaxy would have experienced intense ram-pressure stripping owing to the high density of the intracluster medium along with the high velocity of its orbit, especially near pericenter.

Alternatively, the gas that fuelled intense star formation in the central spiral about 0.15\,Gyr ago could have slowly accumulated over time from a residual cooling flow.  This model can simultaneously explain the semicircular arcs around the central spiral as the historical signature of a minor merger, as well as the large quantities of gas it would have had to possess so as to fuel a recent starburst.  Of course, the requirement of having separate origins for the stellar and gaseous bodies can be alleviated by supposing that a gaseous disk first formed from a residual cooling flow, and then underwent a starburst about 0.15\,Gyr ago to form the stellar body observed today.  In this simpler scenario, the semicircular arcs beyond the central spiral do not comprise the signature of a past merger.  Rather, being bluer than the main body of NGC\,1275, these arcs comprise the trails of relatively young stars generated, perhaps, by the tidal disruption of SSCs that wandered too close to the center of NGC\,1275 along their orbits.

In Paper II of this series \citep{paper2}, we show that an especially luminous population of SSCs is spatially coincident with the central spiral.  By comparison with the even more numerous (by over an order of magnitude) population of SSCs farther out up to a radial distance of $\sim$30\,kpc from the center of NGC\,1275, the central population of SSCs has (i) a maximal luminosity ($\sim$$10^7 \, L_\sun$) and mass ($\sim$$10^7 \, M_\sun$) that is an order of magnitude higher; (ii) a shallower luminosity and mass function; and (iii) an age of $0.5 \pm 0.1 \, \rm Gyr$, whereas the outer population of SSCs span ages of $\sim$1\,Myr up to at least $\sim$1\,Gyr.  Informed by the theoretical simulations of \citet{Brockamp} for the survivability of newly formed globular clusters in present-day galaxies, we argue that the SSCs hosted by the central spiral are those that remain after an early brief but violent phase of tidal disruption that destroyed relatively low-mass and relatively large SSCs.   The remaining population are those that are able to withstand the strong tidal fields near the center of NGC\,1275, and are losing mass only gently over time owing to two-body relaxation.   Their total initial stellar mass is about one-tenth the initial stellar mass of the central spiral, but could have been about one-third after accounting for disrupted SSCs.  In just $\gtrsim 0.5$\,Gyr, these SSCs will exhibit broadband colors resembling globular clusters, and may be difficult to distinguish from genuine globular clusters.

In conclusion, a spiral disk hosting progenitor globular clusters, visibly lacking only a bulge and halo, appears to have recently formed at the center of the BCG in the Perseus cluster.

\begin{acknowledgments}

This work was supported by the Research Grants Council of Hong Kong through a General Research Fund (17300620) to J.L., that also provided partial support for work by M.C.H.Y. toward an MPhil thesis.  Y.O. acknowledges the support by the Ministry of Science and Technology (MOST) of Taiwan through the grant No. MOST 109-2112-M-001-021-.  This work was initiated at the 2019 ASIAA Summer Student Program. This research made use of the Keck Observatory Archive (KOA), which is operated by the W. M. Keck Observatory and the NASA Exoplanet Science Institute (NExScI), under contract with the National Aeronautics and Space Administration.  This research also employed observations made with the NASA/ESA Hubble Space Telescope and made use of archival data from the Hubble Legacy Archive, which is a collaboration between the Space Telescope Science Institute (STScI/NASA), the Space Telescope European Coordinating Facility (ST-ECF/ESAC/ESA), and the Canadian Astronomy Data Centre (CADC/NRC/CSA).

\end{acknowledgments}

\facilities{Keck:I (LRIS), HST (ACS WFC)}

\software{IRAF v2.16 \citep{IRAF,IRAF2}, RVSAO v2.7.8 \citep{rvsao}, INDEXF v4.3 \citep{indexf}}

\appendix

\section{Nebula Contamination}\label{appendix:nebula_contami}

\begin{figure}
\centering
\includegraphics[width=0.45\textwidth]{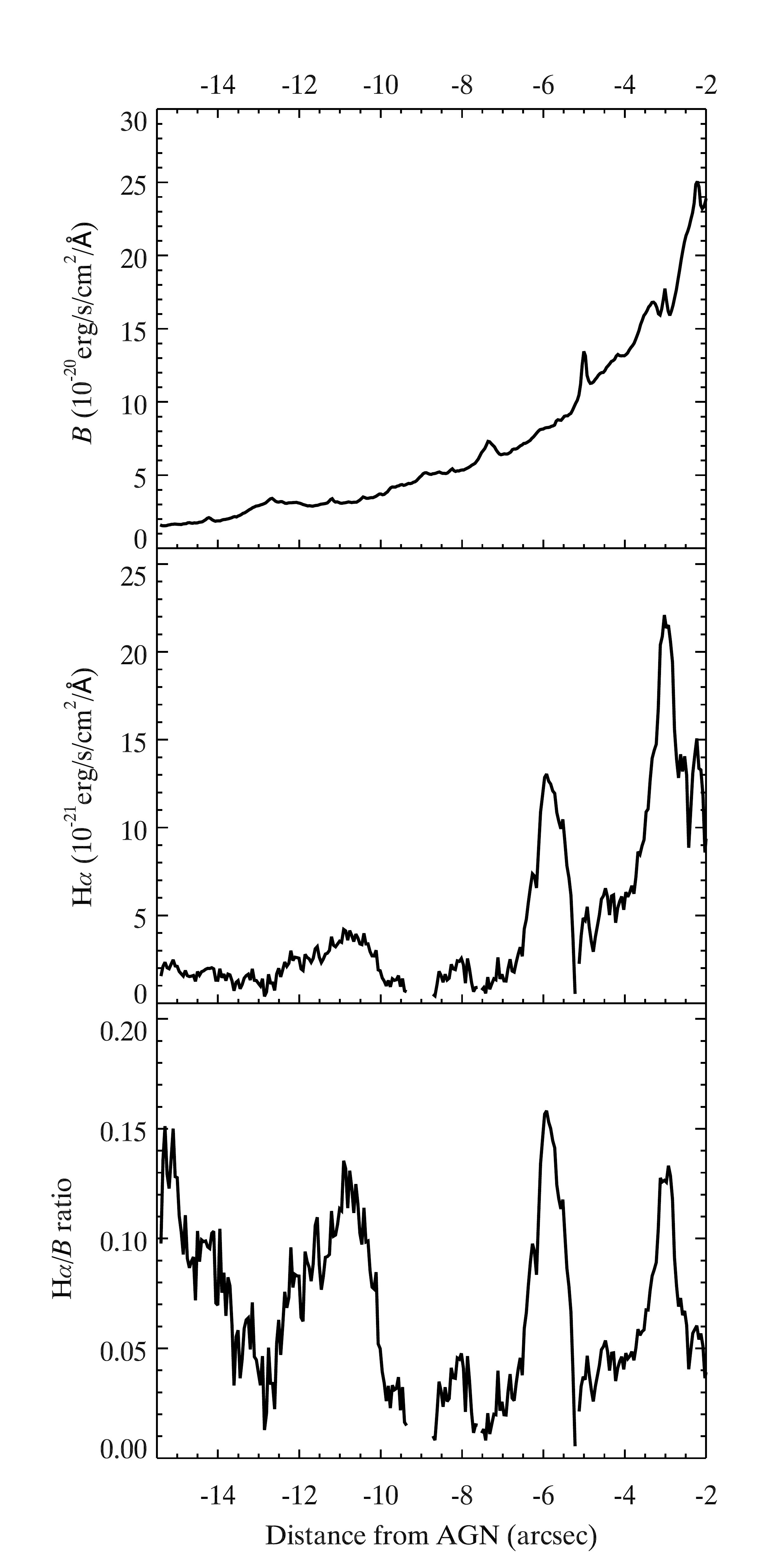}
\caption{One-dimensional extractions along the slit of the Hubble Space Telescope images ($B$ and H$\alpha$; Figure\,\ref{fig:HST_img}) and their ratio.
\label{fig:nebula_contami}}
\end{figure}

To evaluate the degree of contamination by emission lines from the nebula associated with NGC\,1275, we extracted, from the HST image, the brightness of the galaxy in the $B$ band as a function of radius along an aperture equivalent to the long-slit spectrosopic observation.   In addition, we extracted the brightness of the H$\alpha$ emission line as a function of radius, and computed the ratio in brightness between the H$\alpha$ line and the continuum in $B$ band.  The results are plotted in Figure\,\ref{fig:nebula_contami}.  The $B$-band continuum (top panel) decreases almost monotonically with radius outwards, exhibiting local enhancements that are much weaker than in H$\alpha$.
By contrast, the H$\alpha$ emission (middle panel) is strongly enhanced around $-3\arcsec$ (region referred to as \quotes{inner central spiral}), the innermost region measurable for the central spiral (within which light from the AGN dominates the measured spectra), as well as $-6\arcsec$ (region referred to as \quotes{nebula}), where the tip of a bright nebulae filament lies in the spectroscopic slit (Figure\,\ref{fig:HST_img}).  A weaker local enhancement can be seen at $-11\arcsec$ (region referred to as \quotes{red gap}), where the $B-V$ color is locally redder by $\sim0.1$\,mag (Figure\,\ref{fig:pos_index_color}).  

The flux ratio between H$\alpha$ and $B$-band continuum (bottom panel) is a good indicator of the relative degree of potential contamination by nebular emission lines to the measured line indices.  As can be seen, this ratio is relatively high at the \quotes{nebula} and \quotes{red gap}, indicating a relatively large contamination from the nebula at these regions. At the \quotes{inner central spiral}, the H$\alpha$/$B$-band ratio is smaller than at the \quotes{nebula} but still comparable to that at the \quotes{red gap}, indicating that nebular line emission can still affect the stellar Balmer absorption lines even at high order in all these regions.  These precautions are taken into account when deriving physical quantities from the measured line indices.

\section{Calibration Difficulties and Analytical Mitigation}\label{appendix:calibration_issue}

As mentioned in Section\,\ref{subsec:calibration_issue}, we encountered difficulties in calibrating the measured spectra for two reasons: (i) the lack of a reliable standard star spectrum from which to calibrate the spectral response of the instrument; and (ii) an optical ghost from the red channel of the LRIS that contaminates the measured spectra in the blue channel.  To mitigate against our inability to correct for the instrumental spectral response, we applied a continuum normalization to the measured spectra and adopted an analysis technique that could be directly be applied to such spectra.  In addition, we examined the degree to which the optical ghost contaminates the spectra in the blue channel as a function of both wavelength and position along the slit, and limited ourselves to just those slit positions and wavelength ranges where this contamination is negligible.

\subsection{Instrumental Spectral Response}

The standard star was observed at a rather high air mass of 1.8, and at a slit PA, far away from its parallactic angle ($d$(PA)$=108\degr$) without using an ADC.  As a consequence, there was an increasing loss of stellar flux at progressively shorter wavelengths perpendicular to the slit, of width $1\farcs5$, owing to large differential atmospheric dispersion.
Without a reliable standard star spectrum, we could not correct for the instrumental spectral response;
%
instead, we chose to work with continuum-normalized spectra, and employed analytical methods applicable to such data for determining the stellar velocity field (Section\,\ref{sec:stellar kinematics}), as well as for measuring line indices to determine the ages of stellar populations (Section\,\ref{sec:stellar population}).  Specifically, we cross-correlated the measured spectra with synthetic spectral templates for determining stellar velocities, a technique that relies only on localized spectral features and therefore does not require (nor is sensitive to any errors in the) calibration of the instrumental spectral response.  To derive line indices, we measured the integrated flux within a predefined wavelength range corresponding to the line, and divided the integrated flux thus derived with that of the continuum as measured on both sides of the line.  As the continuum flux is measured over wavelength ranges immediately bracketing the line of interest, the line indices thus derived are relatively insensitive to any errors in the overall spectral shape.  Furthermore, to avoid small systematic errors, we compared line indices measured from the continuum-normalized spectra with those derived from the continuum-normalized stellar syntheses templates.  Unfortunately with this technique, however, we could not measure $D$(4000), the continuum break index at 4000\,{\AA}, owing to the loss of this information after continuum normalization.

\subsection{Optical Ghost}

Toward compact objects like the standard star or the AGN in NGC\,1275, a secondary (unexpected) spectrum can be seen superposed on the real (expected) spectrum in the blue channel of the LRIS.  By identifying spectral features in the secondary spectrum, we found that this spectrum is simply the ghost image of the spectrum in the red channel.  Although the dichroic beam splitter should direct all light over a selected wavelength range to the red channel of the LRIS, obviously some of this light leaked inadvertently into the blue channel.  We determined the degree to which the blue channel is contaminated by the ghost spectrum from the red channel, in terms of the optics transmission (i.e., after accounting for differences in the dispersion and pixel scales between the real and ghost spectra), by computing the ratio in count rate between arc lines in the red channel and the same (ghost) arc lines in the blue channel (hereafter, referred to as the ghost count ratio).  This ratio rises monotonically toward shorter wavelengths ($\ll 0.1$\% at the G4300 line and rising to 0.3\% at the H9 line), and is especially severe shortward of 4000\,{\AA}.

The especially bright ghost spectrum of the AGN from the red channel strongly contaminates the spectrum in the blue channel over a broad range of wavelengths at a slit position $\simeq$8\arcsec\ northwest of the AGN.  At this position, even if the ghost spectrum was absent, dust extinction and emission lines associated with the HVS would have prevented us from deriving sensible results for the central spiral anyway.  Away from the AGN, the ghost spectrum from the red channel corresponding to the bright inner regions of the central spiral constitutes the next brightest source of contamination.  Once again, this contamination is largely restricted to slit positions northwest of the nucleus, where we would have had difficulty deriving sensible results for the central spiral in any case, owing to the presence of the HVS.

We evaluated the degree of contamination in the blue channel by the ghost spectrum from the red channel in the following manner.  Taking the spectrum in the red channel at the position of the AGN, we applied the necessary conversions such as the ghost count ratio (measured in the manner described above), the wavelength and pixel scale transformations from the red to blue channels, and the spatial displacement between the ghost and real spectrum, to compute how bright the ghost spectrum should appear in the blue channel at a slit position of $\simeq$8\arcsec\ northwest of the AGN (i.e., where the ghost spectrum of the AGN in the red channel appears, and also where the ghost spectrum at any slit position is brightest).  In this way, we found the measured line indices to be more affected by the ghost spectrum at shorter wavelengths and toward the outer regions of the central spiral: over the wavelength range selected for our analysis (see Figure\,\ref{fig:index_window}), the ghost flux fraction is highest at the H9 line (the shortest wavelength stellar absorption line considered in our analysis) and at the outer arm (the outermost region of the central spiral considered in our analysis at $-13\arcsec$).  Even then, however, the amount of contamination by the ghost spectrum is no more than $\sim$1.0\% in brightness over the entire spectrum in the blue channel.

To mitigate against the ghost spectrum, we confined our attention to slit positions on the southeastern side of the AGN in NGC\,1275 for the purpose of deriving stellar ages from line indices (see Appendix\,\ref{subsubsec:population:indices:measurement}).  In addition, we excluded slit positions between $-2\farcs0$ and $+2\farcs5$ from our analysis as the spectrum there is strongly contaminated by both continuum and line emission from the AGN.  As mentioned above, for deriving stellar kinematics, we cross-correlated the measured spectra with synthetic spectral templates, a method that is more immune to contamination from the ghost spectrum than simply fitting individual line profiles.
Thus, we were able to determine the velocity field of the central spiral not just on the southeastern but also on the northwestern side of the AGN, so long as the correlation significance, $R$, at a given location is sufficiently high (Section\,\ref{subsec:kinematics:crosscorrelation}).


\section{Line Indices}\label{subsec:line indices}


\subsection{Measurements}\label{subsubsec:population:indices:measurement}

Line indices corresponding or closely equivalent to Lick indices \citep{Lick_ori,Lick,Trager98} were measured for the stellar absorption lines listed in Table\,\ref{tab:indices_def} (refer also to Figure\,\ref{fig:index_window}).
Because Lick indices for Balmer absorption lines were originally defined only for transitions as high as H$\delta$, for the higher-order H8 and H9 lines we used line indices as defined by \citet{H8H9window}.
In addition, as the Lick index is not defined for Ca\,K, we define our own line index for Ca\,K as described below.
Line indices were measured only for absorption lines not coincident with or adjacent to bright emission lines from the nebula associated with NGC\,1275, with the exception of the higher-order H8 and H9 lines, provided that they are much less contaminated than their lower-order counterparts and place strong constraints on stellar age. Those for which flanking bandpasses, used to measure the pseudo-continuum, were difficult to define (e.g., the H10 and H11 lines) were also excluded.  We redefined the wavelength ranges for the H9, H8, and G4300 lines to minimize contamination by relatively weak nebular emission lines (see below).
For the remaining metal absorption lines comprising Mg\,$b$, Fe5270, and Fe5335, we employed the wavelength ranges normally defined for measuring Lick indices in these spectral lines.  Confusingly, different line indices can be derived for Mg\,$b$ depending on the wavelength ranges used for the flanking bandpasses; here, we use the flanking bandpasses for which the index thus derived is referred to as the Mg2 index.  The wavelength ranges over which all these line indices were measured are listed in Table\,\ref{tab:indices_def}.  In this table, the central bandpass corresponds to the wavelength range spanning a given absorption line, and the red and blue bandpasses to the wavelength ranges over which the pseudo-continuum for this line is defined.  All these wavelength ranges are defined at the rest wavelength, which for the H8, H9, and Ca\,K lines are  corrected for the measured radial velocity of the central spiral at each slit position (Fig\,\ref{fig:rv}, top-left panel), and which for the remaining lines are corrected for the systemic velocity of NGC\,1275.  An example of the bandpasses employed at a slit position  corresponding to the inner central spiral are shown by the shaded color regions in Figure\,\ref{fig:index_window}.  

The modified wavelength ranges over which line indices were measured in H8 and H9 minimize contamination by the [$\rm{Ne\,III}$] nebular emission line, which lies between H8 and H9 as is apparent (albeit not labeled) in Figure\,\ref{fig:index_window} and Figure\,\ref{fig:velocity_fields:profile_comp} (left column).  Nonetheless, as mentioned earlier, both these stellar Balmer absorption lines may be partially filled in by their emission-line counterparts from the nebula associated with NGC\,1275:
the values for both these line indices should therefore be regarded as lower limits.  In our work, the H8 and H9 line indices are particularly sensitive to a relatively young population having an age over the range 0.1--1\,Gyr, and despite the problems mentioned are therefore retained in our analyses.  The modified wavelength range for the G4300 line (produced by CH molecules) minimizes contamination by the H$\delta$ nebular emission line.  

As mentioned above, we measured a line index for Ca\,K that is free of contamination from nebular emission lines.  The Ca\,H line, however, overlaps with the H$\epsilon$ nebular emission line, and so we do not measure a line index for Ca\,H.  
Nonetheless, a comparison between the line strengths of Ca\,K and Ca\,H$+$H$\epsilon$ provided a useful diagnostic of stellar age owing to the different age evolution of the Ca and Balmer absorption line strengths \citep{Rose,Leonardi}.  For the same reason, a comparison between the line strengths of Ca\,K and H8 (the nearest unblended Balmer absorption) also provided a useful diagnostic of stellar age.  

Although we measured line indices for Fe5270 and Fe5335 separately, to improve the S/N we adopted the combined iron index $\langle$Fe$\rangle=($Fe5270$+$Fe5335$)/2$ \citep[e.g.,][]{Trager1,mFe}.  As mentioned in Section\,\ref{subsec:calibration_issue}, we could not measure the continuum break index at 4000\,{\AA}, $D(4000)$, owing to problems with the spectral-response calibration.
Instead, we make use of the $D(4000)$ index as measured by \citet{johnstone} and listed in Table\,\ref{tab:indices_def} for diagnostic purposes.

\begin{deluxetable}{lccc}
\tablecaption{Bandpass definitions at rest wavelengths for measuring line indices.\label{tab:indices_def}}
\tablehead{
\colhead{} & \colhead{Blue Bandpass} & \colhead{Central Bandpass} & \colhead{Red Bandpass} \\
\colhead{Index} & \colhead{(\AA)} & \colhead{(\AA)} & \colhead{(\AA)}
}
\startdata
H9\tablenotemark{a} & 3810.000 \text{--} 3820.000 & 3825.000 \text{--} 3845.000 & 3850.000 \text{--} 3860.000 \\
H8\tablenotemark{a} & 3850.000 \text{--} 3860.000 & 3875.000 \text{--} 3900.000 & 3905.000 \text{--} 3915.000 \\
Ca\,K\tablenotemark{b} & 3910.000 \text{--} 3920.000 & 3920.000 \text{--} 3947.000 & 4000.000 \text{--} 4020.000 \\
G4300\tablenotemark{c} & 4266.375 \text{--} 4282.625 & 4281.375 \text{--} 4316.375 & 4400.000 \text{--} 4416.000 \\
Mg2 & 4895.125 \text{--} 4957.625 & 5154.125 \text{--} 5196.625 & 5301.125 \text{--} 5366.125 \\
Fe5270 & 5233.150 \text{--} 5248.150 & 5245.650 \text{--} 5285.650 & 5285.650 \text{--} 5318.150 \\
Fe5335 & 5304.625 \text{--} 5315.875 & 5312.125 \text{--} 5352.125 & 5353.375 \text{--} 5363.375 \\
$D$(4000)\tablenotemark{d} & 3750.000 \text{--} 3950.000 & \nodata & 4050.000 \text{--} 4250.000 \\
\enddata
\tablenotetext{a}{Revised from the original definitions of \citet{H8H9window}.}
\tablenotetext{b}{Newly defined in this paper.}
\tablenotetext{c}{Revised from the original definition of \citet{Trager98}.}
\tablenotetext{d}{Measured by \citet{johnstone}.}
\end{deluxetable}

The \quotes{indexf} software package \citep{indexf} was used to derive all the aforementioned line indices.
Among the required input parameters are radial velocity and its associated uncertainty: for H8, H9, and Ca\,K, we used the best-fit radial velocity and its associated uncertainty for the central spiral at different slit positions as determined in the manner described in Section\,\ref{subsec:kinematics:velocity_fields}, whereas we use the systemic velocity of NGC\,1275 and adopted an uncertainty of 40\,km\,s$^{-1}$ for the remaining metal lines, G4300, Mg\,$b$, Fe5270, and Fe5335, at all slit positions.
Measurement uncertainties in line indices owing to measurement uncertainties in both flux densities and radial velocities are calculated by the software.  To avoid later confusion, we mention at this stage that the H8, H9, and Ca\,K lines contain a minor contribution from the older stellar population at the systemic velocity of NGC\,1275, whereas the G4300, Mg\,$b$, Fe5270, and Fe5335 lines contain a minor contribution from the younger stellar population at the velocity of the central spiral.  Because these two populations differ little in velocity, as demonstrated in Section\,\ref{subsec:kinematics:velocity_fields}, the line indices thus computed reflect the contributions from both stellar populations.

We confine measurements of line indices to the southeastern side of the center of NGC 1275 at slit positions between $-2\arcsec$ and $-15\farcs5$.  Further outwards ($< -15\farcs5$), the S/N becomes too low for sensible measurements of line indices in many, if not all, of the spectral lines considered.   Like for the velocity field, line indices were not measured between $-2\farcs0$ and $+2\farcs5$ owing to strong contamination by bright emission lines associated with the AGN in NGC\,1275.  Farther outwards along the northwestern side of the center ($> +2\farcs5$), we did not measure line indices owing to contamination by the optical ghost from the red channel (see Section\,\ref{subsec:calibration_issue}), as well as spectral lines produced by the HVS.  Unlike the cross-correlation analysis that helps minimize contamination by emission lines when determining radial velocities, such contamination causes systematic errors in line index measurements of, especially, relatively faint spectral lines.   

\begin{figure*}
\centering
\includegraphics[width=\textwidth]{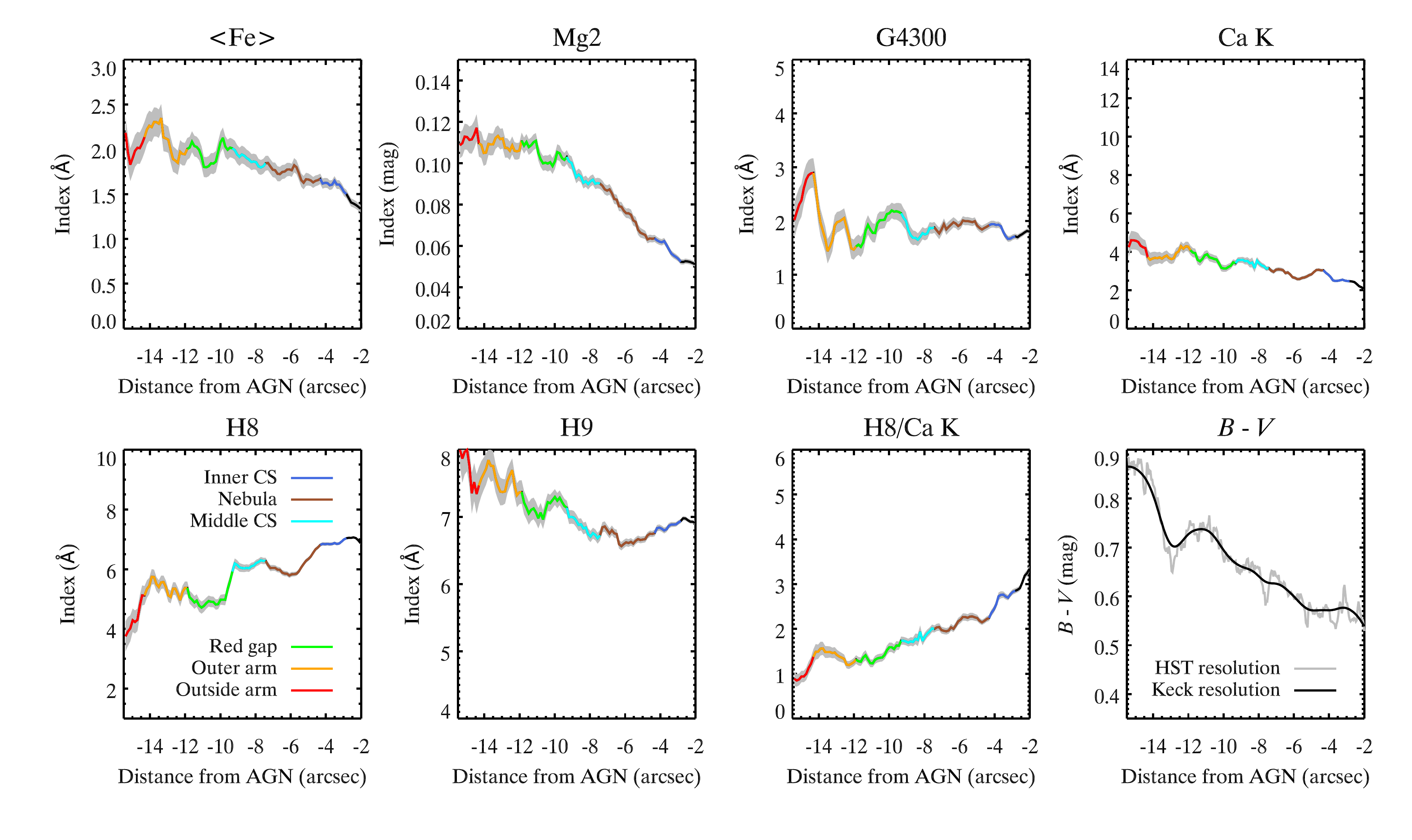}
\vspace{-1.0cm}
\caption{Radial variations in the measured line indices, H8/Ca\,K index ratio, and $B-V$ color.  The Balmer and Ca\,K line indices are measured at the radial velocity of the central spiral at each position, whereas the remaining line indices are measured at the systemic velocity of NGC\,1275 (Section\,\ref{subsec:kinematics:velocity_fields}).  All the measured line indices are smoothed to $\simeq1\arcsec$, the natural seeing.  The $\pm 1 \sigma$ uncertainty in the line index or index ratio measurements are depicted by the gray or color band. All indices are shown by different colors for the six different subregions (see Figure\,\ref{fig:HST_img} and Table\,\ref{tab:region}).  The $B-V$ color is shown at both the original angular resolution of the Hubble Space Telescope (gray) and after smoothing to a seeing of 1\arcsec\ (black).}
\label{fig:pos_index_color}
\end{figure*}

To suppress fluctuations in the measured line indices at spatial scales smaller than the seeing, all measurements of line indices at each slit position were smoothed by taking a weighted mean of 7 pixels (corresponding to the seeing of $\simeq1\arcsec$).  We also measured the $B-V$ color along the slit from Figure\,\ref{fig:HST_img} (middle panel).  The results for both the line indices and $B-V$ color as a function of slit position are shown in Figure\,\ref{fig:pos_index_color}.  We also averaged the measured spectra over different subregions to derive an average line index for each subregion as listed in Table\,\ref{tab:indices_measured}.

\begin{deluxetable}{lCCCCCC}
\tablecaption{Measured line indices and $B-V$ color at the six subregions (see Figure\,\ref{fig:HST_img}).}
\label{tab:indices_measured}
\tablehead{
\colhead{} & \colhead{Outside} & \colhead{Outer} & \colhead{Red} & \colhead{Middle} & \colhead{Nebula} & \colhead{Inner}\\
\colhead{Index} & \colhead{Outer arm} & \colhead{Arm} & \colhead{Gap} & \colhead{Central spiral} & \colhead{} & \colhead{Central spiral}
}
\startdata
\rm{H8} (\AA) & 4.33\pm0.34 & 5.30\pm0.14 & 4.98\pm0.12 & 6.16\pm0.09 & 6.19\pm0.04 & 6.88\pm0.04\\
\rm{H9} (\AA) & 7.66\pm0.28 & 7.49\pm0.12 & 7.15\pm0.08 & 6.86\pm0.06 & 6.71\pm0.03 & 6.87\pm0.03\\
\rm{Ca\,K} (\AA) & 4.32\pm0.41 & 3.92\pm0.17 & 3.45\pm0.12 & 3.42\pm0.11 & 2.88\pm0.05 & 2.58\pm0.05\\
\rm{G4300} (\AA) & 2.52\pm0.26 & 1.80\pm0.13 & 1.97\pm0.10 & 1.78\pm0.09 & 1.91\pm0.04 & 1.78\pm0.04\\
\rm{Mg2} (mag)& 0.113\pm0.006 & 0.109\pm0.003 & 0.104\pm0.002 & 0.093\pm0.002 & 0.073\pm0.001 & 0.058\pm0.001\\
$\langle\rm{Fe}\rangle$ (\AA) & 2.01\pm0.16 & 2.07\pm0.09 & 1.97\pm0.07 & 1.87\pm0.06 & 1.71\pm0.04 & 1.61\pm0.04\\
$B-V$ (mag) & 0.86 & 0.75 & 0.72 & 0.65 & 0.60 & 0.57 \\
\enddata
\end{deluxetable}

\begin{figure*}
\centering
\vspace{-0.5cm}
\includegraphics[width=\textwidth]{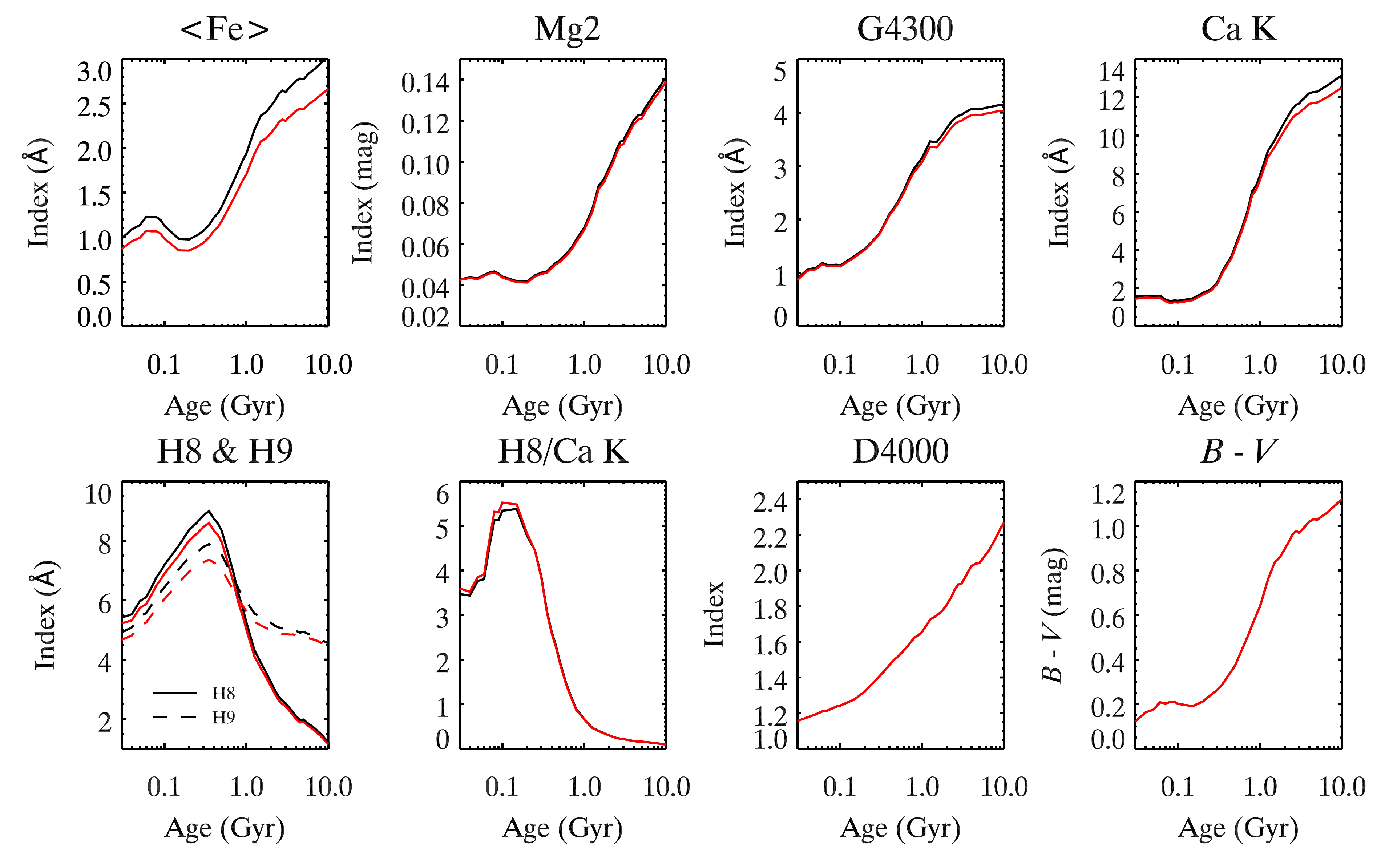}
\caption{Age dependences of the line indices, index ratio, and color measurements shown in Figure\,\ref{fig:pos_index_color}, and in addition the D(4000) line index, as predicted by the model solar-metallicity single stellar population.  The predictions depicted by the red and black lines take into account the instrumental spectral resolution, and for the red line also a stellar velocity dispersion of $246 {\rm \, km \,s^{-1}}$.  Only one line is plotted for both $D$(4000) and $B-V$ as smoothing by the stellar velocity dispersion has a negligible effect on both.}
\label{fig:sing_calib}
\end{figure*}

To make a meaningful comparison between model and measured line indices, we constructed line indices from model spectra in exactly the same manner.  After continuum normalization (Section\,\ref{subsec:calibration_issue}), we applied a Gaussian smoothing to the model spectra commensurate with the spectral resolution at the individual lines as measured with the LRIS (Section\,\ref{subsec:obs}).  An additional Gaussian smoothing was also applied to the Mg\,$b$, Fe5270, Fe5335 and G4300 lines, but not the H8, H9, or Ca\,K lines (see reasons in Section\,\ref{subsec:kinematics:velocity_fields}), to account for a stellar velocity dispersion of $246 {\rm \, km \,s^{-1}}$ as measured by \citet{Heckman}.  Figure\,\ref{fig:sing_calib} shows the model line indices thus generated for solar-metallicity SSPs having different ages and either accounting (red curves) or not (black curves) for the aforementioned stellar velocity dispersion.


\subsection{Spatial Trends}\label{subsubsec:population:indices:spatial_trends}

At the inner central spiral, the H8 index is as large as $\sim$7\,{\AA} (Figure\,\ref{fig:pos_index_color} and Table\,\ref{tab:indices_measured}), a value that should be considered a lower limit owing to possible contamination by the corresponding nebula emission line.  Even then, this value is close to the largest among the post-starburst galaxies simulated by \citet{H8H9window} of $\sim$10\,{\AA}.  The H9 index at this position is also large, having a similar value of $\sim$7\,{\AA}.  The Balmer absorption depths are a strong function of age: enhanced at 0.1--1\,Gyr during the post-starburst phase, they reach their greatest depths at an age of 0.35\,Gyr when main-sequence A-type stars dominate the integrated SSP spectrum (see Figure\,\ref{fig:sing_calib}, and also Appendix\,\ref{appendix:h9} for more details about H9).
The large (lower limits in) H8 and H9 indices that we measure at the inner central spiral therefore indicate that the dominant stellar population there is only multiple 0.1\,Gyr old.  Similarly, the very small Mg2, $\langle$Fe$\rangle$, and Ca\,K indices that we measure at the inner central spiral can naturally be explained by a relatively young stellar population (compare the measurements in Figure\,\ref{fig:pos_index_color} with the model predictions in Figure\,\ref{fig:sing_calib}).  

Very deep Balmer absorption at higher orders has previously been reported toward the inner regions of NGC\,1275 \citep{Minkowski_first,Sidney,Rubin,johnstone,Crawfod93,brodie}.
Such deep Balmer absorption, without any accompanying evidence for \ion{H}{2} regions (e.g., \citealt{star_clus_ID,Norgaard2,ferruit,conselice,Fabian}), is consistent with a stellar population having an age somewhere in the range $\sim$0.1--1\,Gyr.

The H8 index gradually decreases outwards\footnote{We found two regions within the central spiral showing locally shallower Balmer absorption on top of an otherwise almost continuously decreasing trend.  One is the \quotes{nebula} region (at $-6\arcsec$), and another is the \quotes{red gap} showing locally redder $B-V$ color (at $-11\arcsec$).  See Appendix\,\ref{appendix:nebula_contami} for more about these regions.}, and is $\sim$5.4\,{\AA} at about $-13\arcsec$ on the outer arm.  Even at this position, however, the H8 absorption is still deeper than that of old stellar populations in elliptical galaxies \citep{H8H9window}.  Similarly, the H9 index decreases outwards until about $-6\arcsec$, beyond which this index increases while the H8 index continues to decrease.  Such an anti-correlation cannot be simply explained by changing the SSP age as a function of radius.  In Appendix\,\ref{appendix:h9}, we explain that this anticorrelation is likely caused by contamination of the H9 line by the $\rm{Mg\,I}$ line, and so we preferentially focus more on H8 than H9 in our analysis.  

By contrast with H8, the Mg2, $\langle$Fe$\rangle$, and Ca\,K indices increase gradually outwards, with a more gradual rise beyond about $-10\arcsec$.  The ratio between the H8 and Ca\,K indices (H8/Ca\,K index ratio) decreases rapidly outwards by a factor of $\sim$3 over just $10\arcsec$ from the center.  As shown in Figure\,\ref{fig:sing_calib}, the H8/Ca\,K index ratio decreases monotonically with increasing stellar age beyond 0.1\,Gyr, unlike the Balmer indices that continue to increase with age until about 0.35\,Gyr before decreasing.  Thus, the H8/Ca\,K index ratio indicates an increasing SSP age with radius, corresponding to an age of 0.3\,Gyr at the inner central spiral at $-4\arcsec$ and 0.8\,Gyr on the outer arm at $-13\arcsec$ (Figure\,\ref{fig:sing_calib}).
The approximate ages that satisfy both the H8 index and the H8/Ca\,K index ratio measurements at the different subregions are listed in Table\,\ref{tab:sing_best-fit}, along with the model-predicted strengths of the different line indices considered and $B-V$ color.

While providing more stringent constraints on ages, an analysis confined to individual line indices does not tell the full story; that is, as demonstrated in Section\,\ref{subsec:kinematics:velocity_fields}, different stellar populations having different ages, radial velocities, and velocity dispersions dominate the H8, H9, and Ca\,K lines by comparison with the Mg\,$b$ and Fe lines.  As we will show next in Appendix\,\ref{subsubsec:population:indices:index_index}, invoking just one SSP at any given location, albeit with different ages at different locations, cannot reproduce all the line indices simultaneously.

\begin{deluxetable}{lCCCCCC}
\tablecaption{Best-fit solar-metallicity SSP age based on H8 line index and H8/Ca\,K index ratio.\label{tab:sing_best-fit}}
\tablehead{
\colhead{} & \colhead{Outside}& \colhead{Outer} & \colhead{Red} & \colhead{Middle} & \colhead{Nebula} & \colhead{Inner} \\
\colhead{} & \colhead{Outer arm}& \colhead{Arm} & \colhead{Gap} & \colhead{Central spiral} & \colhead{} & \colhead{Central spiral} 
}
\startdata
Age (Gyr) & 1.25 & 1.00 & 1.00 & 0.80 & 0.80 & 0.70 \\
H8 ({\AA}) & 4.33 & 5.26 & 5.26 & 6.21 & 6.21 & 6.89 \\
H9 ({\AA}) & 5.56 & 5.95 & 5.95 & 6.39 & 6.39 & 6.73 \\
Ca\,K ({\AA})& 9.19 & 7.92 & 7.92 & 7.07 & 7.07 & 6.02 \\
G4300 ({\AA}) & 3.36 & 3.07 & 3.07 & 2.89 & 2.89 & 2.73 \\
Mg2 (mag) & 0.076 & 0.067 & 0.067 & 0.064 & 0.064 & 0.057 \\
$\langle\rm{Fe}\rangle$ ({\AA}) & 1.93 & 1.71 & 1.71 & 1.54 & 1.54 & 1.44 \\
$D$(4000) & 1.72 & 1.66 & 1.66 & 1.62 & 1.62 & 1.59 \\
$B-V$ (mag) & 0.76 & 0.64 & 0.64 & 0.55 & 0.55 & 0.50 \\
\enddata
\end{deluxetable}

Although we could not measure the $D$(4000) index from our continuum-normalized spectra (Section\,\ref{subsec:calibration_issue}), \citet{johnstone} have previously measured this line index for the central spiral.  Their spectroscopic observations utilized a slit crossing the nucleus at a $PA=103.9{\degr}$, only somewhat different from that utilized in the observations reported here of $PA=128{\degr}$.  They reported a shallow and almost constant $D$(4000) index of $\sim$1.5 at most off-nuclear positions along their slit, except within $\pm 1\arcsec$ of the nucleus where the spectra are contaminated by AGN-related emission.  This value for the $D$(4000) index is much smaller than that typically measured for massive elliptical galaxies of $D(4000) = 2.4$ \citep{johnstone} or $D_{\rm n}(4000) \sim 2$ \citep{Kauffmann}\footnote{$D_{\rm n}$(4000) measures the continuum break at 4000\,{\AA} in a very similar manner as $D$(4000), but with a slightly modified bandpass as defined by \citet{Kauffmann} compared with that originally defined for $D$(4000) by \citet{Bruzual}.}.  \citet{johnstone} argued that the $D$(4000) index is smaller than that expected for old stellar populations in massive elliptical galaxies owing to the presence of hot stars (see also \citealt{Johnstone87}).  Indeed, a young stellar population having an age of $\sim$0.5\,Gyr is predicted to have $D(4000)=1.5$ (Figure\,\ref{fig:sing_calib}), and can therefore naturally explain the $D$(4000) index measured for the central spiral.

\subsection{Index--index Trends}\label{subsubsec:population:indices:index_index}

We now analyze index--index trends (the manner in which one line index varies with another) to extract mutual relationships between spectral features.  
This analysis has the advantage that we do not have to simultaneously fit for multiple line indices, but simply examine how well model predictions reproduce the observed trends.
%

\begin{figure*}
\centering
\includegraphics[width=\textwidth]{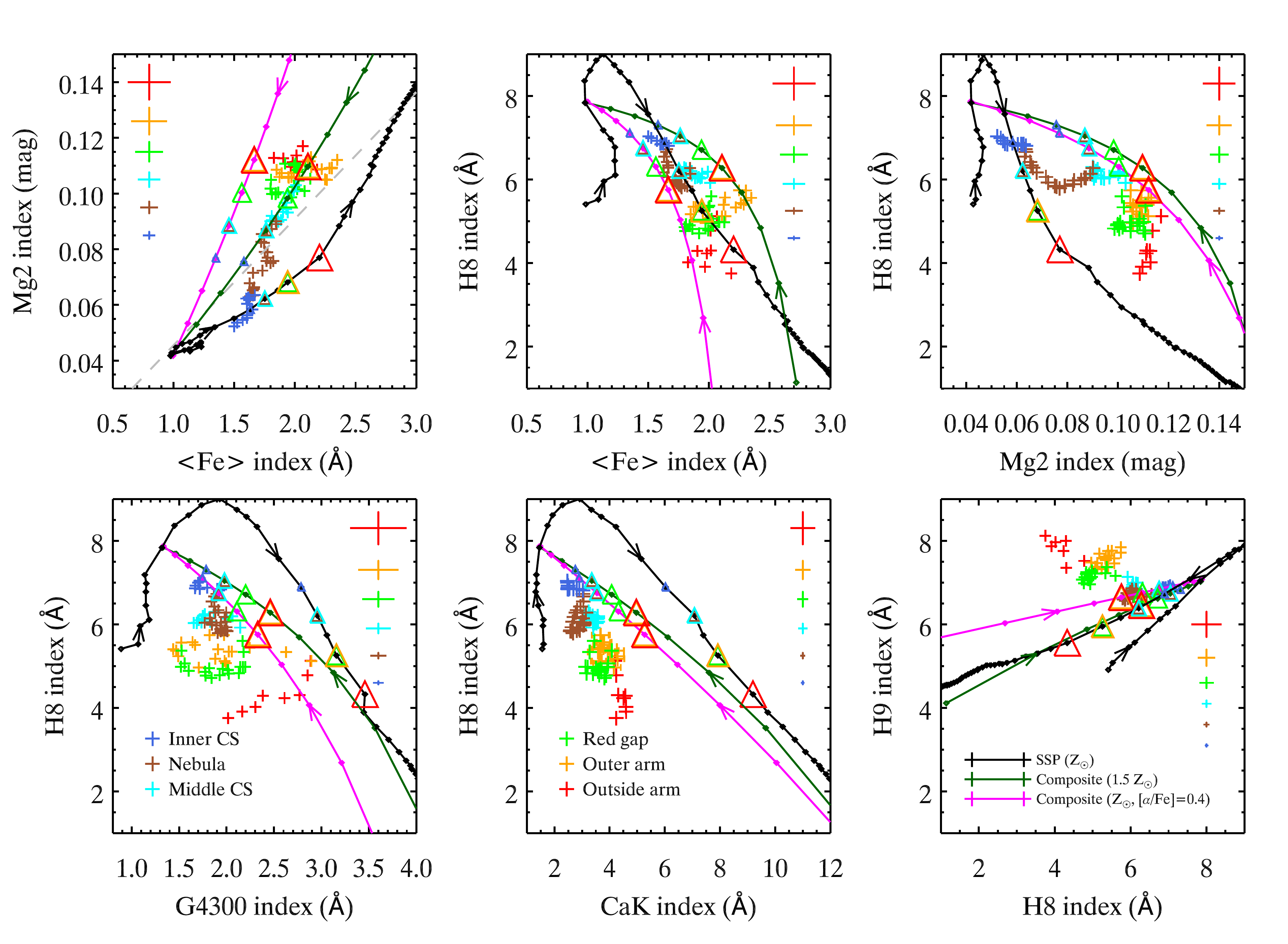}
\caption{Measured line indices (+ symbols) in the six subregions (see Table\,\ref{tab:region}), plotted as one line index versus another.  Measurements over each subregion are colored differently (see legends in the lower-left and lower-middle panels for subregion identification).  The arm lengths of the + symbols stacked vertically in each panel indicate the typical $\pm 1\sigma$ measurement uncertainty for the individual line indices measured in each subregion.  Overlaid curves are model-predicted index--index loci for (i) a solar-metallicity single stellar population (SSP) ranging in age from 30\,Myr--14\,Gyr (black); two-SSP composite models having ages of 0.15\,Gyr for the younger and 10\,Gyr for the older SSP, along with (ii) $Z=1.5\,Z_\odot$ (dark green) or (iii) $Z=Z_\odot$ $\alpha$-enhanced (magenta), ranging in fractional contribution to the total flux from the younger SSP from 0\%--100\%. Arrows are drawn on each loci to indicate the direction of increasing age for the solar-metallicity SSP or increasing flux contribution from the younger SSP in the composite models. Triangles (colored according to subregions and plotted with different sizes purely for clarity) along each locus correspond to the closest representative values to the measured line indices averaged each subregion, which for the lone SSP model is based primarily on agreement with the H8 index (Section\,\ref{subsec:population:SSP}), and for the two-SSP composite models are based on agreement with all the line indices considered as a whole  (Section\,\ref{subsec:population:composite}). The dashed line in the upper-left panel corresponds to a constant Mg2/$\langle$Fe$\rangle$ index ratio of $=0.045$.}
\label{fig:in-in}
\end{figure*}

In Figure\,\ref{fig:in-in}, we plot measurements of one line index versus another for the six subregions defined in Table\,\ref{tab:region} in different colors.  Significant correlations or anticorrelations can be seen for each set of line indices, suggesting that a simple model involving only a small number of parameters can explain the observed trends.  Specifically, we found clear anticorrelations between the H8 versus the $\langle$Fe$\rangle$ (upper-central panel), Mg2 (upper-right panel), and Ca\,K (lower-central panel) indices.  The H8 versus the G4300 indices also show a similar anticorrelation (lower-left panel), albeit with a larger scatter than those between the H8 versus the $\langle$Fe$\rangle$, Mg2, or Ca\,K indices.  Conversely, there is a clear positive correlation between the Mg2 versus the $\langle$Fe$\rangle$ indices (upper-left panel).

The model age locus for the adopted solar-metallicity SSP is indicated by the black curve in the individual panels of Figure\,\ref{fig:in-in}.  As can be seen, the index--index trends can be qualitatively explained by an increasing SSP age outwards, starting from an age of about 0.35\,Gyr at the innermost region where the H8 index is largest (refer to Figure\,\ref{fig:sing_calib})---the same conclusion reached considering the H8/Ca\,K index ratio alone (Appendix\,\ref{subsubsec:population:indices:spatial_trends}).  
In the index--index plots of Figure\,\ref{fig:in-in}, however, the deficiencies of such a simple explanation become apparent.  The SSP model indicated by the black locus in this figure significantly and systematically overestimates, for a given H8, the G4300 (lower-left panel) and Ca\,K (lower-central panel) indices over the entire central spiral, as well as significantly and systematically underestimates the Mg2 index (upper-right panel) at the middle central spiral and on the outer arm.
In addition, this SSP model systematically underestimates the Mg2 index for a given $\langle$Fe$\rangle$ index (upper-left panel) beyond the inner central spiral.  Thus, invoking just one SSP at a given location, albeit with different ages at different locations, cannot reproduce all the line indices simultaneously. 
%

As mentioned earlier, and as can be seen also in Figure\,\ref{fig:in-in} (lower right panel), the H8 and H9 indices exhibit positive and negative correlations, respectively, at the inner and outer regions of the central spiral.  The negative correlation is caused by an increasingly weaker H8 index outwards, whereas the H9 index initially decreases outwards before reversing at about $-6\arcsec$ (Figure\,\ref{fig:pos_index_color}).  This reversal is attributed to contamination of the H9 line by the $\rm{Mg\,I}$ line (Appendix\,\ref{appendix:h9}), such that the degree of contamination increases outwards as the strength of the H9 line from the younger stellar population decreases more quickly outwards than the strength of the $\rm{Mg\,I}$ line from the old stellar population associated with NGC\,1275.

\subsection{Two Stellar Populations Composite Fits}\label{subsubsec:2-SSP:index_index}
In Section\,\ref{subsec:kinematics:velocity_fields}, we showed that the stellar absorption lines in the blue and red correlation ranges have different radial velocities and velocity dispersions (Figure\,\ref{fig:velocity_fields:profile_comp} and Figure\,\ref{fig:rv}).  Employing therefore two-SSP composite models, we tuned the individual ages of these two stellar populations so as to best match the index--index measurements shown in Figure\,\ref{fig:in-in}.
%
%
To make a direct comparison with the index--index measurements, we convolved the model spectrum for each of the SSPs in the two-SSP composite model by the instrumental spectral resolution, and in addition that for the older SSP by a velocity dispersion of $246 {\rm \, km \, s^{-1}}$ (corresponding to that of the old stellar population comprising the main stellar body of NGC\,1275) based on our analysis of Section\,\ref{subsec:kinematics:velocity_fields}.  As can be seen in Figure\,\ref{fig:sing_calib}, the model line profile and hence line index for $\langle$Fe$\rangle$ is more sensitive to convolution by the stellar velocity dispersion than that of the other lines studied.
These model spectra were then combined over a range of different relative fluxes for the two SSPs, and each normalized by the resulting (summed) continuum spectrum.  Initially just for simplicity, we adopted the same radial velocity for the two SSPs corresponding to the systemic velocity of NGC\,1275 ($5284 \rm \, km \, s^{-1}$).  We determined line indices from the model spectra using the same method as for the corresponding line indices in the measured spectra (see Appendix\,\ref{subsubsec:population:indices:measurement}), but now of course at the systemic velocity of NGC\,1275 for all the spectral lines irrespective of whether the major contributor is the younger or older SSP.  This simplification allows an essentially direct comparison with the measured line indices -- the values for which reflect the contributions from both stellar populations despite their small velocity differences as mentioned in Appendix\,\ref{subsubsec:population:indices:measurement} -- shown in Figure\,\ref{fig:in-in}, without the extra complication of having to produce model spectra having different combinations of velocities for the two SSPs.

\begin{figure*}
\centering
\includegraphics[width=\textwidth]{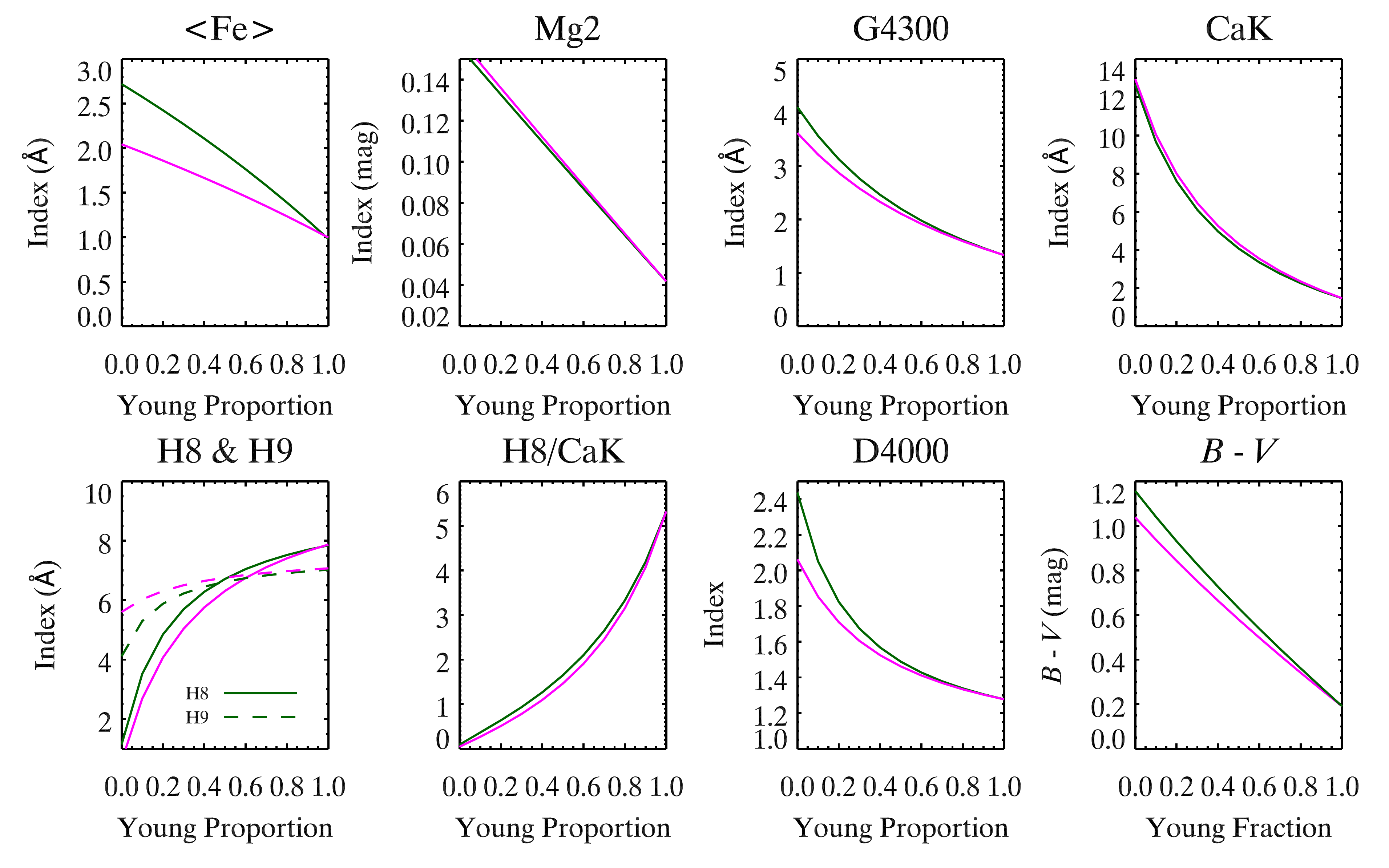}
\caption{Model-predicted line indices as a function of fractional contribution in flux from the younger SSP in the two-single stellar population (SSP) composite model.  This model has an age of 0.15\,Gyr for the younger SSP and an age of 10\,Gyr for the older SSP, and either $Z = 1.5 \,Z_\odot$ (dark green curve) and $Z=Z_\odot$ $\alpha$-enhanced (magenta curve), that we find to best reproduce the index--index plots of Figure\,\ref{fig:in-in}.
\label{fig:comp_calib}}
\end{figure*}

The process of finding a two-SSP composite model that best agrees with the index--index measurements shown in Figure\,\ref{fig:in-in} required extensive trial-and-error to encroach on the optimal match.  In brief, we tried ages for the younger SSP over the range 0.08--0.5\,Gyr in steps of 0.05\,Gyr and 0.5--1\,Gyr in steps of 0.1\,Gyr, and ages for the older SSP over the range 1.5-14.0\,Gyr in steps of 0.5\,Gyr.  At a given age for the younger SSP, we generated a suite of two-SSP composite models comprising different ages for the older SSP.  After applying a continuum normalization to the combined model spectrum for a given two-SSP composite model, we computed the model-predicted line indices: by considering different relative contributions to the total flux from the two SSPs, we generated model loci showing how a given line index varies with a different fractional contributions to the total flux.  An example is shown in Figure\,\ref{fig:comp_calib}, for which the younger SSP has an age of 0.15\,Gyr and the older SSP an age of 10\,Gyr for either $Z = 1.5 \, Z_\sun$ (dark green curve) or $Z = Z_\sun$ $\alpha$-enhanced (magenta curve).   By comparing (overlaying) these model loci with the measured index--index plots shown in Figure\,\ref{fig:in-in}, we determined how well a particular two-SSP composite model reproduced the measured line indices for different relative contributions of the two SSPs to the total flux.  For example, in Figure\,\ref{fig:in-in}, it is clearly apparent that the two-SSP composite models are in better agreement with the index--index measurements than the lone SSP model (recall that model loci in Figure\,\ref{fig:in-in} reflect an age change for the lone SSP, whereas they reflect different relative contributions to the total flux from the two SSPs in the two-SSP composite models).  Furthermore, the model locus for $Z = 1.5 \, Z_\sun$ (dark green curve) is in better overall agreement with the Mg\,2 versus $\langle \rm Fe \rangle$ indices than the model locus for $Z = Z_\sun$ $\alpha$-enhanced (brown curve).  The superior match of the model locus for $Z = 1.5 \, Z_\sun$ (dark green curve) is also apparent for the H8 versus $\langle \rm Fe \rangle$ indices given that the H8 index is likely underestimated owing to a nebular contribution, and the $Z= 1.5\,Z_\odot$ model provides an indistinguishably good match as the $Z = Z_\sun$ $\alpha$-enhanced (magenta curve) for the remaining index--index plots shown in Figure\,\ref{fig:in-in}.

In this way, we found simultaneous best-fit ages of 0.15\,Gyr for the younger stellar population based primarily on its good fit to the relatively shallow Ca\,K line at the inner central spiral ($-4\arcsec$ from the center, where the younger stellar population is most dominant), and an age of 10\,Gyr for the older stellar population regardless of whether $Z = 1.5 \, Z_\odot$ or $Z =\, Z_\odot$ $\alpha$-enhanced based primarily on Mg\,{\it b} and $\langle$Fe$\rangle$ lines beyond the outer arm (at $-15\arcsec$, where the older population is most dominant among the regions considered).  The age of the older population, unlike that of the younger population, is not well constrained by the index--index plots alone, although a visual inspection of how well the two-SSP composite model spectra reproduce the measured spectra indicate a best-fit age close to 10\,Gyr.  The contribution from the older population is not negligible even at the inner central spiral, amounting to $\sim$30\% of the light just shortward of the Mg\,$b$ line to $\sim$10\% of the light around the middle of the blue correlation range where the younger population dominates the light (Figure\,\ref{fig:composite_spec}); if not taken into account, the younger stellar population would have an older age if inferred simply from the measured H8/Ca\,K index ratio (Appendix\,\ref{subsubsec:population:indices:spatial_trends}) at this position.  The best-fit fractional fluxes (to the nearest 10\%) for each subregion and for either $Z=1.5 \, Z_\odot$ or $Z=Z_\odot$ $\alpha$-enhanced, along with their corresponding model-predicted line indices and $B-V$ color, are listed in Table\,\ref{tab:composite_best-fit}.

\section{M\lowercase{g}\,I contamination of H9 line}\label{appendix:h9}

As the SSP calibration curves for the H8 and H9 line indices shown in Figure\,\ref{fig:sing_calib} indicate, the strengths of both Balmer lines are predicted to change in a very similar way with age, and should therefore to be correlated.  As can be seen in Figure\,\ref{fig:in-in}, out to a radius of about $-6\arcsec$, the H8 and H9 indices are indeed correlated.  At larger radii, however, these line indices exhibit an anticorrelation as the strength of the H8 line continues to decrease monotonically outwards, whereas the strength of the H9 line reverses to increase outwards as can be seen in Figure\,\ref{fig:pos_index_color}.  Such an anticorrelation between these two line indices cannot be explained either by changing the age of an SSP or by two-SSP composite models (Section\,\ref{subsec:population:composite}).

In the SSP calibration curves for line indices shown in Figure\,\ref{fig:sing_calib}, both the H8 and H9 line indices decrease beyond an age of $\sim$0.35\,Gyr.  The H9 line index flattens to a moderately large value ($\sim$4\,\AA) at ages beyond $\sim$1\,Gyr, whereas the H8 line index continues to decrease much more steeply with increasing age.  This flattening of the H9 line index is caused by metal absorption, primarily $\rm{Mg\,I}$ \citep[e.g.,][]{Bica,Serote2}, at wavelengths very close to H9.  Absorption by $\rm{Mg\,I}$ can be clearly seen in the spectra of late F-type stars, as showcased by \citet{Kauffmann,Dokkum,Wild2007,Demarco,Nantais} for both stellar and galaxy spectra around these features.  Contamination by metal absorption lines makes H9 less accurate, when compared with H8, as an indicator of stellar age, particularly at the middle central spiral and on the outer arm where contamination by metal absorption (from the old stellar population comprising the main body of NGC\,1275) is expected to be more severe according to the two-SSP composite model (Section\,\ref{subsec:population:composite}).  Hence, we focus more on the H8 than the H9 line in our work, and allow inconsistencies between the best-fit two-SSP composite model for the predicted and the measured H9 line indices when evaluating the overall goodness of fit.
\bibliography{sample63}{}

\begin{thebibliography}{}
\expandafter\ifx\csname natexlab\endcsname\relax\def\natexlab#1{#1}\fi
\providecommand{\url}[1]{\href{#1}{#1}}
\providecommand{\dodoi}[1]{doi:~\href{http://doi.org/#1}{\nolinkurl{#1}}}
\providecommand{\doeprint}[1]{\href{http://ascl.net/#1}{\nolinkurl{http://ascl.net/#1}}}
\providecommand{\doarXiv}[1]{\href{https://arxiv.org/abs/#1}{\nolinkurl{https://arxiv.org/abs/#1}}}

\bibitem[{{Bica} \& {Alloin}(1986)}]{Bica}
{Bica}, E., \& {Alloin}, D. 1986, \aap, 162, 21

\bibitem[{{Brockamp} {et~al.}(2014){Brockamp}, {K{\"u}pper}, {Thies},
  {Baumgardt}, \& {Kroupa}}]{Brockamp}
{Brockamp}, M., {K{\"u}pper}, A.~H.~W., {Thies}, I., {Baumgardt}, H., \&
  {Kroupa}, P. 2014, \mnras, 441, 150, \dodoi{10.1093/mnras/stu562}

\bibitem[{{Brodie} {et~al.}(1998){Brodie}, {Schroder}, {Huchra}, {Phillips},
  {Kissler-Patig}, \& {Forbes}}]{brodie}
{Brodie}, J.~P., {Schroder}, L.~L., {Huchra}, J.~P., {et~al.} 1998, \aj, 116,
  691, \dodoi{10.1086/300442}

\bibitem[{{Bruzual A.}(1983)}]{Bruzual}
{Bruzual A.}, G. 1983, \apj, 273, 105, \dodoi{10.1086/161352}

\bibitem[{{Caldwell} {et~al.}(2011){Caldwell}, {Schiavon}, {Morrison}, {Rose},
  \& {Harding}}]{mFe}
{Caldwell}, N., {Schiavon}, R., {Morrison}, H., {Rose}, J.~A., \& {Harding}, P.
  2011, \aj, 141, 61, \dodoi{10.1088/0004-6256/141/2/61}

\bibitem[{{Cardiel}(2010)}]{indexf}
{Cardiel}, N. 2010, {indexf: Line-strength Indices in Fully Calibrated FITS
  Spectra}, {Astrophysics Source Code Library}.
\newblock \doeprint{1010.046}

\bibitem[{{Carlson} \& {Holtzman}(2001)}]{Carlson2001}
{Carlson}, M.~N., \& {Holtzman}, J.~A. 2001, \pasp, 113, 1522,
  \dodoi{10.1086/324417}

\bibitem[{{Carlson} {et~al.}(1998){Carlson}, {Holtzman}, {Watson}, {Grillmair},
  {Mould}, {Ballester}, {Burrows}, {Clarke}, {Crisp}, {Evans}, {Gallagher},
  {Griffiths}, {Hester}, {Hoessel}, {Scowen}, {Stapelfeldt}, {Trauger}, \&
  {Westphal}}]{Carlson}
{Carlson}, M.~N., {Holtzman}, J.~A., {Watson}, A.~M., {et~al.} 1998, \aj, 115,
  1778, \dodoi{10.1086/300334}

\bibitem[{{Conselice} {et~al.}(2001){Conselice}, {Gallagher}, \&
  {Wyse}}]{conselice}
{Conselice}, C.~J., {Gallagher}, John~S., I., \& {Wyse}, R. F.~G. 2001, \aj,
  122, 2281, \dodoi{10.1086/323534}

\bibitem[{{Crawford} \& {Fabian}(1993)}]{Crawfod93}
{Crawford}, C.~S., \& {Fabian}, A.~C. 1993, \mnras, 265, 431,
  \dodoi{10.1093/mnras/265.2.431}

\bibitem[{{Demarco} {et~al.}(2010){Demarco}, {Gobat}, {Rosati}, {Lidman},
  {Rettura}, {Nonino}, {van der Wel}, {Jee}, {Blakeslee}, {Ford}, \&
  {Postman}}]{Demarco}
{Demarco}, R., {Gobat}, R., {Rosati}, P., {et~al.} 2010, \apj, 725, 1252,
  \dodoi{10.1088/0004-637X/725/1/1252}

\bibitem[{{Fabian} {et~al.}(2008){Fabian}, {Johnstone}, {Sanders}, {Conselice},
  {Crawford}, {Gallagher}, \& {Zweibel}}]{Fabian}
{Fabian}, A.~C., {Johnstone}, R.~M., {Sanders}, J.~S., {et~al.} 2008, \nat,
  454, 968, \dodoi{10.1038/nature07169}

\bibitem[{{Fabian} {et~al.}(2006){Fabian}, {Sanders}, {Taylor}, {Allen},
  {Crawford}, {Johnstone}, \& {Iwasawa}}]{Fabian2006}
{Fabian}, A.~C., {Sanders}, J.~S., {Taylor}, G.~B., {et~al.} 2006, \mnras, 366,
  417, \dodoi{10.1111/j.1365-2966.2005.09896.x}

\bibitem[{{Ferruit} \& {Pecontal}(1994)}]{ferruit}
{Ferruit}, P., \& {Pecontal}, E. 1994, \aap, 288, 65

\bibitem[{{Gendron-Marsolais} {et~al.}(2018){Gendron-Marsolais},
  {Hlavacek-Larrondo}, {Martin}, {Drissen}, {McDonald}, {Fabian}, {Edge},
  {Hamer}, {McNamara}, \& {Morrison}}]{Gendron}
{Gendron-Marsolais}, M., {Hlavacek-Larrondo}, J., {Martin}, T.~B., {et~al.}
  2018, \mnras, 479, L28, \dodoi{10.1093/mnrasl/sly084}

\bibitem[{{Hatch} {et~al.}(2006){Hatch}, {Crawford}, {Johnstone}, \&
  {Fabian}}]{Hatch06}
{Hatch}, N.~A., {Crawford}, C.~S., {Johnstone}, R.~M., \& {Fabian}, A.~C. 2006,
  \mnras, 367, 433, \dodoi{10.1111/j.1365-2966.2006.09970.x}

\bibitem[{{Heckman} {et~al.}(1985){Heckman}, {Illingworth}, {Miley}, \& {van
  Breugel}}]{Heckman}
{Heckman}, T.~M., {Illingworth}, G.~D., {Miley}, G.~K., \& {van Breugel},
  W.~J.~M. 1985, \apj, 299, 41, \dodoi{10.1086/163681}

\bibitem[{{Ho} {et~al.}(2009){Ho}, {Lim}, \& {Trung}}]{Ho_jeremy}
{Ho}, I.~T., {Lim}, J., \& {Trung}, D.-V. 2009, \apj, 698, 1191,
  \dodoi{10.1088/0004-637X/698/2/1191}

\bibitem[{{Holtzman} {et~al.}(1992){Holtzman}, {Faber}, {Shaya}, {Lauer},
  {Groth}, {Hunter}, {Baum}, {Ewald}, {Hester}, {Light}, {Lynds}, {O'Neil}, \&
  {Westphal}}]{star_clus_ID}
{Holtzman}, J.~A., {Faber}, S.~M., {Shaya}, E.~J., {et~al.} 1992, \aj, 103,
  691, \dodoi{10.1086/116094}

\bibitem[{{Johnstone} \& {Fabian}(1988)}]{johnstone}
{Johnstone}, R.~M., \& {Fabian}, A.~C. 1988, \mnras, 233, 581,
  \dodoi{10.1093/mnras/233.3.581}

\bibitem[{{Johnstone} {et~al.}(1987){Johnstone}, {Fabian}, \&
  {Nulsen}}]{Johnstone87}
{Johnstone}, R.~M., {Fabian}, A.~C., \& {Nulsen}, P.~E.~J. 1987, \mnras, 224,
  75, \dodoi{10.1093/mnras/224.1.75}

\bibitem[{{Kauffmann} {et~al.}(2003){Kauffmann}, {Heckman}, {White}, {Charlot},
  {Tremonti}, {Brinchmann}, {Bruzual}, {Peng}, {Seibert}, {Bernardi},
  {Blanton}, {Brinkmann}, {Castander}, {Cs{\'a}bai}, {Fukugita}, {Ivezic},
  {Munn}, {Nichol}, {Padmanabhan}, {Thakar}, {Weinberg}, \& {York}}]{Kauffmann}
{Kauffmann}, G., {Heckman}, T.~M., {White}, S. D.~M., {et~al.} 2003, \mnras,
  341, 33, \dodoi{10.1046/j.1365-8711.2003.06291.x}

\bibitem[{{Koss} {et~al.}(2017){Koss}, {Trakhtenbrot}, {Ricci}, {Lamperti},
  {Oh}, {Berney}, {Schawinski}, {Balokovi{\'c}}, {Baronchelli}, {Crenshaw},
  {Fischer}, {Gehrels}, {Harrison}, {Hashimoto}, {Hogg}, {Ichikawa}, {Masetti},
  {Mushotzky}, {Sartori}, {Stern}, {Treister}, {Ueda}, {Veilleux}, \&
  {Winter}}]{Koss}
{Koss}, M., {Trakhtenbrot}, B., {Ricci}, C., {et~al.} 2017, \apj, 850, 74,
  \dodoi{10.3847/1538-4357/aa8ec9}

\bibitem[{{Kroupa}(2001)}]{Kroupa}
{Kroupa}, P. 2001, \mnras, 322, 231, \dodoi{10.1046/j.1365-8711.2001.04022.x}

\bibitem[{{Kurtz} \& {Mink}(1998)}]{rvsao}
{Kurtz}, M.~J., \& {Mink}, D.~J. 1998, \pasp, 110, 934, \dodoi{10.1086/316207}

\bibitem[{{Leonardi} \& {Rose}(1996)}]{Leonardi}
{Leonardi}, A.~J., \& {Rose}, J.~A. 1996, \aj, 111, 182, \dodoi{10.1086/117772}

\bibitem[{{Lim} {et~al.}(2008){Lim}, {Ao}, \& {Trung}}]{SMA_Jeremy}
{Lim}, J., {Ao}, Y., \& {Trung}, D.-V. 2008, \apj, 672, 252,
  \dodoi{10.1086/523664}

\bibitem[{{Lim} {et~al.}(2012){Lim}, {Ohyama}, {Chi-Hung}, {Trung}, \&
  {Shiang-Yu}}]{Jeremy2012}
{Lim}, J., {Ohyama}, Y., {Chi-Hung}, Y., {Trung}, D.-V., \& {Shiang-Yu}, W.
  2012, \apj, 744, 112, \dodoi{10.1088/0004-637X/744/2/112}

\bibitem[{{Lim} {et~al.}(2021){Lim}, {Ohyama}, {Wong}, \& {Yeung}}]{paper2}
{Lim}, J., {Ohyama}, Y., {Wong}, E., \& {Yeung}, M. C.~H. 2021, \apj|in press

\bibitem[{{Lim} {et~al.}(2017){Lim}, {Trung}, {Vrtilek}, {David}, \&
  {Forman}}]{Jeremy2017}
{Lim}, J., {Trung}, D.-V., {Vrtilek}, J., {David}, L.~P., \& {Forman}, W. 2017,
  \apj, 850, 31, \dodoi{10.3847/1538-4357/aa9275}

\bibitem[{{Lim} {et~al.}(2020){Lim}, {Wong}, {Ohyama}, {Broadhurst}, \&
  {Medezinski}}]{Jeremy}
{Lim}, J., {Wong}, E., {Ohyama}, Y., {Broadhurst}, T., \& {Medezinski}, E.
  2020, \natas, 4, 153, \dodoi{10.1038/s41550-019-0909-6}

\bibitem[{{Loubser} {et~al.}(2009){Loubser}, {S{\'a}nchez-Bl{\'a}zquez},
  {Sansom}, \& {Soechting}}]{loubser09}
{Loubser}, S.~I., {S{\'a}nchez-Bl{\'a}zquez}, P., {Sansom}, A.~E., \&
  {Soechting}, I.~K. 2009, \mnras, 398, 133,
  \dodoi{10.1111/j.1365-2966.2009.15171.x}

\bibitem[{{Marcillac} {et~al.}(2006){Marcillac}, {Elbaz}, {Charlot}, {Liang},
  {Hammer}, {Flores}, {Cesarsky}, \& {Pasquali}}]{H8H9window}
{Marcillac}, D., {Elbaz}, D., {Charlot}, S., {et~al.} 2006, \aap, 458, 369,
  \dodoi{10.1051/0004-6361:20064996}

\bibitem[{{McNamara} \& {O'Connell}(1992)}]{McNamara}
{McNamara}, B.~R., \& {O'Connell}, R.~W. 1992, \apj, 393, 579,
  \dodoi{10.1086/171529}

\bibitem[{{Minkowski}(1957)}]{Minkowski_first}
{Minkowski}, R. 1957, in Radio astronomy, ed. H.~C. {van de Hulst}, Vol.~4, 107

\bibitem[{Nagai {et~al.}(2019)Nagai, Onishi, Kawakatu, Fujita, Kino, Fukazawa,
  Lim, Forman, Vrtilek, Nakanishi, Noda, Asada, Wajima, Ohyama, David, \&
  Daikuhara}]{Nagai}
Nagai, H., Onishi, K., Kawakatu, N., {et~al.} 2019, \apj, 883, 193,
  \dodoi{10.3847/1538-4357/ab3e6e}

\bibitem[{{Nantais} {et~al.}(2013){Nantais}, {Rettura}, {Lidman}, {Demarco},
  {Gobat}, {Rosati}, \& {Jee}}]{Nantais}
{Nantais}, J.~B., {Rettura}, A., {Lidman}, C., {et~al.} 2013, \aap, 556, A112,
  \dodoi{10.1051/0004-6361/201321877}

\bibitem[{{Nelson} \& {Whittle}(1995)}]{nelson}
{Nelson}, C.~H., \& {Whittle}, M. 1995, \apjs, 99, 67, \dodoi{10.1086/192179}

\bibitem[{{Norgaard-Nielsen} {et~al.}(1993){Norgaard-Nielsen}, {Goudfrooij},
  {Jorgensen}, \& {Hansen}}]{Norgaard2}
{Norgaard-Nielsen}, H.~U., {Goudfrooij}, P., {Jorgensen}, H.~E., \& {Hansen},
  L. 1993, \aap, 279, 61

\bibitem[{{Oke} {et~al.}(1995){Oke}, {Cohen}, {Carr}, {Cromer}, {Dingizian},
  {Harris}, {Labrecque}, {Lucinio}, {Schaal}, {Epps}, \& {Miller}}]{LRIS}
{Oke}, J.~B., {Cohen}, J.~G., {Carr}, M., {et~al.} 1995, \pasp, 107, 375,
  \dodoi{10.1086/133562}

\bibitem[{{Onori} {et~al.}(2017){Onori}, {Ricci}, {La Franca}, {Bianchi},
  {Bongiorno}, {Brusa}, {Fiore}, {Maiolino}, {Marconi}, {Sani}, \&
  {Vignali}}]{Onori}
{Onori}, F., {Ricci}, F., {La Franca}, F., {et~al.} 2017, \mnras, 468, L97,
  \dodoi{10.1093/mnrasl/slx032}

\bibitem[{{Penny} {et~al.}(2012){Penny}, {Forbes}, \& {Conselice}}]{penny}
{Penny}, S.~J., {Forbes}, D.~A., \& {Conselice}, C.~J. 2012, \mnras, 422, 885,
  \dodoi{10.1111/j.1365-2966.2012.20669.x}

\bibitem[{{Pietrinferni} {et~al.}(2004){Pietrinferni}, {Cassisi}, {Salaris}, \&
  {Castelli}}]{Pietrinferni1}
{Pietrinferni}, A., {Cassisi}, S., {Salaris}, M., \& {Castelli}, F. 2004, \apj,
  612, 168, \dodoi{10.1086/422498}

\bibitem[{{Pietrinferni} {et~al.}(2006){Pietrinferni}, {Cassisi}, {Salaris}, \&
  {Castelli}}]{Pietrinferni2}
---. 2006, \apj, 642, 797, \dodoi{10.1086/501344}

\bibitem[{{Riffel} {et~al.}(2020){Riffel}, {Storchi-Bergmann}, {Zakamska}, \&
  {Riffel}}]{Riffel}
{Riffel}, R.~A., {Storchi-Bergmann}, T., {Zakamska}, N.~L., \& {Riffel}, R.
  2020, \mnras, 496, 4857, \dodoi{10.1093/mnras/staa1922}

\bibitem[{{Romanishin}(1986)}]{Romanishin1}
{Romanishin}, W. 1986, \apj, 301, 675, \dodoi{10.1086/163933}

\bibitem[{{Romanishin}(1987)}]{Romanishin2}
---. 1987, \apjl, 323, L113, \dodoi{10.1086/185068}

\bibitem[{{Rose}(1984)}]{Rose}
{Rose}, J.~A. 1984, \aj, 89, 1238, \dodoi{10.1086/113618}

\bibitem[{{Rubin} {et~al.}(1977){Rubin}, {Ford}, {Peterson}, \& {Oort}}]{Rubin}
{Rubin}, V.~C., {Ford}, W.~K., J., {Peterson}, C.~J., \& {Oort}, J.~H. 1977,
  \apj, 211, 693, \dodoi{10.1086/154979}

\bibitem[{{Salom{\'e}} {et~al.}(2011){Salom{\'e}}, {Combes}, {Revaz}, {Downes},
  {Edge}, \& {Fabian}}]{salome2011}
{Salom{\'e}}, P., {Combes}, F., {Revaz}, Y., {et~al.} 2011, \aap, 531, A85,
  \dodoi{10.1051/0004-6361/200811333}

\bibitem[{{Salom{\'e}} {et~al.}(2008{\natexlab{a}}){Salom{\'e}}, {Combes},
  {Revaz}, {Edge}, {Hatch}, {Fabian}, \& {Johnstone}}]{Salome08b}
---. 2008{\natexlab{a}}, \aap, 484, 317, \dodoi{10.1051/0004-6361:200809493}

\bibitem[{{Salom{\'e}} {et~al.}(2008{\natexlab{b}}){Salom{\'e}}, {Revaz},
  {Combes}, {Pety}, {Downes}, {Edge}, \& {Fabian}}]{Salome08a}
{Salom{\'e}}, P., {Revaz}, Y., {Combes}, F., {et~al.} 2008{\natexlab{b}}, \aap,
  483, 793, \dodoi{10.1051/0004-6361:200809412}

\bibitem[{{Salom{\'e}} {et~al.}(2006){Salom{\'e}}, {Combes}, {Edge},
  {Crawford}, {Erlund}, {Fabian}, {Hatch}, {Johnstone}, {Sanders}, \&
  {Wilman}}]{Salome06}
{Salom{\'e}}, P., {Combes}, F., {Edge}, A.~C., {et~al.} 2006, \aap, 454, 437,
  \dodoi{10.1051/0004-6361:20054745}

\bibitem[{{S{\'a}nchez-Bl{\'a}zquez} {et~al.}(2006){S{\'a}nchez-Bl{\'a}zquez},
  {Gorgas}, \& {Cardiel}}]{Sanchez}
{S{\'a}nchez-Bl{\'a}zquez}, P., {Gorgas}, J., \& {Cardiel}, N. 2006, \aap, 457,
  823, \dodoi{10.1051/0004-6361:20064846}

\bibitem[{{Sanders} \& {Fabian}(2007)}]{Sanders}
{Sanders}, J.~S., \& {Fabian}, A.~C. 2007, \mnras, 381, 1381,
  \dodoi{10.1111/j.1365-2966.2007.12347.x}

\bibitem[{{Sani} {et~al.}(2018){Sani}, {Ricci}, {La Franca}, {Bianchi},
  {Bongiorno}, {Brusa}, {Marconi}, {Onori}, {Shankar}, \& {Vignali}}]{sani}
{Sani}, E., {Ricci}, F., {La Franca}, F., {et~al.} 2018, Frontiers in Astronomy
  and Space Sciences, 5, 2, \dodoi{10.3389/fspas.2018.00002}

\bibitem[{{Scharw{\"a}chter} {et~al.}(2013){Scharw{\"a}chter}, {McGregor},
  {Dopita}, \& {Beck}}]{scharwchter}
{Scharw{\"a}chter}, J., {McGregor}, P.~J., {Dopita}, M.~A., \& {Beck}, T.~L.
  2013, \mnras, 429, 2315, \dodoi{10.1093/mnras/sts502}

\bibitem[{{Serote Roos} \& {Gon{\c{c}}alves}(2004)}]{Serote2}
{Serote Roos}, M., \& {Gon{\c{c}}alves}, A.~C. 2004, \aap, 413, 91,
  \dodoi{10.1051/0004-6361:20031493}

\bibitem[{{Thomas} {et~al.}(2003){Thomas}, {Maraston}, \& {Bender}}]{Thomas}
{Thomas}, D., {Maraston}, C., \& {Bender}, R. 2003, \mnras, 339, 897,
  \dodoi{10.1046/j.1365-8711.2003.06248.x}

\bibitem[{{Tody}(1986)}]{IRAF}
{Tody}, D. 1986, in Society of Photo-Optical Instrumentation Engineers (SPIE)
  Conference Series, Vol. 627, Instrumentation in astronomy VI, ed. D.~L.
  {Crawford}, 733, \dodoi{10.1117/12.968154}

\bibitem[{{Tody}(1993)}]{IRAF2}
{Tody}, D. 1993, in Astronomical Society of the Pacific Conference Series,
  Vol.~52, Astronomical Data Analysis Software and Systems II, ed. R.~J.
  {Hanisch}, R.~J.~V. {Brissenden}, \& J.~{Barnes}, 173

\bibitem[{{Trager} {et~al.}(2000{\natexlab{a}}){Trager}, {Faber}, {Worthey}, \&
  {Gonz{\'a}lez}}]{Trager1}
{Trager}, S.~C., {Faber}, S.~M., {Worthey}, G., \& {Gonz{\'a}lez}, J.~J.
  2000{\natexlab{a}}, \aj, 119, 1645, \dodoi{10.1086/301299}

\bibitem[{{Trager} {et~al.}(2000{\natexlab{b}}){Trager}, {Faber}, {Worthey}, \&
  {Gonz{\'a}lez}}]{Trager2}
---. 2000{\natexlab{b}}, \aj, 120, 165, \dodoi{10.1086/301442}

\bibitem[{{Trager} {et~al.}(1998){Trager}, {Worthey}, {Faber}, {Burstein}, \&
  {Gonz{\'a}lez}}]{Trager98}
{Trager}, S.~C., {Worthey}, G., {Faber}, S.~M., {Burstein}, D., \&
  {Gonz{\'a}lez}, J.~J. 1998, \apjs, 116, 1, \dodoi{10.1086/313099}

\bibitem[{{van den Bergh}(1972)}]{Sidney}
{van den Bergh}, S. 1972, \jrasc, 66, 237

\bibitem[{{van Dokkum} \& {Stanford}(2003)}]{Dokkum}
{van Dokkum}, P.~G., \& {Stanford}, S.~A. 2003, \apj, 585, 78,
  \dodoi{10.1086/345989}

\bibitem[{{Vazdekis} {et~al.}(2010){Vazdekis}, {S{\'a}nchez-Bl{\'a}zquez},
  {Falc{\'o}n-Barroso}, {Cenarro}, {Beasley}, {Cardiel}, {Gorgas}, \&
  {Peletier}}]{miles_first}
{Vazdekis}, A., {S{\'a}nchez-Bl{\'a}zquez}, P., {Falc{\'o}n-Barroso}, J.,
  {et~al.} 2010, \mnras, 404, 1639, \dodoi{10.1111/j.1365-2966.2010.16407.x}

\bibitem[{{Vazdekis} {et~al.}(2015){Vazdekis}, {Coelho}, {Cassisi},
  {Ricciardelli}, {Falc{\'o}n-Barroso}, {S{\'a}nchez-Bl{\'a}zquez}, {La
  Barbera}, {Beasley}, \& {Pietrinferni}}]{miles}
{Vazdekis}, A., {Coelho}, P., {Cassisi}, S., {et~al.} 2015, \mnras, 449, 1177,
  \dodoi{10.1093/mnras/stv151}

\bibitem[{{von der Linden} {et~al.}(2007){von der Linden}, {Best}, {Kauffmann},
  \& {White}}]{von_der_Linden}
{von der Linden}, A., {Best}, P.~N., {Kauffmann}, G., \& {White}, S. D.~M.
  2007, \mnras, 379, 867, \dodoi{10.1111/j.1365-2966.2007.11940.x}

\bibitem[{{Wild} {et~al.}(2007){Wild}, {Kauffmann}, {Heckman}, {Charlot},
  {Lemson}, {Brinchmann}, {Reichard}, \& {Pasquali}}]{Wild2007}
{Wild}, V., {Kauffmann}, G., {Heckman}, T., {et~al.} 2007, \mnras, 381, 543,
  \dodoi{10.1111/j.1365-2966.2007.12256.x}

\bibitem[{{Worthey} {et~al.}(1994){Worthey}, {Faber}, {Gonzalez}, \&
  {Burstein}}]{Lick_ori}
{Worthey}, G., {Faber}, S.~M., {Gonzalez}, J.~J., \& {Burstein}, D. 1994,
  \apjs, 94, 687, \dodoi{10.1086/192087}

\bibitem[{{Worthey} \& {Ottaviani}(1997)}]{Lick}
{Worthey}, G., \& {Ottaviani}, D.~L. 1997, \apjs, 111, 377,
  \dodoi{10.1086/313021}

\bibitem[{{Yu} {et~al.}(2015){Yu}, {Lim}, {Ohyama}, {Chan}, \&
  {Broadhurst}}]{HVS}
{Yu}, A. P.~Y., {Lim}, J., {Ohyama}, Y., {Chan}, J. C.~C., \& {Broadhurst}, T.
  2015, \apj, 814, 101, \dodoi{10.1088/0004-637X/814/2/101}

\end{thebibliography}
\bibliographystyle{aasjournal}
\end{document}